\documentclass[12pt]{article}
\usepackage{amsfonts}
\usepackage{amsmath}
\usepackage{amssymb}

\allowdisplaybreaks
\begin{document}

\author{C. Bizdadea\thanks{%
E-mail address: bizdadea@central.ucv.ro}, E. M. Cioroianu\thanks{%
E-mail address: manache@central.ucv.ro}, A. Danehkar\thanks{%
Present address: School of Mathematics and Physics, Queen's University,
Belfast BT7 1NN, UK. E-mail address: adanehkar01@qub.ac.uk}, \and M.
Iordache, S. O. Saliu\thanks{%
E-mail address: osaliu@central.ucv.ro}, S. C. S\u{a}raru\thanks{%
E-mail address: scsararu@central.ucv.ro} \\
Faculty of Physics, University of Craiova\\
13 Al. I. Cuza Str., Craiova 200585, Romania}
\title{Consistent interactions of dual linearized gravity in $D=5$:
couplings with a topological BF model}
\date{}
\maketitle

\begin{abstract}
Under some plausible assumptions, we find that the dual formulation of
linearized gravity in $D=5$ can be nontrivially coupled to the topological
BF model in such a way that the interacting theory exhibits a deformed gauge
algebra and some deformed, on-shell reducibility relations. Moreover, the
tensor field with the mixed symmetry $(2,1)$ gains some shift gauge
transformations with parameters from the BF sector.


\end{abstract}

\section{Introduction}

Topological field theories~\cite{birmingham91,labastida97} are important in
view of the fact that certain interacting, non-Abelian versions are related
to a Poisson structure algebra~\cite{stroblspec} present in various versions
of Poisson sigma models~\cite{psmikeda94}--\cite{psmcattaneo2001}, which are
known to be useful at the study of two-dimensional gravity~\cite{grav2teit83}%
--\cite{grav2grumvassil02} (for a detailed approach, see~\cite{grav2strobl00}%
). It is well known that pure three-dimensional gravity is just a BF theory.
Moreover, in higher dimensions general relativity and supergravity in
Ashtekar formalism may also be formulated as topological BF theories with
some extra constraints~\cite{ezawa}--\cite{ling}. In view of these results,
it is important to know the self-interactions in BF theories as well as the
couplings between BF models and other theories. This problem has been
considered in literature in relation with self-interactions in various
classes of BF models~\cite{defBFizawa2000}--\cite{defBFijmpajuvi06} and
couplings to other (matter or gauge) fields~\cite{defBFikeda03}--\cite%
{defBFRSepjc} by using the powerful BRST cohomological reformulation of the
problem of constructing consistent interactions within the Lagrangian~\cite%
{deflag,17and5} or the Hamiltonian~\cite{defham} setting, based on the
computation of local BRST cohomology~\cite{gen1}--\cite{gen}. Other aspects
concerning interacting, topological BF models can be found in~\cite%
{otherBFikeda02} and~\cite{otherBFikedaizawa04}.

On the other hand, tensor fields in \textquotedblleft
exotic\textquotedblright\ representations of the Lorentz group,
characterized by a mixed Young symmetry type~\cite{curt2}--\cite{zinov1},
held the attention lately on some important issues, like the dual
formulation of field theories of spin two or higher~\cite{dualsp1}--\cite%
{dualsp2II}, the impossibility of consistent cross-interactions in the dual
formulation of linearized gravity~\cite{lingr}, a Lagrangian first-order
approach~\cite{zinov2,zinov3} to some classes of massless or partially
massive mixed symmetry type tensor gauge fields, suggestively resembling to
the tetrad formalism of General Relativity, or the derivation of some exotic
gravitational interactions~\cite{boulangerCQG,ancoPRD}. An important matter
related to mixed symmetry type tensor fields is the study of their
consistent interactions, among themselves as well as with other gauge
theories~\cite{high1}--\cite{dcjpa}.

The purpose of this paper is to investigate the consistent interactions in $%
D=5$ between a massless tensor gauge field with the mixed symmetry of a
two-column Young diagram of the type $(2,1)$ and an Abelian BF model with a
maximal field spectrum (a scalar field, two sorts of one-forms, two types of
two-forms and a three-form). It is worth mentioning the duality of a free
massless tensor gauge field with the mixed symmetry $(2,1)$ to the
Pauli--Fierz theory in $D=5$ dimensions. In view of this feature, we can
state that our paper searches the consistent couplings in $D=5$ between the
dual formulation of linearized gravity and a topological BF model. Our
analysis relies on the deformation of the solution to the master equation by
means of cohomological techniques with the help of the local BRST
cohomology. We mention that the self-interactions in the $(2,1)$ sector have
been investigated in~\cite{lingr} and the couplings in $D=5$ that can be
added to an Abelian BF model with a maximal field spectrum have been
constructed in~\cite{juviBFD5}.

Under the hypotheses of analyticity in the coupling constant, spacetime
locality, Lorentz covariance, and Poincar\'{e} invariance of the
deformations, combined with the preservation of the number of derivatives on
each field, we find a deformation of the solution to the master equation
that provides nontrivial cross-couplings. The emerging Lagrangian action
contains mixing-component terms of order one in the coupling constant that
couple the massless tensor field with the mixed symmetry $(2,1)$ mainly to
one of the two-forms and to the three-form from the BF sector. Also, it is
interesting to note the appearance of some self-interactions in the BF
sector at order two in the coupling constant that are strictly due to the
presence of the tensor field with the mixed symmetry $(2,1)$ (they all
vanish in its absence). The gauge transformations of all fields are deformed
and, in addition, some of them include gauge parameters from the
complementary sector. This is the first known case where the gauge
transformations of the tensor field with the mixed symmetry $(2,1)$ do
change with respect to the free ones (by shifts in some of the BF gauge
parameters). The gauge algebra and the reducibility structure of the coupled
model are strongly modified during the deformation procedure, becoming open
and respectively on-shell, by contrast to the free theory, whose gauge
algebra is Abelian and the reducibility relations hold off-shell. Our result
is important because dual formulations of linearized gravity have proved to
be extremely rigid in allowing consistent interactions to themselves as well
as to many matter or gauge theories. Actually, we think that this is the
first time when a massless tensor field with the mixed symmetry $(k,1)$
allows consistent interactions that fulfill all the working hypotheses
precisely in the dimension $D=k+3$ where it becomes dual to the Pauli--Fierz
theory.

\section{The free theory: Lagrangian, gauge symmetries and BRST differential
\label{free}}

The starting point is a free theory in $D=5$, whose Lagrangian action is
written as the sum between the Lagrangian action of an Abelian BF model with
a maximal field spectrum (a single scalar field $\varphi $, two types of
one-forms $H^{\mu }$ and $V_{\mu }$, two kinds of two-forms $B^{\mu \nu }$
and $\phi _{\mu \nu }$, and one three-form $K^{\mu \nu \rho }$) and the
Lagrangian action of a free, massless tensor field with the mixed symmetry $%
(2,1)$ $t_{\mu \nu |\alpha }$ (meaning it is antisymmetric in its first two
indices $t_{\mu \nu |\alpha }=-t_{\nu \mu |\alpha }$ and fulfills the
identity $t_{[\mu \nu |\alpha ]}\equiv 0$)%
\begin{eqnarray}
S_{0}^{\mathrm{L}}\left[ \Phi ^{\alpha _{0}}\right] &=&\int d^{5}x\left[
H^{\mu }\partial _{\mu }\varphi +\tfrac{1}{2}B^{\mu \nu }\partial _{\lbrack
\mu }V_{\nu ]}+\tfrac{1}{3}K^{\mu \nu \rho }\partial _{\lbrack \mu }\phi
_{\nu \rho ]}\right.  \notag \\
&&\left. -\tfrac{1}{12}\left( F_{\mu \nu \rho |\alpha }F^{\mu \nu \rho
|\alpha }-3F_{\mu \nu }F^{\mu \nu }\right) \right]  \notag \\
&\equiv &\int d^{5}x\left( \mathcal{L}_{0}^{\mathrm{BF}}+\mathcal{L}_{0}^{%
\mathrm{t}}\right) ,  \label{f1}
\end{eqnarray}%
where we used the notations%
\begin{eqnarray}
\Phi ^{\alpha _{0}} &=&\left( \varphi ,H^{\mu },V_{\mu },B^{\mu \nu },\phi
_{\mu \nu },K^{\mu \nu \rho },t_{\mu \nu |\alpha }\right) ,  \label{f10a} \\
F_{\mu \nu \rho |\alpha } &=&\partial _{\lbrack \mu }t_{\nu \rho ]|\alpha
},\qquad F_{\mu \nu }=\sigma ^{\rho \alpha }F_{\mu \nu \rho |\alpha }.
\label{nott1}
\end{eqnarray}%
Everywhere in this paper the notations $[\mu \nu \ldots \rho ]$ and $(\mu
\nu \ldots \rho )$ signify complete antisymmetry and respectively complete
symmetry with respect to the (Lorentz) indices between brackets, with the
conventions that the minimum number of terms is always used and the result
is never divided by the number of terms. It is convenient to work with the
Minkowski metric tensor of `mostly plus' signature $\sigma _{\mu \nu
}=\sigma ^{\mu \nu }=\mathrm{diag}\left( -++++\right) $ and with the
five-dimensional Levi--Civita symbol $\varepsilon ^{\mu \nu \rho \lambda
\sigma }$ defined according to the convention $\varepsilon
^{01234}=-\varepsilon _{01234}=-1$.

Action (\ref{f1}) is found invariant under the gauge transformations
\begin{gather}
\delta _{\Omega }\varphi =0,\qquad \delta _{\Omega }H^{\mu }=2\partial _{\nu
}\epsilon ^{\mu \nu },  \label{xf2a} \\
\delta _{\Omega }V_{\mu }=\partial _{\mu }\epsilon ,\qquad \delta _{\Omega
}B^{\mu \nu }=-3\partial _{\rho }\epsilon ^{\mu \nu \rho },  \label{xf2b} \\
\delta _{\Omega }\phi _{\mu \nu }=\partial _{\lbrack \mu }\xi _{\nu
]},\qquad \delta _{\Omega }K^{\mu \nu \rho }=4\partial _{\lambda }\xi ^{\mu
\nu \rho \lambda },  \label{xf2c} \\
\delta _{\Omega }t_{\mu \nu |\alpha }=\partial _{\lbrack \mu }\theta _{\nu
]\alpha }+\partial _{\lbrack \mu }\chi _{\nu ]\alpha }-2\partial _{\alpha
}\chi _{\mu \nu },  \label{xf2d}
\end{gather}%
where all the gauge parameters are bosonic, with $\epsilon ^{\mu \nu }$, $%
\epsilon ^{\mu \nu \rho }$, $\xi ^{\mu \nu \rho \lambda }$, and $\chi _{\mu
\nu }$ completely antisymmetric and $\theta _{\mu \nu }$ symmetric. By $%
\Omega $ we denoted collectively all the gauge parameters as
\begin{equation}
\Omega ^{\alpha _{1}}\equiv \left( \epsilon ^{\mu \nu },\epsilon ,\epsilon
^{\mu \nu \rho },\xi _{\mu },\xi ^{\mu \nu \rho \lambda },\theta _{\mu \nu
},\chi _{\mu \nu }\right) .  \label{Omegaa1}
\end{equation}%
The gauge transformations given by (\ref{xf2a})--(\ref{xf2d}) are off-shell
reducible of order three (the reducibility relations hold everywhere in the
space of field history, and not only on the stationary surface of field
equations). This means that:

\begin{enumerate}
\item there exist some transformations of the gauge parameters (\ref{Omegaa1}%
)
\begin{equation}
\Omega ^{\alpha _{1}}\rightarrow \Omega ^{\alpha _{1}}=\Omega ^{\alpha
_{1}}\left( \bar{\Omega}^{\alpha _{2}}\right) ,  \label{order1}
\end{equation}%
such that the gauge transformations of all fields vanish strongly
(first-order reducibility relations)%
\begin{equation}
\delta _{\Omega \left( \bar{\Omega}\right) }\Phi ^{\alpha _{0}}=0;
\label{1red}
\end{equation}

\item there exist some transformations of the first-order reducibility
parameters $\bar{\Omega}^{\alpha _{2}}$
\begin{equation}
\bar{\Omega}^{\alpha _{2}}\rightarrow \bar{\Omega}^{\alpha _{2}}=\bar{\Omega}%
^{\alpha _{2}}\left( \check{\Omega}^{\alpha _{3}}\right) ,  \label{order2}
\end{equation}%
such that the gauge parameters vanish strongly (second-order reducibility
relations)%
\begin{equation}
\Omega ^{\alpha _{1}}\left( \bar{\Omega}^{\alpha _{2}}\left( \check{\Omega}%
^{\alpha _{3}}\right) \right) =0;  \label{2red}
\end{equation}

\item there exist some transformations of the second-order reducibility
parameters $\check{\Omega}^{\alpha _{3}}$%
\begin{equation}
\check{\Omega}^{\alpha _{3}}\rightarrow \check{\Omega}^{\alpha _{3}}=\check{%
\Omega}^{\alpha _{3}}\left( \hat{\Omega}^{\alpha _{4}}\right) ,
\label{order3}
\end{equation}%
such that the first-order reducibility parameters vanish strongly
(third-order reducibility relations)%
\begin{equation}
\bar{\Omega}^{\alpha _{2}}\left( \check{\Omega}^{\alpha _{3}}\left( \hat{%
\Omega}^{\alpha _{4}}\right) \right) =0;  \label{3red}
\end{equation}

\item there is no nontrivial transformation of the third-order reducibility
parameters $\hat{\Omega}^{\alpha _{4}}$ that annihilates all the
second-order reducibility parameters%
\begin{equation}
\check{\Omega}^{\alpha _{3}}\left( \hat{\Omega}^{\alpha _{4}}\right)
=0\Leftrightarrow \hat{\Omega}^{\alpha _{4}}=0.  \label{finalred}
\end{equation}
\end{enumerate}

\noindent This is indeed the case for the model under study. In this
situation a complete set of first-order reducibility parameters $\bar{\Omega}%
^{\alpha _{2}}$ is given by
\begin{equation}
\bar{\Omega}^{\alpha _{2}}\equiv \left( \bar{\epsilon}^{\mu \nu \rho },\bar{%
\epsilon}^{\mu \nu \rho \lambda },\bar{\xi},\bar{\xi}^{\mu \nu \rho \lambda
\sigma },\bar{\theta}_{\mu }\right) ,  \label{Omegaa2}
\end{equation}%
and transformations (\ref{order1}) have the form%
\begin{gather}
\epsilon ^{\mu \nu }\left( \bar{\Omega}^{\alpha _{2}}\right) =-3\partial
_{\rho }\bar{\epsilon}^{\mu \nu \rho },\qquad \epsilon \left( \bar{\Omega}%
^{\alpha _{2}}\right) =0,\qquad \epsilon ^{\mu \nu \rho }\left( \bar{\Omega}%
^{\alpha _{2}}\right) =4\partial _{\lambda }\bar{\epsilon}^{\mu \nu \rho
\lambda },  \label{xr1a} \\
\xi _{\mu }\left( \bar{\Omega}^{\alpha _{2}}\right) =\partial _{\mu }\bar{\xi%
},\qquad \xi ^{\mu \nu \rho \lambda }\left( \bar{\Omega}^{\alpha
_{2}}\right) =-5\partial _{\sigma }\bar{\xi}^{\mu \nu \rho \lambda \sigma },
\label{xr1b} \\
\theta _{\mu \nu }\left( \bar{\Omega}^{\alpha _{2}}\right) =3\partial _{(\mu
}\bar{\theta}_{\nu )},\qquad \chi _{\mu \nu }\left( \bar{\Omega}^{\alpha
_{2}}\right) =\partial _{\lbrack \mu }\bar{\theta}_{\nu ]},  \label{xr1c}
\end{gather}%
with $\bar{\epsilon}^{\mu \nu \rho }$, $\bar{\epsilon}^{\mu \nu \rho \lambda
}$, and $\bar{\xi}^{\mu \nu \rho \lambda \sigma }$ completely antisymmetric.
Further, a complete set of second-order reducibility parameters $\check{%
\Omega}^{\alpha _{3}}$ can be taken as%
\begin{equation}
\check{\Omega}^{\alpha _{3}}\equiv \left( \check{\epsilon}^{\mu \nu \rho
\lambda },\check{\epsilon}^{\mu \nu \rho \lambda \sigma }\right) ,
\label{Omegaa3}
\end{equation}%
and transformations (\ref{order2}) are
\begin{gather}
\bar{\epsilon}^{\mu \nu \rho }\left( \check{\Omega}^{\alpha _{3}}\right)
=4\partial _{\lambda }\check{\epsilon}^{\mu \nu \rho \lambda },\qquad \bar{%
\epsilon}^{\mu \nu \rho \lambda }\left( \check{\Omega}^{\alpha _{3}}\right)
=-5\partial _{\sigma }\check{\epsilon}^{\mu \nu \rho \lambda \sigma },
\label{xr2a} \\
\bar{\xi}\left( \check{\Omega}^{\alpha _{3}}\right) =0,\qquad \bar{\xi}^{\mu
\nu \rho \lambda \sigma }\left( \check{\Omega}^{\alpha _{3}}\right)
=0,\qquad \bar{\theta}_{\mu }\left( \check{\Omega}^{\alpha _{3}}\right) =0,
\label{xr2b}
\end{gather}%
where both $\check{\epsilon}^{\mu \nu \rho \lambda }$ and $\check{\epsilon}%
^{\mu \nu \rho \lambda \sigma }$ are some arbitrary, bosonic, completely
antisymmetric tensors. Next, a complete set of third-order reducibility
parameters $\hat{\Omega}^{\alpha _{4}}$ is represented by%
\begin{equation}
\hat{\Omega}^{\alpha _{4}}\equiv \left( \hat{\epsilon}^{\mu \nu \rho \lambda
\sigma }\right) ,  \label{Omegaa4}
\end{equation}%
and transformations (\ref{order3}) can be chosen of the form%
\begin{equation}
\check{\epsilon}^{\mu \nu \rho \lambda }\left( \hat{\Omega}^{\alpha
_{4}}\right) =-5\partial _{\sigma }\hat{\epsilon}^{\mu \nu \rho \lambda
\sigma },\qquad \check{\epsilon}^{\mu \nu \rho \lambda \sigma }\left( \hat{%
\Omega}^{\alpha _{4}}\right) =0,  \label{xr3}
\end{equation}%
with $\hat{\epsilon}^{\mu \nu \rho \lambda \sigma }$ an arbitrary,
completely antisymmetric tensor. Finally, it is easy to check (\ref{finalred}%
). Indeed, we work in $D=5$, such that $\partial _{\sigma }\hat{\epsilon}%
^{\mu \nu \rho \lambda \sigma }=0$ implies $\hat{\epsilon}^{\mu \nu \rho
\lambda \sigma }=const.$.Since $\hat{\epsilon}^{\mu \nu \rho \lambda \sigma
} $ are arbitrary smooth functions that effectively depend on the spacetime
coordinates, it follows that the only possible choice is $\hat{\epsilon}%
^{\mu \nu \rho \lambda \sigma }=0$.

We observe that the free theory under study is a usual linear gauge theory
(its field equations are linear in the fields), whose generating set of
gauge transformations is third-order reducible, such that we can define in a
consistent manner its Cauchy order, which is found to be equal to five.

In order to construct the BRST symmetry of this free theory, we introduce
the field/ghost and antifield spectra (\ref{f10a}) and%
\begin{gather}
\eta ^{\alpha _{1}}=\left( C^{\mu \nu },\eta ,\eta ^{\mu \nu \rho },C_{\mu },%
\mathcal{G}^{\mu \nu \rho \lambda },S_{\mu \nu },A_{\mu \nu }\right) ,
\label{f10b} \\
\eta ^{\alpha _{2}}=\left( C^{\mu \nu \rho },\eta ^{\mu \nu \rho \lambda },C,%
\mathcal{G}^{\mu \nu \rho \lambda \sigma },S_{\mu }\right) ,  \label{f10c} \\
\eta ^{\alpha _{3}}=\left( C^{\mu \nu \rho \lambda },\eta ^{\mu \nu \rho
\lambda \sigma }\right) ,\qquad \eta ^{\alpha _{4}}=\left( C^{\mu \nu \rho
\lambda \sigma }\right) ,  \label{f10d} \\
\Phi _{\alpha _{0}}^{\ast }=\left( \varphi ^{\ast },H_{\mu }^{\ast },V^{\ast
\mu },B_{\mu \nu }^{\ast },\phi ^{\ast \mu \nu },K_{\mu \nu \rho }^{\ast
},t^{\ast \mu \nu |\alpha }\right) ,  \label{f10a'} \\
\eta _{\alpha _{1}}^{\ast }=\left( C_{\mu \nu }^{\ast },\eta ^{\ast },\eta
_{\mu \nu \rho }^{\ast },C^{\ast \mu },\mathcal{G}_{\mu \nu \rho \lambda
}^{\ast },S^{\ast \mu \nu },A^{\ast \mu \nu }\right) ,  \label{f10b'} \\
\eta _{\alpha _{2}}^{\ast }=\left( C_{\mu \nu \rho }^{\ast },\eta _{\mu \nu
\rho \lambda }^{\ast },C^{\ast },\mathcal{G}_{\mu \nu \rho \lambda \sigma
}^{\ast },S^{\ast \mu }\right) ,  \label{f10c'} \\
\eta _{\alpha _{3}}^{\ast }=\left( C_{\mu \nu \rho \lambda }^{\ast },\eta
_{\mu \nu \rho \lambda \sigma }^{\ast }\right) ,\qquad \eta _{\alpha
_{4}}^{\ast }=\left( C_{\mu \nu \rho \lambda \sigma }^{\ast }\right) .
\label{f10d'}
\end{gather}%
The fermionic ghosts (\ref{f10b}) correspond to the bosonic gauge parameters
(\ref{Omegaa1}), and therefore $C^{\mu \nu }$, $\eta ^{\mu \nu \rho }$, $%
\mathcal{G}^{\mu \nu \rho \lambda }$, and $A_{\mu \nu }$ are completely
antisymmetric and $S_{\mu \nu }$ is symmetric. The bosonic ghosts for ghosts
(\ref{f10c}) are respectively associated with the first-order reducibility
parameters (\ref{Omegaa2}), such that $C^{\mu \nu \rho }$, $\eta ^{\mu \nu
\rho \lambda }$, and $\mathcal{G}^{\mu \nu \rho \lambda \sigma }$ are
completely antisymmetric. Along the same line, the fermionic ghosts for
ghosts for ghosts $\eta ^{\alpha _{3}}$ from (\ref{f10d}) correspond to the
second-order reducibility parameters (\ref{Omegaa3}). As a consequence, the
ghost fields $C^{\mu \nu \rho \lambda }$ and $\eta ^{\mu \nu \rho \lambda
\sigma }$ are again completely antisymmetric. Finally, the bosonic ghosts
for ghosts for ghosts for ghosts $\eta ^{\alpha _{4}}$ from (\ref{f10d}) are
associated with the third-order reducibility parameters (\ref{Omegaa4}), so $%
C^{\mu \nu \rho \lambda \sigma }$ is also completely antisymmetric. The star
variables represent the antifields of the corresponding fields/ghosts. Their
Grassmann parities are obtained via the usual rule $\varepsilon \left( \chi
_{\Delta }^{\ast }\right) =\left( \varepsilon \left( \chi ^{\Delta }\right)
+1\right) \mathrm{mod\,}2$, where we employed the notations
\begin{equation}
\chi ^{\Delta }=\left( \Phi ^{\alpha _{0}},\eta ^{\alpha _{1}},\eta ^{\alpha
_{2}},\eta ^{\alpha _{3}},\eta ^{\alpha _{4}}\right) ,\qquad \chi _{\Delta
}^{\ast }=\left( \Phi _{\alpha _{0}}^{\ast },\eta _{\alpha _{1}}^{\ast
},\eta _{\alpha _{2}}^{\ast },\eta _{\alpha _{3}}^{\ast },\eta _{\alpha
_{4}}^{\ast }\right) .  \label{notat}
\end{equation}%
It is understood that the antifields are endowed with the same
symmetry/antisymmetry properties like those of the corresponding
fields/ghosts.

Since both the gauge generators and the reducibility functions are
field-independent, it follows that the BRST differential reduces to $%
s=\delta +\gamma $, where $\delta $ is the Koszul--Tate differential, and $%
\gamma $ means the exterior longitudinal derivative. The Koszul--Tate
differential is graded in terms of the antighost number ($\mathrm{agh}$, $%
\mathrm{agh}\left( \delta \right) =-1$, $\mathrm{agh}\left( \gamma \right)
=0 $) and enforces a resolution of the algebra of smooth functions defined
on the stationary surface of field equations for action (\ref{f1}), $%
C^{\infty }\left( \Sigma \right) $, $\Sigma :\delta S_{0}^{\mathrm{L}%
}/\delta \Phi ^{\alpha _{0}}=0$. The exterior longitudinal derivative is
graded in terms of the pure ghost number ($\mathrm{pgh}$, $\mathrm{pgh}%
\left( \gamma \right) =1$, $\mathrm{pgh}\left( \delta \right) =0$) and is
correlated with the original gauge symmetry via its cohomology in pure ghost
number zero computed in $C^{\infty }\left( \Sigma \right) $, which is
isomorphic to the algebra of physical observables for this free theory.
These two degrees of generators (\ref{f10a}) and (\ref{f10b})--(\ref{f10d'})
from the BRST complex are valued like%
\begin{gather}
\mathrm{pgh}\left( \Phi ^{\alpha _{0}}\right) =0,\qquad \mathrm{pgh}\left(
\eta ^{\alpha _{m}}\right) =m,\qquad \mathrm{pgh}\left( \Phi _{\alpha
_{0}}^{\ast }\right) =\mathrm{pgh}\left( \eta _{\alpha _{m}}^{\ast }\right)
=0,  \label{f11a} \\
\mathrm{agh}\left( \Phi ^{\alpha _{0}}\right) =\mathrm{agh}\left( \eta
^{\alpha _{m}}\right) =0,\qquad \mathrm{agh}\left( \Phi _{\alpha _{0}}^{\ast
}\right) =1,\qquad \mathrm{agh}\left( \eta _{\alpha _{m}}^{\ast }\right)
=m+1,  \label{f11b}
\end{gather}%
for $m=\overline{1,4}$. The actions of the differentials $\delta $ and $%
\gamma $ on the above generators read as
\begin{gather}
\left( \delta \Phi ^{\alpha _{0}}=0,\qquad \delta \eta ^{\alpha
_{m}}=0,\qquad m=\overline{1,4}\right) \Longleftrightarrow \delta \chi
^{\Delta }=0,  \label{f12a} \\
\delta \varphi ^{\ast }=\partial _{\mu }H^{\mu },\qquad \delta H_{\mu
}^{\ast }=-\partial _{\mu }\varphi ,\qquad \delta V^{\ast \mu }=-\partial
_{\nu }B^{\mu \nu },  \label{f12b} \\
\delta B_{\mu \nu }^{\ast }=-\tfrac{1}{2}\partial _{\lbrack \mu }V_{\nu
]},\qquad \delta \phi ^{\ast \mu \nu }=\partial _{\rho }K^{\mu \nu \rho
},\qquad \delta K_{\mu \nu \rho }^{\ast }=-\tfrac{1}{3}\partial _{\lbrack
\mu }\phi _{\nu \rho ]},  \label{f12c} \\
\delta t^{\ast \mu \nu |\alpha }=-\tfrac{1}{2}\partial _{\rho }\left(
F^{\rho \mu \nu |\alpha }-\sigma ^{\alpha \lbrack \mu }F^{\nu \rho ]}\right)
,\qquad \delta C_{\mu \nu }^{\ast }=\partial _{\lbrack \mu }H_{\nu ]}^{\ast
},  \label{f12d} \\
\delta \eta ^{\ast }=-\partial _{\mu }V^{\ast \mu },\qquad \delta \eta _{\mu
\nu \rho }^{\ast }=\partial _{\lbrack \mu }B_{\nu \rho ]}^{\ast },\qquad
\delta C^{\ast \mu }=2\partial _{\nu }\phi ^{\ast \mu \nu },  \label{f12e} \\
\delta \mathcal{G}_{\mu \nu \rho \lambda }^{\ast }=\partial _{\lbrack \mu
}K_{\nu \rho \lambda ]}^{\ast },\qquad \delta S^{\ast \mu \nu }=-\partial
_{\rho }t^{\ast \rho (\mu |\nu )},\qquad \delta A^{\ast \mu \nu }=3\partial
_{\rho }t^{\ast \mu \nu |\rho },  \label{f12f} \\
\delta C_{\mu \nu \rho }^{\ast }=-\partial _{\lbrack \mu }C_{\nu \rho
]}^{\ast },\qquad \delta \eta _{\mu \nu \rho \lambda }^{\ast }=-\partial
_{\lbrack \mu }\eta _{\nu \rho \lambda ]}^{\ast },\qquad \delta C^{\ast
}=\partial _{\mu }C^{\ast \mu },  \label{f12g} \\
\delta \mathcal{G}_{\mu \nu \rho \lambda \sigma }^{\ast }=-\partial
_{\lbrack \mu }\mathcal{G}_{\nu \rho \lambda \sigma ]}^{\ast },\qquad \delta
S^{\ast \mu }=2\partial _{\rho }\left( 3S^{\ast \rho \mu }+A^{\ast \rho \mu
}\right) \equiv 2\partial _{\rho }\mathcal{C}^{\ast \rho \mu },  \label{f12h}
\\
\delta C_{\mu \nu \rho \lambda }^{\ast }=\partial _{\lbrack \mu }C_{\nu \rho
\lambda ]}^{\ast },\qquad \delta \eta _{\mu \nu \rho \lambda \sigma }^{\ast
}=\partial _{\lbrack \mu }\eta _{\nu \rho \lambda \sigma ]}^{\ast },\qquad
\delta C_{\mu \nu \rho \lambda \sigma }^{\ast }=-\partial _{\lbrack \mu
}C_{\nu \rho \lambda \sigma ]}^{\ast },  \label{f12i}
\end{gather}%
and respectively%
\begin{gather}
\left( \gamma \Phi _{\alpha _{0}}^{\ast }=0,\qquad \gamma \eta _{\alpha
_{m}}^{\ast }=0,\qquad m=\overline{1,4}\right) \Longleftrightarrow \gamma
\chi _{\Delta }^{\ast }=0,  \label{f13a} \\
\gamma \varphi =0,\qquad \gamma H^{\mu }=2\partial _{\nu }C^{\mu \nu
},\qquad \gamma V_{\mu }=\partial _{\mu }\eta ,  \label{f13b} \\
\gamma B^{\mu \nu }=-3\partial _{\rho }\eta ^{\mu \nu \rho },\qquad \gamma
\phi _{\mu \nu }=\partial _{\lbrack \mu }C_{\nu ]},\qquad \gamma K^{\mu \nu
\rho }=4\partial _{\lambda }\mathcal{G}^{\mu \nu \rho \lambda },
\label{f13c} \\
\gamma t_{\mu \nu |\alpha }=\partial _{\lbrack \mu }S_{\nu ]\alpha
}+\partial _{\lbrack \mu }A_{\nu ]\alpha }-2\partial _{\alpha }A_{\mu \nu
},\qquad \gamma C^{\mu \nu }=-3\partial _{\rho }C^{\mu \nu \rho },
\label{f13d} \\
\gamma \eta =0,\qquad \gamma \eta ^{\mu \nu \rho }=4\partial _{\lambda }\eta
^{\mu \nu \rho \lambda },\qquad \gamma C_{\mu }=\partial _{\mu }C,
\label{f13e} \\
\gamma \mathcal{G}^{\mu \nu \rho \lambda }=-5\partial _{\sigma }\mathcal{G}%
^{\mu \nu \rho \lambda \sigma },\qquad \gamma S_{\mu \nu }=3\partial _{(\mu
}S_{\nu )},\qquad \gamma A_{\mu \nu }=\partial _{\lbrack \mu }S_{\nu ]},
\label{f13f} \\
\gamma C^{\mu \nu \rho }=4\partial _{\lambda }C^{\mu \nu \rho \lambda
},\qquad \gamma \eta ^{\mu \nu \rho \lambda }=-5\partial _{\sigma }\eta
^{\mu \nu \rho \lambda \sigma },\qquad \gamma C=0,  \label{f13g} \\
\gamma \mathcal{G}^{\mu \nu \rho \lambda \sigma }=0,\qquad \gamma S_{\mu
}=0,\qquad \gamma C^{\mu \nu \rho \lambda }=-5\partial _{\sigma }C^{\mu \nu
\rho \lambda \sigma },  \label{f13h} \\
\gamma \eta ^{\mu \nu \rho \lambda \sigma }=0,\qquad \gamma C^{\mu \nu \rho
\lambda \sigma }=0.  \label{f13i}
\end{gather}%
The overall degree that grades the BRST complex is named ghost number ($%
\mathrm{gh}$) and is defined like the difference between the pure ghost
number and the antighost number, such that $\mathrm{gh}\left( \delta \right)
=\mathrm{gh}\left( \gamma \right) =\mathrm{gh}\left( s\right) =1$.

The BRST symmetry admits a canonical action $s\cdot =\left( \cdot ,\bar{S}%
\right) $, where its canonical generator ($\mathrm{gh}\left( \bar{S}\right)
=0$, $\varepsilon \left( \bar{S}\right) =0$) satisfies the classical master
equation $\left( \bar{S},\bar{S}\right) =0$. The symbol $(,)$ denotes the
antibracket, defined by decreeing the fields/ghosts conjugated with the
corresponding antifields. In the case of the free theory under discussion
the solution to the master equation takes the form%
\begin{eqnarray}
\bar{S}=S_{0}^{\mathrm{L}} &+&\int d^{5}x\left[ 2H_{\mu }^{\ast }\partial
_{\nu }C^{\mu \nu }+V^{\ast \mu }\partial _{\mu }\eta -3B_{\mu \nu }^{\ast
}\partial _{\rho }\eta ^{\mu \nu \rho }+\phi ^{\ast \mu \nu }\partial
_{\lbrack \mu }C_{\nu ]}\right.  \notag \\
&&+4K_{\mu \nu \rho }^{\ast }\partial _{\lambda }\mathcal{G}^{\mu \nu \rho
\lambda }+t^{\ast \mu \nu |\alpha }\left( \partial _{\lbrack \mu }S_{\nu
]\alpha }+\partial _{\lbrack \mu }A_{\nu ]\alpha }-2\partial _{\alpha
}A_{\mu \nu }\right)  \notag \\
&&-3C_{\mu \nu }^{\ast }\partial _{\rho }C^{\mu \nu \rho }+4\eta _{\mu \nu
\rho }^{\ast }\partial _{\lambda }\eta ^{\mu \nu \rho \lambda }+C^{\ast \mu
}\partial _{\mu }C-5\mathcal{G}_{\mu \nu \rho \lambda }^{\ast }\partial
_{\sigma }\mathcal{G}^{\mu \nu \rho \lambda \sigma }  \notag \\
&&+3S^{\ast \mu \nu }\partial _{(\mu }S_{\nu )}+A^{\ast \mu \nu }\partial
_{\lbrack \mu }S_{\nu ]}+4C_{\mu \nu \rho }^{\ast }\partial _{\lambda
}C^{\mu \nu \rho \lambda }  \notag \\
&&\left. -5\eta _{\mu \nu \rho \lambda }^{\ast }\partial _{\sigma }\eta
^{\mu \nu \rho \lambda \sigma }-5C_{\mu \nu \rho \lambda }^{\ast }\partial
_{\sigma }C^{\mu \nu \rho \lambda \sigma }\right] .  \label{f14}
\end{eqnarray}%
The solution to the master equation encodes all the information on the gauge
structure of a given theory. We remark that in our case solution (\ref{f14})
decomposes into terms with antighost numbers ranging from zero to four. Let
us briefly recall the significance of the various terms present in the
solution to the master equation. Thus, the part with the antighost number
equal to zero is nothing but the Lagrangian action of the gauge model under
study. The components of antighost number equal to one are always
proportional with the gauge generators. If the gauge algebra were
non-Abelian, then there would appear terms simultaneously linear in the
antighost number two antifields and quadratic in the pure ghost number one
ghosts. The absence of such terms in our case shows that the gauge
transformations are Abelian. The terms from (\ref{f14}) with higher
antighost numbers give us information on the reducibility functions. If the
reducibility relations held on-shell, then there would appear components
linear in the ghosts for ghosts (ghosts of pure ghost number strictly
greater than one) and quadratic in the various antifields. Such pieces are
not present in (\ref{f14}) since the reducibility relations (\ref{1red}), (%
\ref{2red}), and (\ref{3red}) hold off-shell. Other possible components in
the solution to the master equation offer information on the higher-order
structure functions related to the tensor gauge structure of the theory.
There are no such terms in (\ref{f14}) as a consequence of the fact that all
higher-order structure functions vanish for the theory under study.

\section{Strategy\label{def}}

We begin with a \textquotedblleft free\textquotedblright\ gauge theory,
described by a Lagrangian action $S_{0}^{\mathrm{L}}\left[ \Phi ^{\alpha
_{0}}\right] $, invariant under some gauge transformations
\begin{equation}
\delta _{\epsilon }\Phi ^{\alpha _{0}}=Z_{\;\;\alpha _{1}}^{\alpha
_{0}}\epsilon ^{\alpha _{1}},\qquad \frac{\delta S_{0}^{\mathrm{L}}}{\delta
\Phi ^{\alpha _{0}}}Z_{\;\;\alpha _{1}}^{\alpha _{0}}=0,  \label{bfa2.1}
\end{equation}%
and consider the problem of constructing consistent interactions among the
fields $\Phi ^{\alpha _{0}}$ such that the couplings preserve both the field
spectrum and the original number of gauge symmetries. This matter is
addressed by means of reformulating the problem of constructing consistent
interactions as a deformation problem of the solution to the master equation
corresponding to the \textquotedblleft free\textquotedblright\ theory~\cite%
{deflag,17and5}. Such a reformulation is possible due to the fact that the
solution to the master equation contains all the information on the gauge
structure of the theory. If a consistent interacting gauge theory can be
constructed, then the solution $\bar{S}$ to the master equation associated
with the \textquotedblleft free\textquotedblright\ theory, $\left( \bar{S},%
\bar{S}\right) =0$, can be deformed into a solution $S$,
\begin{eqnarray}
\bar{S}\rightarrow S &=&\bar{S}+\lambda S_{1}+\lambda ^{2}S_{2}+\cdots
\notag \\
&=&\bar{S}+\lambda \int d^{D}x\,a+\lambda ^{2}\int d^{D}x\,b+\lambda
^{3}\int d^{D}x\,c+\cdots  \label{bfa2.2}
\end{eqnarray}%
of the master equation for the deformed theory
\begin{equation}
\left( S,S\right) =0,  \label{bfa2.3}
\end{equation}%
such that both the ghost and antifield spectra of the initial theory are
preserved. The symbol $\left( ,\right) $ denotes the antibracket. Equation (%
\ref{bfa2.3}) splits, according to the various orders in the coupling
constant (or deformation parameter) $\lambda $, into the equivalent tower of
equations
\begin{eqnarray}
\left( \bar{S},\bar{S}\right) &=&0,  \label{bfa2.4} \\
2\left( S_{1},\bar{S}\right) &=&0,  \label{bfa2.5} \\
2\left( S_{2},\bar{S}\right) +\left( S_{1},S_{1}\right) &=&0,  \label{bfa2.6}
\\
\left( S_{3},\bar{S}\right) +\left( S_{1},S_{2}\right) &=&0,  \label{bfa2.7}
\\
2\left( S_{4},\bar{S}\right) +\left( S_{2},S_{2}\right) +2\left(
S_{1},S_{3}\right) &=&0  \label{bfa2.8} \\
&&\vdots  \notag
\end{eqnarray}

Equation (\ref{bfa2.4}) is fulfilled by hypothesis. The next one requires
that the first-order deformation of the solution to the master equation, $%
S_{1}$, is a cocycle of the \textquotedblleft free\textquotedblright\ BRST
differential $s\cdot =\left( \cdot ,\bar{S}\right) $. However, only
cohomologically nontrivial solutions to (\ref{bfa2.5}) should be taken into
account, as the BRST-exact ones can be eliminated by (in general nonlinear)
field redefinitions. This means that $S_{1}$ pertains to the ghost number
zero cohomological space of $s$, $H^{0}\left( s\right) $, which is
generically nonempty due to its isomorphism to the space of physical
observables of the \textquotedblleft free\textquotedblright\ theory. It has
been shown in~\cite{deflag,17and5} (on behalf of the triviality of the
antibracket map in the cohomology of the BRST differential) that there are
no obstructions in finding solutions to the remaining equations, namely, (%
\ref{bfa2.6}), (\ref{bfa2.7}) and so on. However, the resulting interactions
may be nonlocal, and there might even appear obstructions if one insists on
their locality. The analysis of these obstructions can be done with the help
of cohomological techniques. As it will be seen below, all the interactions
in the case of the model under study turn out to be local.

\section{Standard results\label{stand}}

In the sequel we determine all consistent Lagrangian interactions that can
be added to the free\ theory described by (\ref{f1}) and (\ref{xf2a})--(\ref%
{xf2d}). This is done by means of solving the deformation equations (\ref%
{bfa2.5})--(\ref{bfa2.8}), etc., with the help of specific cohomological
techniques. The interacting theory and its gauge structure are then deduced
from the analysis of the deformed solution to the master equation that is
consistent to all orders in the deformation parameter.

For obvious reasons, we consider only analytical, local, Lorentz covariant,
and Poincar\'{e} invariant deformations (i.e., we do not allow explicit
dependence on the spacetime coordinates). The analyticity of deformations
refers to the fact that the deformed solution to the master equation, (\ref%
{bfa2.2}), is analytical in the coupling constant $\lambda $ and reduces to
the original solution, (\ref{f14}), in the free limit $\lambda =0$. In
addition, we require that the overall interacting Lagrangian satisfies two
further restrictions related to the derivative order of its vertices:

\begin{enumerate}
\item[i)] the maximum derivative order of each interaction vertex is equal
to two;

\item[ii)] the differential order of each interacting field equation is
equal to that of the corresponding free equation (meaning that at most one
spacetime derivative can act on each field from the BF sector and at most
two spacetime derivatives on the tensor field $t_{\mu \nu \vert \alpha }$).
\end{enumerate}

If we make the notation $S_{1}=\int d^{5}x\,a$, with $a$ local, then
equation (\ref{bfa2.5}) (which controls the first-order deformation) takes
the local form
\begin{equation}
sa=\partial _{\mu }m^{\mu },\qquad \mathrm{gh}\left( a\right) =0,\qquad
\varepsilon \left( a\right) =0,  \label{3.1}
\end{equation}%
for some local $m^{\mu }$. It shows that the nonintegrated density of the
first-order deformation pertains to the local cohomology of $s$ in ghost
number zero, $a\in H^{0}\left( s\vert d\right) $, where $d$ denotes the
exterior spacetime differential. The solution to (\ref{3.1}) is unique up to
$s$-exact pieces plus divergences
\begin{equation}
a\rightarrow a+sb+\partial _{\mu }n^{\mu }.  \label{3.1a}
\end{equation}%
If the general solution to (\ref{3.1}) is trivial, $a=sb+\partial _{\mu
}n^{\mu }$, then it can be made to vanish, $a=0$.

In order to analyze equation (\ref{3.1}) we develop $a$ according to the
antighost number
\begin{equation}
a=\sum\limits_{i=0}^{I}a_{i},\qquad \mathrm{agh}\left( a_{i}\right)
=i,\qquad \mathrm{gh}\left( a_{i}\right) =0,\qquad \varepsilon \left(
a_{i}\right) =0,  \label{3.2}
\end{equation}%
and assume, without loss of generality, that the above decomposition stops
at some finite value of $I$. This can be shown for instance like in~\cite%
{gen2} (Section 3), under the sole assumption that the interacting
Lagrangian at order one in the coupling constant, $a_{0}$, has a finite, but
otherwise arbitrary derivative order. Inserting (\ref{3.2}) into (\ref{3.1})
and projecting it on the various values of the antighost number, we obtain
the tower of equations (equivalent to (\ref{3.1}))
\begin{eqnarray}
\gamma a_{I} &=&\partial _{\mu }\overset{\left( I\right) }{m}^{\mu },
\label{3.3} \\
\delta a_{I}+\gamma a_{I-1} &=&\partial _{\mu }\overset{\left( I-1\right) }{m%
}^{\mu },  \label{3.4} \\
\delta a_{i}+\gamma a_{i-1} &=&\partial _{\mu }\overset{\left( i-1\right) }{m%
}^{\mu },\qquad I-1\geq i\geq 1,  \label{3.5}
\end{eqnarray}%
for some local $\left( \overset{\left( i\right) }{m}^{\mu }\right) _{i=%
\overline{0,I}}$. Equation (\ref{3.3}) can always be replaced in strictly
positive values of the antighost number by
\begin{equation}
\gamma a_{I}=0,\qquad I>0.  \label{3.6}
\end{equation}%
Due to the second-order nilpotency of $\gamma $ ($\gamma ^{2}=0$), the
solution to (\ref{3.6}) is unique up to $\gamma $-exact contributions
\begin{equation}
a_{I}\rightarrow a_{I}+\gamma b_{I}.  \label{r68}
\end{equation}%
If $a_{I}$ reduces only to $\gamma $-exact terms, $a_{I}=\gamma b_{I}$, then
it can be made to vanish, $a_{I}=0$. The nontriviality of the first-order
deformation $a$ is translated at its highest antighost number component into
the requirement that $a_{I}\in H^{I}\left( \gamma \right) $, where $%
H^{I}\left( \gamma \right) $ denotes the cohomology of the exterior
longitudinal derivative $\gamma $ in pure ghost number equal to $I$. So, in
order to solve equation (\ref{3.1}) (equivalent with (\ref{3.6}) and (\ref%
{3.4})--(\ref{3.5})), we need to compute the cohomology of $\gamma $, $%
H\left( \gamma \right) $, and, as it will be made clear below, also the
local homology of $\delta $, $H\left( \delta \vert d\right) $.

From definitions (\ref{f13a})--(\ref{f13i}) it is posible to show that $%
H\left( \gamma \right) $ is spanned by
\begin{equation}
F_{\bar{A}}=\left( \varphi ,\partial _{\mu }H^{\mu },\partial _{\lbrack \mu
}V_{\nu ]},\partial _{\mu }B^{\mu \nu },\partial _{\lbrack \mu }\phi _{\nu
\rho ]},\partial _{\mu }K^{\mu \nu \rho },R_{\mu \nu \rho |\alpha \beta
}\right) ,  \label{3.7}
\end{equation}%
the antifields $\chi _{\Delta }^{\ast }$, and all of their spacetime
derivatives as well as by the undifferentiated objects
\begin{equation}
\eta ^{\bar{\Upsilon}}=\left( \eta ,D_{\mu \nu \rho },C,\mathcal{G}^{\mu \nu
\rho \lambda \sigma },S_{\mu },\eta ^{\mu \nu \rho \lambda \sigma },C^{\mu
\nu \rho \lambda \sigma }\right) .  \label{notat1}
\end{equation}%
In (\ref{3.7}) and (\ref{notat1}) we respectively used the notations%
\begin{equation}
R_{\mu \nu \rho |\alpha \beta }=-\tfrac{1}{2}F_{\mu \nu \rho |[\alpha ,\beta
]},\qquad D_{\mu \nu \rho }=\partial _{\lbrack \mu }A_{\nu \rho ]},
\label{fie}
\end{equation}%
with $f_{,\beta }\equiv \partial _{\beta }f$. It is useful to denote by $%
R_{\mu \nu |\alpha }$ and $R_{\mu }$ the trace and respectively double trace
of $R_{\mu \nu \rho |\alpha \beta }$%
\begin{equation}
R_{\mu \nu |\alpha }=\sigma ^{\rho \beta }R_{\mu \nu \rho |\alpha \beta
},\qquad R_{\mu }=\sigma ^{\rho \beta }\sigma ^{\nu \alpha }R_{\mu \nu \rho
|\alpha \beta }.  \label{urmer}
\end{equation}%
The spacetime derivatives (of any order) of all the objects from (\ref%
{notat1}) are removed from $H\left( \gamma \right) $ since they are $\gamma $%
-exact. This can be seen directly from the last definition in (\ref{f13b}),
the last present in (\ref{f13e}), the first from (\ref{f13f}), the second in
(\ref{f13g}), the last from (\ref{f13h}), and also using the relations%
\begin{equation}
\partial _{\alpha }D_{\mu \nu \rho }=\gamma \left[ -\tfrac{1}{2}F_{\mu \nu
\rho |\alpha }\right] ,\qquad \partial _{\mu }S_{\nu }=\gamma \left[ \tfrac{1%
}{2}\left( \tfrac{1}{3}S_{\mu \nu }+A_{\mu \nu }\right) \right] \equiv
\gamma \left[ \tfrac{1}{2}\mathcal{C}_{\mu \nu }\right] .  \label{relgamex}
\end{equation}

Let $e^{M}\left( \eta ^{\bar{\Upsilon}}\right) $ be the elements with pure
ghost number $M$ of a basis in the space of polynomials in the objects (\ref%
{notat1}). Then, the general solution to (\ref{3.6}) takes the form (up, to
trivial, $\gamma $-exact contributions)
\begin{equation}
a_{I}=\alpha _{I}\left( \left[ F_{\bar{A}}\right] ,\left[ \chi _{\Delta
}^{\ast }\right] \right) e^{I}\left( \eta ^{\bar{\Upsilon}}\right) ,
\label{3.8}
\end{equation}%
where $\mathrm{agh}\left( \alpha _{I}\right) =I$ and $\mathrm{pgh}\left(
e^{I}\right) =I$. The notation $f([q])$ means that $f$ depends on $q$ and
its spacetime derivatives up to a finite order. The objects $\alpha _{I}$
(obviously nontrivial in $H^{0}\left( \gamma \right) $) will be called
invariant `polynomials'. They are true polynomials with respect to all
variables (\ref{notat1}) and their spacetime derivatives, excepting the
undifferentiated scalar field $\varphi $, with respect to which $\alpha _{I}$
may be series. This is why we will keep the quotation marks around the word
polynomial(s). The result that we can replace equation (\ref{3.3}) with the
less obvious one (\ref{3.6}) for $I>0$ is a nice consequence of the fact
that the cohomology of the exterior spacetime differential is trivial in the
space of invariant `polynomials' in strictly positive antighost numbers.
These results on $H\left( \gamma \right) $ can be synthesized in the
following array%
\begin{equation}
\begin{array}{cccc}
\begin{array}{c}
\mathrm{BRST} \\
\mathrm{generator}%
\end{array}
& \mathrm{pgh} & \mathrm{Grassmann~parity} &
\begin{array}{c}
\mathrm{Nontrivial~object} \\
\mathrm{from}~H\left( \gamma \right)%
\end{array}
\\
\chi _{\Delta }^{\ast } & 0 & \left( \varepsilon \left( \chi ^{\Delta
}\right) +1\right) \mathrm{mod\,}2 & \left[ \chi _{\Delta }^{\ast }\right]
\\
\Phi ^{\alpha _{0}} & 0 & 0 & \left[ F_{\bar{A}}\right] \\
\eta ^{\alpha _{1}} & 1 & 1 & \eta ,D_{\mu \nu \rho }\equiv \partial
_{\lbrack \mu }A_{\nu \rho ]} \\
\eta ^{\alpha _{2}} & 2 & 0 & C,\mathcal{G}^{\mu \nu \rho \lambda \sigma
},S_{\mu } \\
\eta ^{\alpha _{3}} & 3 & 1 & \eta ^{\mu \nu \rho \lambda \sigma } \\
\eta ^{\alpha _{4}} & 4 & 0 & C^{\mu \nu \rho \lambda \sigma }%
\end{array}
\label{hgamma}
\end{equation}%
where notations (\ref{f10a}), (\ref{f10b})--(\ref{f10d'}), (\ref{notat}),
and (\ref{3.7}) should be taken into account.

Inserting (\ref{3.8}) in (\ref{3.4}) we obtain that a necessary (but not
sufficient) condition for the existence of (nontrivial) solutions $a_{I-1}$
is that the invariant `polynomials' $\alpha _{I}$ are (nontrivial) objects
from the local cohomology of Koszul--Tate differential $H\left( \delta
|d\right) $ in antighost number $I>0$ and in pure ghost number zero,
\begin{equation}
\delta \alpha _{I}=\partial _{\mu }\overset{\left( I-1\right) }{j}^{\mu
},\qquad \mathrm{agh}\left( \overset{\left( I-1\right) }{j}^{\mu }\right)
=I-1,\qquad \mathrm{pgh}\left( \overset{\left( I-1\right) }{j}^{\mu }\right)
=0.  \label{3.10a}
\end{equation}%
We recall that $H\left( \delta |d\right) $ is completely trivial in both
strictly positive antighost \textit{and} pure ghost numbers (for instance,
see~\cite{gen1}, Theorem 5.4, and~\cite{gen2}), so from now on it is
understood that by $H\left( \delta |d\right) $ we mean the local cohomology
of $\delta $ at pure ghost number zero. Using the fact that the free model
under study is a linear gauge theory of Cauchy order equal to five and the
general result from the literature~\cite{gen1,gen2} according to which the
local cohomology of the Koszul--Tate differential is trivial in antighost
numbers strictly greater than its Cauchy order, we can state that
\begin{equation}
H_{J}\left( \delta |d\right) =0\quad \mathrm{for\quad all}\quad J>5,
\label{3.11}
\end{equation}%
where $H_{J}\left( \delta |d\right) $ represents the local cohomology of the
Koszul--Tate differential in antighost number $J$. Moreover, it can be shown
that if the invariant `polynomial' $\alpha _{J}$, with $\mathrm{agh}\left(
\alpha _{J}\right) =J\geq 5$, is trivial in $H_{J}\left( \delta |d\right) $,
then it can be taken to be trivial also in $H_{J}^{\mathrm{inv}}\left(
\delta |d\right) $%
\begin{equation}
\left( \alpha _{J}=\delta b_{J+1}+\partial _{\mu }\overset{(J)}{c}^{\mu },%
\mathrm{agh}\left( \alpha _{J}\right) =J\geq 5\right) \Rightarrow \alpha
_{J}=\delta \beta _{J+1}+\partial _{\mu }\overset{(J)}{\gamma }^{\mu },
\label{3.12ax}
\end{equation}%
with both $\beta _{J+1}$ and $\overset{(J)}{\gamma }^{\mu }$ invariant
`polynomials'. Here, $H_{J}^{\mathrm{inv}}\left( \delta |d\right) $ denotes
the invariant characteristic cohomology in antighost number $J$ (the local
cohomology of the Koszul--Tate differential in the space of invariant
`polynomials'). An element of $H_{I}^{\text{\textrm{inv}}}\left( \delta
|d\right) $ is defined via an equation like (\ref{3.10a}), but with the
corresponding current an invariant `polynomial'. This result together with (%
\ref{3.11}) ensures that the entire invariant characteristic cohomology in
antighost numbers strictly greater than five is trivial
\begin{equation}
H_{J}^{\mathrm{inv}}\left( \delta |d\right) =0\quad \mathrm{for\quad all}%
\quad J>5.  \label{3.12x}
\end{equation}

It is possible to show that no nontrivial representative of $H_{J}(\delta
|d) $ or $H_{J}^{\mathrm{inv}}(\delta |d)$ for $J\geq 2$ is allowed to
involve the spacetime derivatives of the fields \cite{juviBFD5} and \cite%
{lingr}. Such a representative may depend at most on the undifferentiated
scalar field $\varphi $. With the help of relations (\ref{f12a})--(\ref{f12i}%
), it can be shown that $H^{\mathrm{inv}}\left( \delta |d\right) $ and $%
H\left( \delta |d\right) $ are spanned by the elements%
\begin{equation}
\begin{array}{ccc}
\mathrm{agh} &
\begin{array}{c}
\mathrm{Nontrivial~representative} \\
\mathrm{spanning}~H_{J}^{\mathrm{inv}}(\delta |d)%
\end{array}
&
\begin{array}{c}
\mathrm{Grassmann} \\
\mathrm{parity}%
\end{array}
\\
>5 & \mathrm{none} & - \\
5 & \left( W\right) _{\mu \nu \rho \lambda \sigma } & 1 \\
4 & \left( W\right) _{\mu \nu \rho \lambda },\eta _{\mu \nu \rho \lambda
\sigma }^{\ast } & 0 \\
3 & \left( W\right) _{\mu \nu \rho },\eta _{\mu \nu \rho \lambda }^{\ast
},C^{\ast },\mathcal{G}_{\mu \nu \rho \lambda \sigma }^{\ast },S^{\ast \mu }
& 1 \\
2 & \left( W\right) _{\mu \nu },\eta ^{\ast },\eta _{\mu \nu \rho }^{\ast
},C^{\ast \mu },\mathcal{G}_{\mu \nu \rho \lambda }^{\ast },S^{\ast \mu \nu
},A^{\ast \mu \nu } & 0%
\end{array}
\label{hinvdelta}
\end{equation}%
where%
\begin{eqnarray}
\left( W\right) _{\mu \nu \rho \lambda \sigma } &=&\frac{dW}{d\varphi }%
C_{\mu \nu \rho \lambda \sigma }^{\ast }+\frac{d^{2}W}{d\varphi ^{2}}\left(
H_{[\mu }^{\ast }C_{\nu \rho \lambda \sigma ]}^{\ast }+C_{[\mu \nu }^{\ast
}C_{\rho \lambda \sigma ]}^{\ast }\right)  \notag \\
&&+\frac{d^{3}W}{d\varphi ^{3}}\left( H_{[\mu }^{\ast }H_{\nu }^{\ast
}C_{\rho \lambda \sigma ]}^{\ast }+H_{[\mu }^{\ast }C_{\nu \rho }^{\ast
}C_{\lambda \sigma ]}^{\ast }\right)  \notag \\
&&+\frac{d^{4}W}{d\varphi ^{4}}H_{[\mu }^{\ast }H_{\nu }^{\ast }H_{\rho
}^{\ast }C_{\lambda \sigma ]}^{\ast }+\frac{d^{5}W}{d\varphi ^{5}}H_{\mu
}^{\ast }H_{\nu }^{\ast }H_{\rho }^{\ast }H_{\lambda }^{\ast }H_{\sigma
}^{\ast },  \label{3.13}
\end{eqnarray}%
\begin{eqnarray}
\left( W\right) _{\mu \nu \rho \lambda } &=&\frac{dW}{d\varphi }C_{\mu \nu
\rho \lambda }^{\ast }+\frac{d^{2}W}{d\varphi ^{2}}\left( H_{[\mu }^{\ast
}C_{\nu \rho \lambda ]}^{\ast }+C_{[\mu \nu }^{\ast }C_{\rho \lambda
]}^{\ast }\right)  \notag \\
&&+\frac{d^{3}W}{d\varphi ^{3}}H_{[\mu }^{\ast }H_{\nu }^{\ast }C_{\rho
\lambda ]}^{\ast }+\frac{d^{4}W}{d\varphi ^{4}}H_{\mu }^{\ast }H_{\nu
}^{\ast }H_{\rho }^{\ast }H_{\lambda }^{\ast },  \label{3.14}
\end{eqnarray}%
\begin{equation}
\left( W\right) _{\mu \nu \rho }=\frac{dW}{d\varphi }C_{\mu \nu \rho }^{\ast
}+\frac{d^{2}W}{d\varphi ^{2}}H_{[\mu }^{\ast }C_{\nu \rho ]}^{\ast }+\frac{%
d^{3}W}{d\varphi ^{3}}H_{\mu }^{\ast }H_{\nu }^{\ast }H_{\rho }^{\ast },
\label{3.15}
\end{equation}%
\begin{equation}
\left( W\right) _{\mu \nu }=\frac{dW}{d\varphi }C_{\mu \nu }^{\ast }+\frac{%
d^{2}W}{d\varphi ^{2}}H_{\mu }^{\ast }H_{\nu }^{\ast },  \label{3.16}
\end{equation}%
whit $W=W\left( \varphi \right) $ an arbitrary, smooth function depending
only on the undifferentiated scalar field $\varphi $.

In contrast to the spaces $\left( H_{J}(\delta \vert d)\right) _{J\geq 2}$
and $\left( H_{J}^{\mathrm{inv}}(\delta \vert d)\right) _{J\geq 2}$, which
are finite-dimensional, the cohomology $H_{1}(\delta \vert d)$ (known to be
related to global symmetries and ordinary conservation laws) is
infinite-dimensional since the theory is free. Fortunately, it will not be
needed in the sequel.

The previous results on $H(\delta |d)$ and $H^{\mathrm{inv}}(\delta |d)$ in
strictly positive antighost numbers are important because they control the
obstructions to removing the antifields from the first-order deformation.
More precisely, we can successively eliminate all the pieces of antighost
number strictly greater that five from the nonintegrated density of the
first-order deformation by adding solely trivial terms, so we can take,
without loss of nontrivial objects, the condition $I\leq 5$ into (\ref{3.2}%
). In addition, the last representative is of the form (\ref{3.8}), where
the invariant `polynomial' is necessarily a nontrivial object from $H_{5}^{%
\mathrm{inv}}(\delta |d)$.

\section{Computation of first-order deformation\label{first}}

In the case $I=5$ the nonintegrated density of the first-order deformation
(see (\ref{3.2})) becomes
\begin{equation}
a=a_{0}+a_{1}+a_{2}+a_{3}+a_{4}+a_{5}.  \label{fo1}
\end{equation}%
We can further decompose $a$ in a natural manner as a sum between two kinds
of deformations
\begin{equation}
a=a^{\mathrm{BF}}+a^{\mathrm{int}},  \label{fo2}
\end{equation}%
where $a^{\mathrm{BF}}$ contains only fields/ghosts/antifields from the BF
sector and $a^{\mathrm{int}}$ describes the cross-interactions between the
two theories.\footnote{%
Decomposition (\ref{fo2}) does not include a component responsible for the
self-interactions of the tensor field with the mixed symmetry $(2,1)$ since
any such component has been proved in~\cite{lingr} to be trivial.} The piece
$a^{\mathrm{BF}}$ is completely known~\cite{juviBFD5}. It is parameterized
by seven smooth, but otherwise arbitrary functions of the undifferentiated
scalar field, $\left( W_{a}\left( \varphi \right) \right) _{a=\overline{1,6}%
} $ and $\bar{M}(\varphi )$. In the sequel we analyze the cross-interacting
piece, $a^{\mathrm{int}}$.

Due to the fact that $a^{\mathrm{BF}}$ and $a^{\mathrm{int}}$ involve
different types of fields and that $a^{\mathrm{BF}}$ separately satisfies an
equation of the type (\ref{3.1}), it follows that $a^{\mathrm{int}}$ is
subject to the equation
\begin{equation}
sa^{\mathrm{int}}=\partial ^{\mu }m_{\mu }^{\mathrm{int}},  \label{fo3}
\end{equation}%
for some local current $m_{\mu }^{\mathrm{int}}$. In the sequel we determine
the general solution to (\ref{fo3}) that complies with all the hypotheses
mentioned in the beginning of section \ref{stand}.

In agreement with (\ref{fo1}), the general solution to the equation $sa^{%
\mathrm{int}}=\partial ^{\mu }m_{\mu }^{\mathrm{int}}$ can be chosen to stop
at antighost number $I=5$
\begin{equation}
a^{\mathrm{int}}=a_{0}^{\mathrm{int}}+a_{1}^{\mathrm{int}}+a_{2}^{\mathrm{int%
}}+a_{3}^{\mathrm{int}}+a_{4}^{\mathrm{int}}+a_{5}^{\mathrm{int}}.
\label{fo1int}
\end{equation}%
We will show in Appendixes \ref{I=5}, \ref{I=4} and \ref{I=3} that we can
always take $a_{5}^{\mathrm{int}}=a_{4}^{\mathrm{int}}=a_{3}^{\mathrm{int}%
}=0 $ into decomposition (\ref{fo1int}), without loss of nontrivial
contributions. Consequently, the first-order deformation of the solution to
the master equation in the interacting case can be taken to stop at
antighost number two%
\begin{equation}
a^{\mathrm{int}}=a_{0}^{\mathrm{int}}+a_{1}^{\mathrm{int}}+a_{2}^{\mathrm{int%
}},  \label{nfo1}
\end{equation}%
where the components on the right-hand side of (\ref{nfo1}) are subject to
equations (\ref{3.6}) and (\ref{3.4})--(\ref{3.5}) for $I=2$.

The piece $a_{2}^{\mathrm{int}}$ as solution to equation (\ref{3.6}) for $I=2
$ has the general form expressed by (\ref{3.8}) for $I=2$, with $\alpha _{2}$
from $H_{2}^{\mathrm{inv}}(\delta |d)$. Looking at formula (\ref{hgamma})
and also at relation (\ref{hinvdelta}) in antighost number two and requiring
that $a_{2}^{\mathrm{int}}$ mixes BRST generators from the BF and $(2,1)$
sectors, we get that the most general solution to (\ref{3.6}) for $I=2$
reads as\footnote{%
In principle, one can add to $a_{2}^{\mathrm{int}}$ the terms $\left( \tilde{%
M}_{2}\right) ^{\mu \nu \rho }\eta D_{\mu \nu \rho }+\tfrac{1}{2}\left(
M_{3}\right) ^{\mu \nu }\sigma ^{\alpha \beta }\tilde{D}_{\mu \alpha }\tilde{%
D}_{\nu \beta }$, where $\left( \tilde{M}_{2}\right) ^{\mu \nu \rho }$ is
the Hodge dual of an expression similar to (\ref{3.16}) with $W\left(
\varphi \right) \rightarrow M_{2}\left( \varphi \right) $, and $\left(
M_{3}\right) ^{\mu \nu }$ reads as in (\ref{3.16}) with $W\left( \varphi
\right) \rightarrow M_{3}\left( \varphi \right) $. Both $M_{2}$ and $M_{3}$
are some arbitrary, real, smooth functions depending on the undifferentiated
scalar field. It can be shown that the above terms finally lead to trivial
interactions, so they can be removed from the first-order deformation.}%
\begin{eqnarray}
a_{2}^{\mathrm{int}} &=&q_{9}\eta ^{\ast \mu \nu \rho }\eta D_{\mu \nu \rho
}+\left( q_{10}\mathcal{\tilde{G}}^{\ast \mu }+q_{11}C^{\ast \mu }\right)
S_{\mu }+q_{12}A^{\ast \mu \nu }\eta \tilde{D}_{\mu \nu }  \notag \\
&&+\tfrac{q_{13}}{2}\tilde{\eta}^{\ast \mu \nu }\sigma ^{\alpha \beta }%
\tilde{D}_{\mu \alpha }\tilde{D}_{\nu \beta }+S^{\ast }\left( k_{1}C+k_{2}%
\mathcal{\tilde{G}}\right) ,  \label{atil2}
\end{eqnarray}%
where all quantities denoted by $q$ or $k\ $are some real, arbitrary
constants.

In the above and from now on we will use a compact writing in terms of the
Hodge duals%
\begin{equation}
\tilde{\Psi}^{\nu _{1}\ldots \nu _{j}}=\tfrac{1}{\left( 5-j\right) !}%
\varepsilon ^{\nu _{1}\ldots \nu _{j}\mu _{1}\ldots \mu _{5-j}}\Psi _{\mu
_{1}\ldots \mu _{5-j}}.  \label{hodge}
\end{equation}%
Consequently $\tilde{\eta}^{\ast \mu \nu }$, $\mathcal{\tilde{G}}^{\ast
\varepsilon }$ and $\mathcal{\tilde{G}}_{\rho }$ are the Hodge duals of $%
\eta _{\rho \lambda \sigma }^{\ast }$, $\mathcal{G}_{\mu \nu \rho \lambda
}^{\ast }$, and respectively $\mathcal{G}^{\mu \nu \lambda \sigma }$.

Substituting (\ref{atil2}) in (\ref{3.4}) for $I=2$ and using definitions (%
\ref{f12a})--(\ref{f13i}), we determine the solution $a_{1}^{\mathrm{int}}$
under the form%
\begin{eqnarray}
a_{1}^{\mathrm{int}} &=&-3q_{9}B^{\ast \mu \nu }\left( V^{\rho }D_{\mu \nu
\rho }+\tfrac{1}{2}\eta F_{\mu \nu }\right) -\left( \tfrac{q_{10}}{2}\tilde{K%
}^{\ast \mu \nu }+q_{11}\phi ^{\ast \mu \nu }\right) A_{\mu \nu }  \notag \\
&&-3q_{12}t^{\ast \mu \nu |\rho }\left( V_{\rho }\tilde{D}_{\mu \nu }+\tfrac{%
1}{2}\eta \tilde{F}_{\mu \nu |\rho }\right) +\tfrac{q_{13}}{2}\tilde{B}%
^{\ast \mu \nu \rho }\sigma ^{\alpha \beta }\tilde{F}_{\mu \alpha |\rho }%
\tilde{D}_{\nu \beta }  \notag \\
&&-2t_{\mu }^{\ast }\left( k_{1}C^{\mu }-\tfrac{k_{2}}{5}\mathcal{\tilde{G}}%
^{\mu }\right) +\bar{a}_{1}^{\mathrm{int}},  \label{ai1n}
\end{eqnarray}%
where $\tilde{F}_{\lambda \mu |\alpha }$ is the Hodge dual of $F_{\qquad
|\alpha }^{\nu \rho \sigma }$ defined in (\ref{nott1}) with respect to its
first three indices
\begin{equation}
\tilde{F}_{\lambda \mu |\alpha }=\tfrac{1}{3!}\varepsilon _{\lambda \mu \nu
\rho \sigma }F_{\qquad |\alpha }^{\nu \rho \sigma }.  \label{dualf}
\end{equation}%
In the last formulas $\tilde{K}_{\lambda \sigma }$ is the dual of the
three-form $K^{\mu \nu \rho }$ from action (\ref{f1}), $\tilde{B}^{\ast \rho
\lambda \sigma }$ and $\tilde{K}^{\ast \lambda \sigma }$ represent the duals
of the antifields $B_{\mu \nu }^{\ast }$ and respectively $K_{\mu \nu \rho
}^{\ast }$ from (\ref{f10a'}).

In the above $\bar{a}_{1}^{\mathrm{int}}$ is the solution to the homogeneous
equation (\ref{3.6}) in antighost number one, meaning that $\bar{a}_{1}^{%
\mathrm{int}}$ is a nontrivial object from $H\left( \gamma \right) $ in pure
ghost number one and in antighost number one. It is useful to decompose $%
\bar{a}_{1}^{\mathrm{int}}$ like in (\ref{descom2})
\begin{equation}
\bar{a}_{1}^{\mathrm{int}}=\hat{a}_{1}^{\mathrm{int}}+\check{a}_{1}^{\mathrm{%
int}},  \label{desca1}
\end{equation}%
with $\hat{a}_{1}^{\mathrm{int}}$ the solution to (\ref{3.6}) for $I=1$ that
ensures the consistency of $a_{1}^{\mathrm{int}}$ in antighost number zero,
namely the existence of $a_{0}^{\mathrm{int}}$ as solution to (\ref{3.5})
for $i=1$ with respect to the terms from $a_{1}^{\mathrm{int}}$ containing
the constants of the type $q$ or $k$, and $\check{a}_{1}^{\mathrm{int}}$ the
solution to (\ref{3.6}) for $I=1$ that is independently consistent in
antighost number zero%
\begin{equation}
\delta \check{a}_{1}^{\mathrm{int}}=-\gamma \check{c}_{0}+\partial _{\mu }%
\check{m}_{0}^{\mu }.  \label{cai1}
\end{equation}%
With the help of definitions (\ref{f12a})--(\ref{f13i}) and taking into
account decomposition (\ref{descom2}), we infer by direct computation%
\begin{eqnarray}
\delta a_{1}^{\mathrm{int}} &=&\delta \left[ \hat{a}_{1}^{\mathrm{int}%
}+\left( 2k_{1}K^{\ast \mu \nu \rho }+\tfrac{k_{2}}{30}\tilde{K}^{\ast \mu
\nu \rho }\right) D_{\mu \nu \rho }\right]  \notag \\
&&+\gamma c_{0}+\partial _{\lambda }j_{0}^{\lambda }+\chi _{0},
\label{ai1na}
\end{eqnarray}%
where%
\begin{equation}
c_{0}=-\check{c}_{0}+\tfrac{q_{13}}{16}\tilde{V}^{\mu \nu \rho \lambda
}\sigma ^{\alpha \beta }\tilde{F}_{\mu \alpha |\rho }\tilde{F}_{\nu \beta
|\lambda }-\left( k_{1}\phi ^{\mu \nu }-\tfrac{k_{2}}{20}\tilde{K}^{\mu \nu
}\right) F_{\mu \nu },  \label{ai1nb}
\end{equation}%
\begin{eqnarray}
\chi _{0} &=&-3q_{9}\left[ \left( \partial ^{\lbrack \mu }V^{\nu ]}\right)
V^{\rho }D_{\mu \nu \rho }+V^{\mu }\left( \partial ^{\nu }\eta \right)
F_{\mu \nu }-V^{\mu }\eta R_{\mu }\right]  \notag \\
&&+\tfrac{1}{18}\left( q_{10}\tilde{\phi}^{\mu \nu \rho }+6q_{11}K^{\mu \nu
\rho }\right) D_{\mu \nu \rho }  \notag \\
&&-\tfrac{3q_{12}}{4}\left[ \partial _{\rho }\left( F^{\rho \mu \nu |\alpha
}-\sigma ^{\alpha \lbrack \mu }F^{\nu \rho ]}\right) \right] \left(
2V_{\alpha }\tilde{D}_{\mu \nu }+\eta \tilde{F}_{\mu \nu |\alpha }\right)
\notag \\
&&+\tfrac{q_{13}}{8}\varepsilon ^{\mu \nu \rho \lambda \sigma }\sigma
^{\alpha \beta }\tilde{R}_{\mu \alpha |\lambda \sigma }\left( 2V_{\rho }%
\tilde{D}_{\nu \beta }+\eta \tilde{F}_{\nu \beta |\rho }\right) ,
\label{ai1nd}
\end{eqnarray}%
and $j_{0}^{\lambda }$ are some local currents. In the above $\tilde{V}^{\mu
\nu \rho \lambda }$ and $\tilde{\phi}^{\mu \nu \lambda }$ represent the
Hodge duals of the one-form $V_{\sigma }$ and respectively of the two-form $%
\phi _{\rho \sigma }$ from (\ref{f10a}) and $\tilde{R}_{\lambda \sigma
|\alpha \beta }$ is nothing but the Hodge dual of the tensor $R_{\qquad
|\alpha \beta }^{\mu \nu \rho }$ defined in (\ref{fie}) with respect to its
first three indices, namely%
\begin{equation}
\tilde{R}_{\lambda \sigma |\alpha \beta }=\tfrac{1}{3!}\varepsilon _{\lambda
\sigma \mu \nu \rho }R_{\qquad |\alpha \beta }^{\mu \nu \rho }.
\label{dualr}
\end{equation}%
Inspecting (\ref{ai1na}), we observe that equation (\ref{3.5}) for $i=1$
possesses solutions if and only if $\chi _{0}$ expressed by (\ref{ai1nd}) is
$\gamma $-exact modulo $d$. A straightforward analysis of $\chi _{0}$ shows
that this is not possible unless%
\begin{equation}
q_{9}=q_{10}=q_{11}=q_{12}=q_{13}=0.  \label{ecconst}
\end{equation}

Now, we insert conditions (\ref{ecconst}) in (\ref{atil2}) and identify the
most general form of the first-order deformation in the interacting sector
at antighost number two
\begin{equation}
a_{2}^{\mathrm{int}}=S^{\ast }\left( k_{1}C+k_{2}\mathcal{\tilde{G}}\right) .
\label{ai2good}
\end{equation}%
The same conditions replaced in (\ref{ai1na}) enable us to write
\begin{equation}
\hat{a}_{1}^{\mathrm{int}}=-\left( 2k_{1}K^{\ast \mu \nu \rho }+\tfrac{k_{2}%
}{30}\tilde{\phi}^{\ast \mu \nu \rho }\right) D_{\mu \nu \rho }.
\label{om1nec}
\end{equation}%
Introducing (\ref{om1nec}) in (\ref{desca1}) and then the resulting result
together with (\ref{ecconst}) in (\ref{ai1n}), we obtain%
\begin{equation}
a_{1}^{\mathrm{int}}=-2t_{\mu }^{\ast }\left( k_{1}C^{\mu }+\tfrac{k_{2}}{5}%
\mathcal{\tilde{G}}^{\mu }\right) -\left( 2k_{1}K^{\ast \mu \nu \rho }+%
\tfrac{k_{2}}{30}\tilde{\phi}^{\ast \mu \nu \rho }\right) D_{\mu \nu \rho }+%
\check{a}_{1}^{\mathrm{int}}.  \label{ai1par}
\end{equation}

Next, we determine $\check{a}_{1}^{\mathrm{int}}$ as the solution to the
homogeneous equation (\ref{3.6}) for $I=1$ that is independently consistent
in antighost number zero, i.e. satisfies equation (\ref{cai1}). According to
(\ref{3.8}) for $I=1$ the general solution to equation (\ref{3.6}) for $I=1$
has the form
\begin{eqnarray}
\check{a}_{1}^{\mathrm{int}} &=&t^{\ast \mu \nu |\rho }\left( L_{\mu \nu
|\rho }\eta +L_{\mu \nu |\rho }^{\alpha \beta \gamma }D_{\alpha \beta \gamma
}\right) +\left( V_{\alpha }^{\ast }M_{\mu \nu \rho }^{\alpha }+\varphi
^{\ast }M_{\mu \nu \rho }\right.  \notag \\
&&\left. +H_{\alpha }^{\ast }\bar{M}_{\mu \nu \rho }^{\alpha }+B_{\alpha
\beta }^{\ast }M_{\mu \nu \rho }^{\alpha \beta }+\phi _{\alpha \beta }^{\ast
}\bar{M}_{\mu \nu \rho }^{\alpha \beta }+K_{\alpha \beta \gamma }^{\ast
}M_{\mu \nu \rho }^{\alpha \beta \gamma }\right) D^{\mu \nu \rho }  \notag \\
&&+\left( V_{\alpha }^{\ast }N^{\alpha }+\varphi ^{\ast }N+H_{\alpha }^{\ast
}\bar{N}^{\alpha }+B_{\alpha \beta }^{\ast }N^{\alpha \beta }+\phi _{\alpha
\beta }^{\ast }\bar{N}^{\alpha \beta }\right.  \notag \\
&&\left. +K_{\alpha \beta \gamma }^{\ast }N^{\alpha \beta \gamma }\right)
\eta ,  \label{solom1}
\end{eqnarray}%
where all the quantities denoted by $L$, $M$, $N$, $\bar{M}$, or $\bar{N}$
are bosonic, gauge-invariant tensors, and therefore they may depend only on $%
F_{\bar{A}}$ given in (\ref{3.7}) and their spacetime derivatives. The
functions $L_{\mu \nu |\rho }$ and $L_{\mu \nu |\rho }^{\alpha \beta \gamma
} $ exhibit the mixed symmetry $(2,1)$ with respect to their lower indices
and, in addition, $L_{\mu \nu |\rho }^{\alpha \beta \gamma }$ is completely
antisymmetric with respect to its upper indices. The remaining functions, $M$%
, $\bar{M}$, $N$, and $\bar{N}$, are separately antisymmetric (where
appropriate) in their upper and respectively lower indices.

In order to determine all possible solutions (\ref{solom1}) we demand that $%
\check{a}_{1}^{\mathrm{int}}$ mixes the BF and $(2,1)$ sectors and (for the
first time) explicitly implement the assumption on the derivative order of
the interacting Lagrangian discussed in the beginning of section \ref{stand}
and structured in requirements i) and ii). Because all the terms involving
the functions $N$ or $\bar{N}$ contain only BRST generators from the BF
sector, it follows that each such function must contain at least one tensor $%
R_{\mu \nu \rho |\alpha \beta }$ defined in (\ref{fie}), with $F$ as in (\ref%
{nott1}). The corresponding terms from $\check{a}_{1}^{\mathrm{int}}$, if
consistent, would produce an interacting Lagrangian that does not agree with
requirement ii) with respect to the BF fields and therefore we must take%
\begin{equation}
N^{\alpha }=N=\bar{N}^{\alpha }=N^{\alpha \beta }=\bar{N}^{\alpha \beta
}=N^{\alpha \beta \gamma }=0.  \label{eccona1a}
\end{equation}%
In the meantime, requirement ii) also restricts all the functions $M$ and $%
\bar{M}$ to be derivative-free. Since the undifferentiated scalar field is
the only element among $F_{\bar{A}}$ and their spacetime derivatives that
contains no derivatives, it follows that all $M$ and $\bar{M}$ may depend at
most on $\varphi $. Due to the fact that we work in $D=5$ and taking into
account the various antisymmetry properties of these functions, it follows
that the only eligible representations are%
\begin{gather}
M_{\mu \nu \rho }^{\alpha }=M_{\mu \nu \rho }=\bar{M}_{\mu \nu \rho
}^{\alpha }=0,  \label{eccona1b} \\
M_{\mu \nu \rho }^{\alpha \beta }=U_{13}\varepsilon _{\quad \mu \nu \rho
}^{\alpha \beta },\qquad \bar{M}_{\mu \nu \rho }^{\alpha \beta
}=U_{14}\varepsilon _{\quad \mu \nu \rho }^{\alpha \beta },\qquad M_{\mu \nu
\rho }^{\alpha \beta \gamma }=\tfrac{1}{6}U_{15}\delta _{\lbrack \mu
}^{\alpha }\delta _{\nu }^{\beta }\delta _{\rho ]}^{\gamma },
\label{eccona1c}
\end{gather}%
with $U_{13}$, $U_{14}$, and $U_{15}$ some real, smooth functions of $%
\varphi $. The same observation stands for $L_{\mu \nu |\rho }$ and $L_{\mu
\nu |\rho }^{\alpha \beta \gamma }$, so their tensorial behaviour can only
be realized via some constant Lorentz tensors. Nevertheless, there is no
such constant tensor in $D=5$ with the required mixed symmetry properties,
and hence we must put
\begin{equation}
L_{\mu \nu |\rho }=0,\qquad L_{\mu \nu |\rho }^{\alpha \beta \gamma }=0.
\label{eccona1e}
\end{equation}%
Inserting results (\ref{eccona1a})--(\ref{eccona1e}) in (\ref{solom1}), it
follows that the most general (nontrivial) solution to equation (\ref{3.6})
for $I=1$ that complies with all the working hypotheses, including that on
the differential order of the interacting Lagrangian, is given by%
\begin{equation}
\check{a}_{1}^{\mathrm{int}}=\varepsilon ^{\mu \nu \rho \lambda \sigma
}\left( U_{13}B_{\mu \nu }^{\ast }+U_{14}\phi _{\mu \nu }^{\ast }\right)
D_{\rho \lambda \sigma }+U_{15}K^{\ast \mu \nu \rho }D_{\mu \nu \rho }.
\label{solom1a}
\end{equation}

By acting with $\delta $ on (\ref{solom1a}) and using definitions (\ref{f12a}%
)--(\ref{f13i}) we infer
\begin{eqnarray}
\delta \check{a}_{1}^{\mathrm{int}} &=&\gamma \left[ \left( -3U_{14}\tilde{K}%
^{\mu \nu }+2U_{15}\phi ^{\mu \nu }\right) F_{\mu \nu }\right]  \notag \\
&&+\partial _{\alpha }\left( \varepsilon ^{\mu \nu \rho \lambda \alpha
}U_{13}V_{\mu }D_{\nu \rho \lambda }-\varepsilon _{\mu \nu \rho \lambda
\sigma }U_{14}K^{\alpha \mu \nu }D^{\rho \lambda \sigma }+U_{14}\phi _{\mu
\nu }D^{\alpha \mu \nu }\right)  \notag \\
&&+\varepsilon _{\mu \nu \rho \lambda \sigma }\left[ -\left( \partial ^{\mu
}U_{13}\right) V^{\nu }+\left( \partial _{\alpha }U_{14}\right) K^{\alpha
\mu \nu }\right] D^{\rho \lambda \sigma }  \notag \\
&&-\left( \partial _{\mu }U_{15}\right) \phi _{\nu \rho }D^{\mu \nu \rho
}+2F^{\mu \nu }\left( 6U_{14}\partial _{\lbrack \mu }\mathcal{\tilde{G}}%
_{\nu ]}-U_{15}\partial _{\lbrack \mu }C_{\nu ]}\right) .  \label{conssolom1}
\end{eqnarray}%
Comparing (\ref{conssolom1}) with (\ref{cai1}), we conclude that function $%
U_{13}$ reduces to a real constant and meanwhile functions $U_{14}$ and $%
U_{15}$ must vanish%
\begin{equation}
U_{13}=u_{13},\qquad U_{14}=0=U_{15},  \label{eccon1f}
\end{equation}%
so (\ref{solom1a}) becomes%
\begin{equation}
\check{a}_{1}^{\mathrm{int}}=\varepsilon ^{\mu \nu \rho \lambda \sigma
}u_{13}B_{\mu \nu }^{\ast }D_{\rho \lambda \sigma },  \label{fina1checkint}
\end{equation}%
wich produces trivial deformations because it is a trivial element from $%
H_{1}(\delta |d)$
\begin{equation}
\check{a}_{1}^{\mathrm{int}}=\delta \left( \varepsilon ^{\mu \nu \rho
\lambda \sigma }u_{13}\eta _{\mu \nu \rho }^{\ast }A_{\lambda \sigma
}\right) +\partial _{\mu }\left( \varepsilon ^{\mu \nu \rho \lambda \sigma
}u_{13}B_{\nu \rho }^{\ast }A_{\lambda \sigma }\right)  \label{eccon1g}
\end{equation}%
and by further taking
\begin{equation}
\check{a}_{1}^{\mathrm{int}}=0.  \label{eccon1h}
\end{equation}

As a consequence, we can safely take the nontrivial part of the first-order
deformation in the interaction sector in antighost number one, (\ref{ai1par}%
), of the form
\begin{equation}
a_{1}^{\mathrm{int}}=-2t_{\mu }^{\ast }\left( k_{1}C^{\mu }+\tfrac{k_{2}}{5}%
\mathcal{\tilde{G}}^{\mu }\right) -\left( 2k_{1}K^{\ast \mu \nu \rho }+%
\tfrac{k_{2}}{30}\tilde{\phi}^{\ast \mu \nu \rho }\right) D_{\mu \nu \rho }.
\label{ai1good}
\end{equation}%
In addition, (\ref{eccon1h}) leads to
\begin{equation}
\check{c}_{0}=0,\qquad \check{m}_{0}^{\mu }=0  \label{eccon1i}
\end{equation}%
in (\ref{cai1}). Replacing now (\ref{ecconst}) and (\ref{eccon1i}) in (\ref%
{ai1na}), we are able to identify the piece of antighost number zero from
the first-order deformation in the interacting sector as%
\begin{equation}
a_{0}^{\mathrm{int}}=\left( k_{1}\phi ^{\mu \nu }-\tfrac{k_{2}}{20}\tilde{K}%
^{\mu \nu }\right) F_{\mu \nu }+\bar{a}_{0}^{\mathrm{int}},  \label{ai0n}
\end{equation}%
where $\bar{a}_{0}^{\mathrm{int}}$ is the solution to the `homogeneous'
equation in antighost number zero
\begin{equation}
\gamma \bar{a}_{0}^{\mathrm{int}}=\partial _{\mu }\bar{m}_{0}^{\mu }.
\label{om0}
\end{equation}%
We will prove in Appendix \ref{I=0} that the only solution to (\ref{om0})
that satisfies all our working hypotheses, including that on the derivative
order of the interacting Lagrangian, is $\bar{a}_{0}^{\mathrm{int}}=0$, such
that the nontrivial part of the first-order deformation in the interaction
sector in antighost number zero reads as
\begin{equation}
a_{0}^{\mathrm{int}}=\left( k_{1}\phi ^{\mu \nu }-\tfrac{k_{2}}{20}\tilde{K}%
^{\mu \nu }\right) F_{\mu \nu }.  \label{ai0good}
\end{equation}

The main conclusion of this section is that the general form of the
first-order deformation of the solution to the master equation as solution
to (\ref{bfa2.5}) for the model under study is expressed by%
\begin{equation}
S_{1}=\int d^{5}x\left( a^{\mathrm{BF}}+a^{\mathrm{int}}\right) ,
\label{s1deffinal}
\end{equation}%
where $a^{\mathrm{BF}}$ can be found in \cite{juviBFD5}\ and%
\begin{eqnarray}
a^{\mathrm{int}} &=&a_{0}^{\mathrm{int}}+a_{1}^{\mathrm{int}}+a_{2}^{\mathrm{%
int}}  \notag \\
&=&S^{\ast }\left( k_{1}C+k_{2}\mathcal{\tilde{G}}\right) -2t_{\mu }^{\ast
}\left( k_{1}C^{\mu }+\tfrac{k_{2}}{5}\mathcal{\tilde{G}}^{\mu }\right)
\notag \\
&&-\left( 2k_{1}K^{\ast \mu \nu \rho }+\tfrac{k_{2}}{30}\tilde{\phi}^{\ast
\mu \nu \rho }\right) D_{\mu \nu \rho }+\left( k_{1}\phi ^{\mu \nu }-\tfrac{%
k_{2}}{20}\tilde{K}^{\mu \nu }\right) F_{\mu \nu }.  \label{aintgood}
\end{eqnarray}%
It is now clear that the first-order deformation is parameterized by seven
arbitrary, smooth functions of the undifferentiated scalar field ($\left(
W_{a}\left( \varphi \right) \right) _{a=\overline{1,6}}$ and $\bar{M}%
(\varphi )$ corresponding to $a^{\mathrm{BF}}$ and by two arbitrary, real
constants ($k_{1}$ and $k_{2}$ from $a^{\mathrm{int}}$). We will see in the
next section that the consistency of the deformed solution to the master
equation in order two in the coupling constant will restrict these functions
and constants to satisfy some specific equations.

\section{Computation of higher-order deformations\label{higher}}

With the first-order deformation at hand, in the sequel we determine the
higher-order deformations of the solution to the master equation, governed
by equations (\ref{bfa2.6})--(\ref{bfa2.8}), etc., which comply with our
working hypotheses.

In the first step we approach the second-order deformation, $S_{2}$, as
(nontrivial) solution to equation (\ref{bfa2.6}). If we denote by $\Delta $
the nonintegrated density of the antibracket $\left( S_{1},S_{1}\right) $
and by $b$ the nonintegrated density associated with $S_{2}$,%
\begin{equation}
\left( S_{1},S_{1}\right) =\int d^{5}x\ \Delta ,\qquad S_{2}=\int d^{5}x\ b,
\label{notts2}
\end{equation}%
then equation (\ref{bfa2.6}) takes the local form%
\begin{equation}
\Delta +2sb=\partial _{\mu }n^{\mu },  \label{lbfa2.6}
\end{equation}%
with $n^{\mu }$ a local current. By direct computation it follows that $%
\Delta $ decomposes as%
\begin{equation}
\Delta =\Delta ^{\mathrm{BF}}+\Delta ^{\mathrm{int}},  \label{desc_d}
\end{equation}%
where $\Delta ^{\mathrm{BF}}$ involves only BRST generators from the BF
sector and each term from $\Delta ^{\mathrm{int}}$ depends simultaneously on
the BRST generators of both sectors (BF and mixed symmetry $(2,1)$), such
that $\Delta ^{\mathrm{int}}$ couples the two theories. Consequently,
decomposition (\ref{desc_d}) induces a similar one at the level of the
second-order deformation%
\begin{equation}
b=b^{\mathrm{BF}}+b^{\mathrm{int}}  \label{desc_b}
\end{equation}%
and equation (\ref{lbfa2.6}) becomes equivalent to two equations, one for
the BF sector and the other for the interacting sector
\begin{eqnarray}
\Delta ^{\mathrm{BF}}+2sb^{\mathrm{BF}} &=&\partial ^{\mu }n_{\mu }^{\mathrm{%
BF}},  \label{ecl2_bf} \\
\Delta ^{\mathrm{int}}+2sb^{\mathrm{int}} &=&\partial ^{\mu }n_{\mu }^{%
\mathrm{int}}.  \label{ecl2_int}
\end{eqnarray}

Equation (\ref{ecl2_bf}) has been completely solved in~\cite{juviBFD5},
where it was shown that it possesses only the trivial solution%
\begin{equation}
b^{\mathrm{BF}}=0  \label{s2BF}
\end{equation}%
and, in addition, the seven functions $\left( W_{a}\right) _{a=\overline{1,6}%
}$ and $\bar{M}(\varphi )$ that parameterize $a^{\mathrm{BF}}$ are subject
to the following equations:
\begin{gather}
\frac{d\bar{M}\left( \varphi \right) }{d\varphi }W_{1}\left( \varphi \right)
=0,\qquad W_{1}\left( \varphi \right) W_{2}\left( \varphi \right) =0,
\label{ecc1} \\
W_{1}\left( \varphi \right) \frac{dW_{2}\left( \varphi \right) }{d\varphi }%
-3W_{2}\left( \varphi \right) W_{3}\left( \varphi \right) +6W_{5}\left(
\varphi \right) W_{6}\left( \varphi \right) =0,  \label{ecc2} \\
W_{2}\left( \varphi \right) W_{3}\left( \varphi \right) +W_{5}\left( \varphi
\right) W_{6}\left( \varphi \right) =0,  \label{ecc3} \\
W_{1}\left( \varphi \right) \frac{dW_{6}\left( \varphi \right) }{d\varphi }%
+3W_{3}\left( \varphi \right) W_{6}\left( \varphi \right) -6W_{2}\left(
\varphi \right) W_{4}\left( \varphi \right) =0,  \label{ecc4} \\
W_{1}\left( \varphi \right) W_{6}\left( \varphi \right) =0,\qquad
W_{2}\left( \varphi \right) W_{4}\left( \varphi \right) +W_{3}\left( \varphi
\right) W_{6}\left( \varphi \right) =0,  \label{ecc5} \\
W_{2}\left( \varphi \right) W_{5}\left( \varphi \right) =0,\qquad
W_{4}\left( \varphi \right) W_{6}\left( \varphi \right) =0.  \label{ecc6}
\end{gather}%
Now, we investigate the latter equation, (\ref{ecl2_int}). By direct
computation $\Delta ^{\mathrm{int}}$ can be brought to the form%
\begin{eqnarray}
\Delta ^{\mathrm{int}} &=&s\left[ -3\left( k_{1}\phi _{\mu \nu }-\tfrac{k_{2}%
}{20}\tilde{K}_{\mu \nu }\right) \left( k_{1}\phi ^{\mu \nu }-\tfrac{k_{2}}{%
20}\tilde{K}^{\mu \nu }\right) \right]  \notag \\
&&+\bar{\Delta}^{\mathrm{int}}+\partial ^{\mu }\bar{n}_{\mu }^{\mathrm{int}},
\label{del_int}
\end{eqnarray}%
where $\bar{n}_{\mu }^{\mathrm{int}}$ is a local current and
\begin{equation}
\bar{\Delta}^{\mathrm{int}}=\sum\limits_{i=1}^{3}\sum\limits_{p=0}^{3}\frac{%
d^{p}\bar{Y}^{\left( i\right) }}{d\varphi ^{p}}\bar{X}_{p}^{\left( i\right)
}.  \label{delbar_int}
\end{equation}%
In $\bar{\Delta}^{\mathrm{int}}$ we used the notations%
\begin{gather}
\bar{Y}^{\left( 1\right) }=k_{1}W_{3}+\tfrac{k_{2}}{60}W_{5},\qquad \bar{Y}%
^{\left( 2\right) }=k_{1}W_{4}+\tfrac{k_{2}}{2\cdot 5!}W_{3},  \label{ybar1}
\\
\bar{Y}^{\left( 3\right) }=k_{1}W_{6}+\tfrac{k_{2}}{5!}W_{2},  \label{ybar3}
\end{gather}%
and the polynomials $\bar{X}_{p}^{\left( i\right) }$ are listed in Appendix %
\ref{struct} (see formulas (\ref{xxt1})--(\ref{xxt25})). It can be shown
that (\ref{delbar_int}) cannot be written as a $s$-exact modulo $d$ element
from local functions and therefore it must vanish%
\begin{equation}
\bar{\Delta}^{\mathrm{int}}=0,  \label{delbarint0}
\end{equation}%
which further restricts the functions and constants that parameterize the
first-order deformation to obey the supplementary equations%
\begin{gather}
k_{1}W_{3}+\tfrac{k_{2}}{60}W_{5}=0,\qquad k_{1}W_{4}+\tfrac{k_{2}}{2\cdot 5!%
}W_{3}=0,  \label{ecc7} \\
k_{1}W_{6}+\tfrac{k_{2}}{5!}W_{2}=0.  \label{ecc9}
\end{gather}

As a consequence, the consistency of the first-order deformation at order
two in the coupling constant (the existence of local solutions to equation (%
\ref{bfa2.6})) on the one hand restricts the functions and constants that
parameterize $S_{1}$ to fulfill equations (\ref{ecc1})--(\ref{ecc6}) and (%
\ref{ecc7})--(\ref{ecc9}) and, on the other hand, enables us (via formulas (%
\ref{notts2}), (\ref{desc_b}), (\ref{ecl2_int}), (\ref{s2BF}), (\ref{del_int}%
), and (\ref{delbarint0})) to infer the second-order deformation as
\begin{equation}
S_{2}=S_{2}^{\mathrm{int}}=\int d^{5}x\left[ \tfrac{3}{2}\left( k_{1}\phi
_{\mu \nu }-\tfrac{k_{2}}{20}\tilde{K}_{\mu \nu }\right) \left( k_{1}\phi
^{\mu \nu }-\tfrac{k_{2}}{20}\tilde{K}^{\mu \nu }\right) \right] .
\label{s2}
\end{equation}

In the second step we solve the equation that governs the third-order
deformation, namely, (\ref{bfa2.7}). If we make the notations
\begin{equation}
\left( S_{1},S_{2}\right) =\int d^{5}x\ \Lambda ,\qquad S_{3}=\int d^{5}x\ c,
\label{notts3}
\end{equation}%
then equation (\ref{bfa2.7}) takes the local form%
\begin{equation}
\Lambda +sc=\partial _{\mu }p^{\mu },  \label{ecl_def3}
\end{equation}%
with $p^{\mu }$ a local current. By direct computation we obtain
\begin{equation}
\Lambda =\partial _{\mu }\bar{p}^{\mu
}+\sum\limits_{i=1}^{3}\sum\limits_{p=0}^{2}\frac{d^{p}\bar{Y}^{\left(
i\right) }}{d\varphi ^{p}}U_{p}^{\left( i\right) },  \label{omega}
\end{equation}%
where $\bar{p}^{\mu }$ is a local current and the functions $U_{p}^{\left(
i\right) }$ appearing in the right-hand side of (\ref{omega}) are listed in
Appendix \ref{struct} (see formulas (\ref{u01})--(\ref{u24})). Taking into
account the result that the functions and constants that parameterize both
the first- and second-order deformations satisfy equations (\ref{ecc1})--(%
\ref{ecc6}) and (\ref{ecc7})--(\ref{ecc9}) and comparing (\ref{omega}) with
equation (\ref{ecl_def3}), it results that the third-order deformation can
be chosen to be completely trivial%
\begin{equation}
S_{3}=0.  \label{def3}
\end{equation}

Related to the equation that governs the fourth-order deformation, namely, (%
\ref{bfa2.8}), we have that%
\begin{equation}
2\left( S_{1},S_{3}\right) +\left( S_{2},S_{2}\right) =0.  \label{mml_def4}
\end{equation}%
From (\ref{mml_def4}) and (\ref{bfa2.8}) we find that $S_{4}$\ is completely
trivial%
\begin{equation}
S_{4}=0.  \label{def4}
\end{equation}%
Along a similar line, it can be shown that all the remaining higher-order
deformations $S_{k}$ ($k\geq 5$) can be taken to vanish%
\begin{equation}
S_{k}=0,\qquad k\geq 5.  \label{sk0}
\end{equation}

The main conclusion of this section is that the deformed solution to the
master equation for the model under study, which is consistent to all orders
in the coupling constant, can be taken as%
\begin{equation}
S=\bar{S}+\lambda S_{1}+\lambda ^{2}S_{2},  \label{soldeformata}
\end{equation}%
where $\bar{S}$ reads as in (\ref{f14}), $S_{1}$ is given in (\ref%
{s1deffinal}) with $a^{\mathrm{int}}$ of the form (\ref{aintgood}), and $%
S_{2}$ is expressed by (\ref{s2}). It represents the most general solution
that complies with all our working hypotheses (see the discussion from the
beginning of section \ref{stand}). We cannot stress enough that the (seven)
functions and (two) constants that parameterize the fully deformed solution
to the master equation are no longer independent. They must obey equations (%
\ref{ecc1})--(\ref{ecc6}) and (\ref{ecc7})--(\ref{ecc9}).

\section{The coupled theory: Lagrangian and gauge structure\label{ident}}

In this section we start from the concrete form of (\ref{soldeformata}) and
identify the entire gauge structure of the Lagrangian model that describes
all consistent interactions in $D=5$ between the BF theory and the massless
tensor field $t_{\mu \nu |\alpha }$. To this end we recall the discussion
from the end of section \ref{free} related to the relationship between the
gauge structure of a given Lagrangian field theory and various terms of
definite antighost number present in the solution of the master equation. Of
course, we assume that the functions $\left( W_{a}\right) _{a=\overline{1,6}}
$, $\bar{M}$ together with the constants $k_{1}$ and $k_{2}$ satisfy
equations (\ref{ecc1})--(\ref{ecc6}) and (\ref{ecc7})--(\ref{ecc9}). The
analysis of solutions that are interesting from the point of view of
cross-couplings (at least one of the constants $k_{1}$ and $k_{2}$ is
nonvanishing) is done in Section \ref{clasesol}.

The piece of antighost number zero from (\ref{soldeformata}) provides
nothing but the Lagrangian action of the interacting theory%
\begin{eqnarray}
S^{\mathrm{L}}\left[ \Phi ^{\alpha _{0}}\right] &=&\int d^{5}x\left\{ H_{\mu
}\partial ^{\mu }\varphi +\tfrac{1}{2}B^{\mu \nu }\partial _{\lbrack \mu
}V_{\nu ]}+\tfrac{1}{3}K^{\mu \nu \rho }\partial _{\lbrack \mu }\phi _{\nu
\rho ]}\right.  \notag \\
&&+\lambda \left[ W_{1}V_{\mu }H^{\mu }+W_{2}B_{\mu \nu }\phi ^{\mu \nu
}-W_{3}\phi _{\lbrack \mu \nu }V_{\rho ]}K^{\mu \nu \rho }+\bar{M}(\varphi
)\right.  \notag \\
&&\left. +\varepsilon ^{\alpha \beta \gamma \delta \varepsilon }\left(
9W_{4}V_{\alpha }\tilde{K}_{\beta \gamma }\tilde{K}_{\delta \varepsilon }+%
\tfrac{1}{4}W_{5}V_{\alpha }\phi _{\beta \gamma }\phi _{\delta \varepsilon
}+W_{6}B_{\alpha \beta }K_{\gamma \delta \varepsilon }\right) \right]  \notag
\\
&&-\tfrac{1}{12}\left( F_{\mu \nu \rho |\alpha }F^{\mu \nu \rho |\alpha
}-3F_{\mu \nu }F^{\mu \nu }\right)  \notag \\
&&\left. +\lambda \left( k_{1}\phi ^{\mu \nu }-\tfrac{k_{2}}{20}\tilde{K}%
^{\mu \nu }\right) \left[ F_{\mu \nu }+\tfrac{3\lambda }{2}\left( k_{1}\phi
_{\mu \nu }-\tfrac{k_{2}}{20}\tilde{K}_{\mu \nu }\right) \right] \right\} ,
\label{acLagdef}
\end{eqnarray}%
where $\Phi ^{\alpha _{0}}$ is the field spectrum (\ref{f10a}). The terms of
antighost number one from the deformed solution of the master equation,
generically written as $\Phi _{\alpha _{0}}^{\ast }Z_{\quad \alpha
_{1}}^{\alpha _{0}}\eta ^{\alpha _{1}}$, allow the identification of the
gauge transformations of action (\ref{acLagdef}) via replacing the ghosts $%
\eta ^{\alpha _{1}}$ with the gauge parameters $\Omega ^{\alpha _{1}}$%
\begin{equation}
\bar{\delta}_{\Omega }\Phi ^{\alpha _{0}}=Z_{\quad \alpha _{1}}^{\alpha
_{0}}\Omega ^{\alpha _{1}}.  \label{defgauge}
\end{equation}%
In our case, taking into account formula (\ref{soldeformata}) and
maintaining the notation (\ref{Omegaa1}) for the gauge parameters, we find
the concrete form of the deformed gauge transformations as
\begin{equation}
\bar{\delta}_{\Omega }\varphi =-\lambda W_{1}\epsilon ,  \label{tdef2}
\end{equation}%
\begin{eqnarray}
\bar{\delta}_{\Omega }H^{\mu } &=&2D_{\nu }\epsilon ^{\mu \nu }+\lambda
\left( \frac{dW_{1}}{d\varphi }H^{\mu }-3\frac{dW_{3}}{d\varphi }K^{\mu \nu
\rho }\phi _{\nu \rho }\right) \epsilon  \notag \\
&&-3\lambda \frac{dW_{2}}{d\varphi }\phi _{\nu \rho }\epsilon ^{\mu \nu \rho
}+2\lambda \left( \frac{dW_{2}}{d\varphi }B^{\mu \nu }-3\frac{dW_{3}}{%
d\varphi }K^{\mu \nu \rho }V_{\rho }\right) \xi _{\nu }  \notag \\
&&+12\lambda \frac{dW_{3}}{d\varphi }V_{\nu }\phi _{\rho \lambda }\xi ^{\mu
\nu \rho \lambda }+2\lambda \frac{dW_{6}}{d\varphi }B^{\mu \nu }\varepsilon
_{\nu \alpha \beta \gamma \delta }\xi ^{\alpha \beta \gamma \delta }  \notag
\\
&&+3\lambda K^{\mu \nu \rho }\left( 4\frac{dW_{4}}{d\varphi }V_{\nu
}\varepsilon _{\rho \alpha \beta \gamma \delta }\xi ^{\alpha \beta \gamma
\delta }-\frac{dW_{6}}{d\varphi }\varepsilon _{\nu \rho \alpha \beta \gamma
}\epsilon ^{\alpha \beta \gamma }\right)  \notag \\
&&+\lambda \varepsilon ^{\mu \nu \rho \lambda \sigma }\left[ \tfrac{1}{4}%
\frac{dW_{4}}{d\varphi }\varepsilon _{\nu \rho \alpha \beta \gamma
}K^{\alpha \beta \gamma }\varepsilon _{\lambda \sigma \alpha ^{\prime }\beta
^{\prime }\gamma ^{\prime }}K^{\alpha ^{\prime }\beta ^{\prime }\gamma
^{\prime }}\epsilon \right.  \notag \\
&&\left. -\frac{dW_{5}}{d\varphi }\phi _{\nu \rho }\left( V_{\lambda }\xi
_{\sigma }-\tfrac{1}{4}\phi _{\lambda \sigma }\epsilon \right) \right] ,
\label{tdef7}
\end{eqnarray}%
\begin{equation}
\bar{\delta}_{\Omega }V_{\mu }=\partial _{\mu }\epsilon -2\lambda W_{2}\xi
_{\mu }-2\lambda \varepsilon _{\mu \nu \rho \lambda \sigma }W_{6}\xi ^{\nu
\rho \lambda \sigma },  \label{tdef1}
\end{equation}%
\begin{eqnarray}
\bar{\delta}_{\Omega }B^{\mu \nu } &=&-3\partial _{\rho }\epsilon ^{\mu \nu
\rho }-2\lambda W_{1}\epsilon ^{\mu \nu }+6\lambda W_{3}\left( 2\phi _{\rho
\lambda }\xi ^{\mu \nu \rho \lambda }+K^{\mu \nu \rho }\xi _{\rho }\right)
\notag \\
&&+\lambda \left( 12W_{4}K^{\mu \nu \rho }\varepsilon _{\rho \alpha \beta
\gamma \delta }\xi ^{\alpha \beta \gamma \delta }-W_{5}\varepsilon ^{\mu \nu
\rho \lambda \sigma }\phi _{\rho \lambda }\xi _{\sigma }\right) ,
\label{tdef3}
\end{eqnarray}%
\begin{eqnarray}
\bar{\delta}_{\Omega }\phi _{\mu \nu } &=&D_{[\mu }^{\left( -\right) }\xi
_{\nu ]}+3\lambda \left( W_{3}\phi _{\mu \nu }\epsilon -2W_{4}V_{[\mu
}\varepsilon _{\nu ]\alpha \beta \gamma \delta }\xi ^{\alpha \beta \gamma
\delta }\right)  \notag \\
&&+3\lambda \varepsilon _{\mu \nu \rho \lambda \sigma }\left( 2W_{4}K^{\rho
\lambda \sigma }\epsilon +W_{6}\epsilon ^{\rho \lambda \sigma }-\tfrac{k_{2}%
}{180}\partial ^{\lbrack \rho }\chi ^{\lambda \sigma ]}\right) ,
\label{tdef4}
\end{eqnarray}%
\begin{eqnarray}
\bar{\delta}_{\Omega }K^{\mu \nu \rho } &=&4D_{\lambda }^{\left( +\right)
}\xi ^{\mu \nu \rho \lambda }-3\lambda \left( W_{2}\epsilon ^{\mu \nu \rho
}+W_{3}K^{\mu \nu \rho }\epsilon \right)  \notag \\
&&-\lambda \varepsilon ^{\mu \nu \rho \lambda \sigma }W_{5}\left( V_{\lambda
}\xi _{\sigma }-\tfrac{1}{2}\phi _{\lambda \sigma }\epsilon \right)
-2\lambda k_{1}\partial ^{\lbrack \mu }\chi ^{\nu \rho ]},  \label{tdef5}
\end{eqnarray}%
\begin{equation}
\bar{\delta}_{\Omega }t_{\mu \nu |\alpha }=\partial _{\lbrack \mu }\theta
_{\nu ]\alpha }+\partial _{\lbrack \mu }\chi _{\nu ]\alpha }-2\partial
_{\alpha }\chi _{\mu \nu }+\lambda k_{1}\sigma _{\alpha \lbrack \mu }\xi
_{\nu ]}-\tfrac{\lambda k_{2}}{5!}\sigma _{\alpha \lbrack \mu }\varepsilon
_{\nu ]\beta \gamma \delta \varepsilon }\xi ^{\beta \gamma \delta
\varepsilon },  \label{tdef6}
\end{equation}%
where, in addition, we used the notations
\begin{equation}
D_{\nu }=\partial _{\nu }-\lambda \frac{dW_{1}}{d\varphi }V_{\nu },\qquad
D_{\nu }^{\left( \pm \right) }=\partial _{\nu }\pm 3\lambda W_{3}V_{\nu }.
\label{derivcov}
\end{equation}%
We observe that the cross-interaction terms,%
\begin{equation*}
\lambda \left( k_{1}\phi ^{\mu \nu }-\tfrac{k_{2}}{20}\tilde{K}^{\mu \nu
}\right) F_{\mu \nu },
\end{equation*}%
are only of order one in the deformation parameter and couple the tensor
field $t_{\lambda \mu |\alpha }$ to the two-form $\phi _{\mu \nu }$ and to
the three-form $K^{\mu \nu \rho }$ from the BF sector. Also, it is
interesting to see that the interaction components%
\begin{equation*}
\tfrac{3\lambda ^{2}}{2}\left( k_{1}\phi ^{\mu \nu }-\tfrac{k_{2}}{20}\tilde{%
K}^{\mu \nu }\right) \left( k_{1}\phi _{\mu \nu }-\tfrac{k_{2}}{20}\tilde{K}%
_{\mu \nu }\right) ,
\end{equation*}%
which describe self-interactions in the BF sector, are strictly due to the
presence of the tensor $t_{\lambda \mu |\alpha }$ (in its absence $%
k_{1}=k_{2}=0$, so they would vanish). The gauge transformations of the BF
fields $\phi _{\mu \nu }$ and $K^{\mu \nu \rho }$ are deformed in such a way
to include gauge parameters from the $(2,1)$ sector. Related to the other BF
fields, $\varphi $, $H^{\mu }$, $V_{\mu }$, and $B^{\mu \nu }$, their gauge
transformations are also modified with respect to the free theory, but only
with terms specific to the BF sector. A remarkable feature is that the gauge
transformations of the tensor $t_{\lambda \mu |\alpha }$ are modified by
shift terms in some of the gauge parameters from the BF sector.

From the components of higher antighost number present in (\ref{soldeformata}%
) we read the entire gauge structure of the interacting theory: the
commutators among the deformed gauge transformations (\ref{tdef2})--(\ref%
{tdef6}), and hence the properties of the deformed gauge algebra, their
associated higher-order structure functions, and also the new reducibility
functions and relations together with their properties. (The reducibility
order itself of the interacting theory is not modified by the deformation
procedure and remains equal to that of the free model, namely, three.) We do
not give here the concrete form of all these deformed structure functions,
which is analyzed in detail in Appendix \ref{gauge}, but only briefly
discuss their main properties by contrast to the gauge features of the free
theory (see section \ref{free}).

The nonvanishing commutators among the deformed gauge transformations result
from the terms quadratic in the ghosts with pure ghost number one present in
(\ref{soldeformata}). Since their form can be generically written as $\tfrac{%
1}{2}(\eta _{\alpha _{1}}^{\ast }C_{\quad \beta _{1}\gamma _{1}}^{\alpha
_{1}}-\tfrac{1}{2}\Phi _{\alpha _{0}}^{\ast }\Phi _{\beta _{0}}^{\ast
}M_{\beta _{1}\gamma _{1}}^{\alpha _{0}\beta _{0}})\eta ^{\beta _{1}}\eta
^{\gamma _{1}}$, it follows that the commutators among the deformed gauge
transformations only close on-shell (on the stationary surface of the
deformed field equations)%
\begin{equation}
\left[ \bar{\delta}_{\Omega _{1}},\bar{\delta}_{\Omega _{2}}\right] \Phi
^{\alpha _{0}}=\bar{\delta}_{\Omega }\Phi ^{\alpha _{0}}+M_{\Omega }^{\alpha
_{0}\beta _{0}}\frac{\delta S^{\mathrm{L}}}{\delta \Phi ^{\beta _{0}}}.
\label{defalgebra}
\end{equation}%
Here, $\delta S^{\mathrm{L}}/\delta \Phi ^{\beta _{0}}$ stand for the
Euler--Lagrange (EL) derivatives of the interacting action (\ref{acLagdef}),
$\Omega _{1}$ and $\Omega _{2}$ represent two independent sets of gauge
parameters of the type (\ref{Omegaa1}), and $\Omega $ is a quadratic
combination of $\Omega _{1}$ and $\Omega _{2}$. The exact form of the
corresponding commutators is included in the Appendix \ref{gauge} (see
formulas (\ref{com1})--(\ref{com7})). In conclusion, the gauge algebra
corresponding to the interacting theory is open (the commutators among the
deformed gauge transformations only close on-shell), by contrast to the free
theory, where the gauge algebra is Abelian.

The first-order reducibility functions and relations follow from the terms
linear in the ghosts for ghosts appearing in (\ref{soldeformata}). Because
they can be generically set in the form ($\eta _{\alpha _{1}}^{\ast
}Z_{\quad \alpha _{2}}^{\alpha _{1}}+\tfrac{1}{2}\Phi _{\alpha _{0}}^{\ast
}\Phi _{\beta _{0}}^{\ast }C_{\alpha _{2}}^{\alpha _{0}\beta _{0}})\eta
^{\alpha _{2}}$, it follows that if we transform the gauge parameters $%
\Omega ^{\alpha _{1}}$ in terms of the first-order reducibility parameters $%
\bar{\Omega}^{\alpha _{2}}$ as in%
\begin{equation}
\Omega ^{\alpha _{1}}\rightarrow \Omega ^{\alpha _{1}}=Z_{\quad \alpha
_{2}}^{\alpha _{1}}\bar{\Omega}^{\alpha _{2}},  \label{red1par}
\end{equation}%
then the transformed gauge transformations (\ref{defgauge}) of all fields
vanish on-shell%
\begin{equation}
\delta _{\Omega \left( \bar{\Omega}\right) }\Phi ^{\alpha _{0}}\equiv
Z_{\quad \alpha _{1}}^{\alpha _{0}}Z_{\quad \alpha _{2}}^{\alpha _{1}}\bar{%
\Omega}^{\alpha _{2}}=C_{\bar{\Omega}}^{\alpha _{0}\beta _{0}}\frac{\delta
S^{\mathrm{L}}}{\delta \Phi ^{\beta _{0}}}\approx 0.  \label{red1def}
\end{equation}%
Along the same line, the second-order reducibility functions and relations
are given by the terms linear in the ghosts for ghosts for ghosts appearing
in (\ref{soldeformata}), which can be generically written as $(\eta _{\alpha
_{2}}^{\ast }Z_{\quad \alpha _{3}}^{\alpha _{2}}-\eta _{\alpha _{1}}^{\ast
}\Phi _{\beta _{0}}^{\ast }C_{\alpha _{3}}^{\alpha _{1}\beta _{0}}+\cdots
)\eta ^{\alpha _{3}}$. Consequently, if we transform the first-order
reducibility parameters $\bar{\Omega}^{\alpha _{2}}$ in terms of the
second-order reducibility parameters $\check{\Omega}^{\alpha _{3}}$ as in%
\begin{equation}
\bar{\Omega}^{\alpha _{2}}\rightarrow \bar{\Omega}^{\alpha _{2}}=Z_{\quad
\alpha _{3}}^{\alpha _{2}}\check{\Omega}^{\alpha _{3}},  \label{red2par}
\end{equation}%
then the transformed gauge parameters (\ref{red1par}) vanish on-shell%
\begin{equation}
\Omega ^{\alpha _{1}}\left( \bar{\Omega}^{\alpha _{2}}\left( \check{\Omega}%
^{\alpha _{3}}\right) \right) \equiv Z_{\quad \alpha _{2}}^{\alpha
_{1}}Z_{\quad \alpha _{3}}^{\alpha _{2}}\check{\Omega}^{\alpha _{3}}=C_{%
\check{\Omega}}^{\alpha _{1}\beta _{0}}\frac{\delta S^{\mathrm{L}}}{\delta
\Phi ^{\beta _{0}}}\approx 0.  \label{red2def}
\end{equation}%
Finally, the third-order reducibility functions and relations are withdrawn
from the terms linear in the ghosts for ghosts for ghosts for ghosts from (%
\ref{soldeformata}), which have the generic form $(\eta _{\alpha _{3}}^{\ast
}Z_{\quad \alpha _{4}}^{\alpha _{3}}+\eta _{\alpha _{2}}^{\ast }\Phi _{\beta
_{0}}^{\ast }C_{\alpha _{4}}^{\alpha _{2}\beta _{0}}+\cdots )\eta ^{\alpha
_{4}}$, such that if we transform the second-order reducibility parameters $%
\check{\Omega}^{\alpha _{3}}$ in terms of the third-order reducibility
parameters $\hat{\Omega}^{\alpha _{4}}$ as in%
\begin{equation}
\check{\Omega}^{\alpha _{3}}\rightarrow \check{\Omega}^{\alpha
_{3}}=Z_{\quad \alpha _{4}}^{\alpha _{3}}\hat{\Omega}^{\alpha _{4}},
\label{red3par}
\end{equation}%
then the transformed first-order reducibility parameters (\ref{red2par})
again vanish on-shell%
\begin{equation}
\bar{\Omega}^{\alpha _{2}}\left( \check{\Omega}^{\alpha _{3}}\left( \hat{%
\Omega}^{\alpha _{4}}\right) \right) \equiv Z_{\quad \alpha _{3}}^{\alpha
_{2}}Z_{\quad \alpha _{4}}^{\alpha _{3}}\hat{\Omega}^{\alpha _{4}}=C_{\check{%
\Omega}}^{\alpha _{2}\beta _{0}}\frac{\delta S^{\mathrm{L}}}{\delta \Phi
^{\beta _{0}}}\approx 0.  \label{red3def}
\end{equation}%
In the above the notations $\Omega ^{\alpha _{1}}$, $\bar{\Omega}^{\alpha
_{2}}$, $\check{\Omega}^{\alpha _{3}}$, and $\hat{\Omega}^{\alpha _{4}}$ are
the same from the free case, namely (\ref{Omegaa1}), (\ref{Omegaa2}), (\ref%
{Omegaa3}), and (\ref{Omegaa4}), while the BRST generators are structured
according to formulas (\ref{f10b})--(\ref{f10d'}). It is now clear that the
reducibility relations associated with the interacting model ((\ref{red1def}%
), (\ref{red2def}), and (\ref{red3def})) only hold on-shell, by contrast to
those corresponding to the free theory ((\ref{1red}), (\ref{2red}), and
respectively (\ref{3red})), which hold off-shell. Their concrete form is
detailed in Appendix \ref{gauge}.

\section{Some solutions to the consistency equations\label{clasesol}}

Equations (\ref{ecc1})--(\ref{ecc6}) and (\ref{ecc7})--(\ref{ecc9}),
required by the consistency of the first-order deformation, possess the
following classes of solutions, interesting from the point of view of
cross-couplings between the BF field sector and the tensor field with the
mixed symmetry $(2,1)$.

\begin{enumerate}
\item[I.] The real constants $k_{1}$ and $k_{2}$ are arbitrary ($%
k_{1}^{2}+k_{2}^{2}>0$), functions $\bar{M}$\ and $W_{2}$ are some
arbitrary, real, smooth functions of the undifferentiated scalar field, and%
\begin{gather}
W_{1}\left( \varphi \right) =W_{3}\left( \varphi \right) =W_{4}\left(
\varphi \right) =W_{5}\left( \varphi \right) =0,  \label{spf1} \\
W_{6}\left( \varphi \right) =-\frac{k_{2}}{5!k_{1}}W_{2}\left( \varphi
\right) .  \label{spf2}
\end{gather}%
The above formulas allow one to infer directly the solution in the general
case $k_{2}=0$. This class of solutions can be equivalently reformulated as:
the real constants $k_{1}$ and $k_{2}$ are arbitrary ($k_{1}^{2}+k_{2}^{2}>0$%
), functions $\bar{M}$\ and $W_{6}$ are some arbitrary, real, smooth
functions of the undifferentiated scalar field, and%
\begin{gather}
W_{1}\left( \varphi \right) =W_{3}\left( \varphi \right) =W_{4}\left(
\varphi \right) =W_{5}\left( \varphi \right) =0,  \label{spg1} \\
W_{2}\left( \varphi \right) =-\frac{5!k_{1}}{k_{2}}W_{6}\left( \varphi
\right) .  \label{spg2}
\end{gather}%
The last formulas are useful at writing down the solution in the particular
case $k_{1}=0$.

\item[II.] The real constants $k_{1}$ and $k_{2}$ are arbitrary ($%
k_{1}^{2}+k_{2}^{2}>0$), functions $\bar{M}$\ and $W_{5}$ are some
arbitrary, real, smooth functions of the undifferentiated scalar field, and
\begin{gather}
W_{1}\left( \varphi \right) =W_{2}\left( \varphi \right) =W_{6}\left(
\varphi \right) =0,  \label{uuuu} \\
W_{3}\left( \varphi \right) =-\frac{k_{2}}{60k_{1}}W_{5}\left( \varphi
\right) ,\qquad W_{4}\left( \varphi \right) =\left( \frac{k_{2}}{5!k_{1}}%
\right) ^{2}W_{5}\left( \varphi \right) .  \label{vvvv}
\end{gather}%
The above formulas allow one to infer directly the solution in the general
case $k_{2}=0$. This class of solutions can be equivalently reformulated as:
the real constants $k_{1}$ and $k_{2}$ are arbitrary ($k_{1}^{2}+k_{2}^{2}>0$%
), functions $\bar{M}$\ and $W_{4}$ are some arbitrary, real, smooth
functions of the undifferentiated scalar field, and%
\begin{gather}
W_{1}\left( \varphi \right) =W_{2}\left( \varphi \right) =W_{6}\left(
\varphi \right) =0,  \label{yyyy} \\
W_{3}\left( \varphi \right) =-2\cdot 5!\frac{k_{1}}{k_{2}}W_{4}\left(
\varphi \right) ,\qquad W_{5}\left( \varphi \right) =\left( \frac{5!k_{1}}{%
k_{2}}\right) ^{2}W_{4}\left( \varphi \right) .  \label{zzzz}
\end{gather}%
The last formulas are useful at writing down the solution in the particular
case $k_{1}=0$.

\item[III.] The real constants $k_{1}$ and $k_{2}$ are arbitrary ($%
k_{1}^{2}+k_{2}^{2}>0$), functions $W_{1}$\ and $W_{5}$ are some arbitrary,
real, smooth functions of the undifferentiated scalar field, and%
\begin{gather}
W_{2}\left( \varphi \right) =W_{6}\left( \varphi \right) =\bar{M}\left(
\varphi \right) =0,  \label{sp31} \\
W_{3}\left( \varphi \right) =-\frac{k_{2}}{60k_{1}}W_{5}\left( \varphi
\right) ,\qquad W_{4}\left( \varphi \right) =\left( \frac{k_{2}}{5!k_{1}}%
\right) ^{2}W_{5}\left( \varphi \right) .  \label{sp32}
\end{gather}%
The above formulas allow one to infer directly the solution in the general
case $k_{2}=0$. This class of solutions can be equivalently reformulated as:
the real constants $k_{1}$ and $k_{2}$ are arbitrary ($k_{1}^{2}+k_{2}^{2}>0$%
), functions $W_{1}$\ and $W_{4}$ are some arbitrary, real, smooth functions
of the undifferentiated scalar field, and%
\begin{gather}
W_{2}\left( \varphi \right) =W_{6}\left( \varphi \right) =\bar{M}\left(
\varphi \right) =0,  \label{sp71} \\
W_{3}\left( \varphi \right) =-2\cdot 5!\frac{k_{1}}{k_{2}}W_{4}\left(
\varphi \right) ,\qquad W_{5}\left( \varphi \right) =\left( \frac{5!k_{1}}{%
k_{2}}\right) ^{2}W_{4}\left( \varphi \right) .  \label{sp72}
\end{gather}%
The last formulas are useful at writing down the solution in the particular
case $k_{1}=0$.
\end{enumerate}

For all classes of solutions the emerging interacting theories display the
following common features:

\begin{enumerate}
\item[1.] there appear nontrivial cross-couplings between the BF fields and
the tensor field with the mixed symmetry $(2,1)$;

\item[2.] the gauge transformations are modified with respect to those of
the free theory and the gauge algebras become open (only close on-shell);

\item[3.] the first-order reducibility functions are changed during the
deformation process and the first-order reducibility relations take place
on-shell.
\end{enumerate}

Nevertheless, there appear the following differences between the above classes
of solutions at the level of the higher-order reducibility:

\begin{enumerate}
\item[a)] for class I the second-order reducibility functions are modified
with respect to the free ones and the corresponding reducibility relations
take place on-shell. The third-order reducibility functions remain those
from the free case and hence the associated reducibility relations hold
off-shell;

\item[b)] for class II both the second- and third-order reducibility
functions remain those from the free case and hence the associated
reducibility relations hold off-shell;

\item[c)] for class III all the second- and third-order reducibility
functions are deformed and the corresponding reducibility relations only
close on-shell.
\end{enumerate}

\section{Conclusion}

The most important conclusion of this paper is that under the hypotheses of
analyticity in the coupling constant, spacetime locality, Lorentz
covariance, and Poincar\'{e} invariance of the deformations, combined with
the preservation of the number of derivatives on each field, the dual
formulation of linearized gravity in $D=5$ allows for the first time
nontrivial couplings to another theory, namely with a topological BF model,
whose field spectrum consists in a scalar field, two sorts of one-forms, two
types of two-forms, and a three-form. The deformed Lagrangian contains
mixing-component terms of order one in the deformation parameter that couple
the massless tensor field with the mixed symmetry $(2,1)$ mainly to one of
the two-forms and to the three-form from the BF sector. There appear some
self-interactions in the BF sector at order two in the coupling constant
that are strictly due to the presence of the tensor field with the mixed
symmetry $(2,1)$. One of the striking features of the deformed model is that
the gauge transformations of all fields are deformed. This is the first case
where the gauge transformations of the tensor field with the mixed symmetry $%
(2,1)$ do change with respect to the free ones (by shifts in some of the BF
gauge parameters). All the ingredients of the gauge structure are modified
by the deformation procedure: the gauge algebra becomes open and the
reducibility relations hold on-shell.

\section*{Acknowledgments}

The work of A.D. is supported by the EU contract MRTN-CT-2004-005104. C.B.,
E.M.C., S.O.S. and S.C.S. acknowledge support from the type A grant 581/2007
with the Romanian National Council for Academic Scientific Research
(C.N.C.S.I.S.) and the Romanian Ministry of Education, Research and Youth
(M.E.C.T.).

\appendix{}

\section{No-go result for $I=5$ in $a^{\mathrm{int}}$\label{I=5}}

In agreement with (\ref{fo1}), the general solution to the equation $sa^{%
\mathrm{int}}=\partial ^{\mu }m_{\mu }^{\mathrm{int}}$ can be chosen to stop
at antighost number $I=5$
\begin{equation}
a^{\mathrm{int}}=a_{0}^{\mathrm{int}}+a_{1}^{\mathrm{int}}+a_{2}^{\mathrm{int%
}}+a_{3}^{\mathrm{int}}+a_{4}^{\mathrm{int}}+a_{5}^{\mathrm{int}},
\label{fo5}
\end{equation}%
where the components on the right-hand side of (\ref{fo5}) are subject to
the equations (\ref{3.6}) and (\ref{3.4})--(\ref{3.5}) for $I=5$.

The piece $a_{5}^{\mathrm{int}}$ as solution to equation (\ref{3.6}) for $%
I=5 $ has the general form expressed by (\ref{3.8}) for $I=5$, with $\alpha
_{5}$ from $H_{5}^{\mathrm{inv}}(\delta |d)$. According to (\ref{hinvdelta})
at antighost number five, it follows that $H_{5}^{\mathrm{inv}}(\delta |d)$
is spanned by the generic representatives (\ref{3.13}). Since $a_{5}^{%
\mathrm{int}}$ should effectively mix the BF and the $(2,1)$ tensor field
sectors in order to produce cross-couplings and (\ref{3.13}) involves only
BF generators, it follows that one should retain from the basis elements $%
e^{5}\left( \eta ^{\bar{\Upsilon}}\right) $ only the objects containing at
least one ghost from the $(2,1)$ tensor field sector, namely $D_{\mu \nu
\rho }$ or $S_{\mu }$. Recalling that we work precisely in $D=5$, we obtain
that the general solution to (\ref{3.6}) for $I=5$ reduces to
\begin{eqnarray}
a_{5}^{\mathrm{int}} &=&\tfrac{1}{3!}\left( \left( \tilde{U}_{1}\right)
C+\left( \tilde{U}_{2}\right) \mathcal{\tilde{G}}\right) \tilde{D}_{\mu
\alpha }\tilde{D}^{\alpha \beta }\tilde{D}_{\beta \nu }\sigma ^{\mu \nu }
\notag \\
&&+\tfrac{1}{2}\left( \left( \tilde{U}_{3}\right) \eta S^{\mu }-\left(
\tilde{U}_{4}\right) D^{\mu \nu \rho }\tilde{D}_{\nu \alpha }\tilde{D}_{\rho
\beta }\sigma ^{\alpha \beta }\right) S_{\mu }.  \label{ai5}
\end{eqnarray}%
Each tilde object from the right-hand side of (\ref{ai5}) means the Hodge
dual of the corresponding non-tilde element, defined in general by formula (%
\ref{hodge}). The elements $\tilde{U}$ are dual to $\left( U\right) _{\mu
_{1}\ldots \mu _{5}}$ as in (\ref{3.13}), with $W\left( \varphi \right) $
respectively replaced by the smooth function $U\left( \varphi \right) $
depending only on the undifferentiated scalar field $\varphi $.

Introducing (\ref{ai5}) in equation (\ref{3.4}) for $I=5$ and recalling
definitions (\ref{f12a})--(\ref{f13i}), we obtain%
\begin{eqnarray}
a_{4}^{\mathrm{int}} &=&-\tfrac{1}{6}\tilde{D}_{\mu \alpha }\tilde{D}%
^{\alpha \beta }\sigma ^{\mu \nu }\left[ \left( \tilde{U}_{1}\right)
^{\lambda }\left( C_{\lambda }\tilde{D}_{\beta \nu }+\tfrac{3}{2}C\tilde{F}%
_{\beta \nu |\lambda }\right) \right.  \notag \\
&&\left. -\left( \tilde{U}_{2}\right) ^{\lambda }\left( \tfrac{1}{5}\mathcal{%
\tilde{G}}_{\lambda }\tilde{D}_{\beta \nu }+\tfrac{3}{2}\mathcal{\tilde{G}}%
\tilde{F}_{\beta \nu |\lambda }\right) \right] +\tfrac{1}{2}\left( \tilde{U}%
_{3}\right) ^{\lambda }\left( V_{\lambda }S_{\mu }+\eta \mathcal{C}_{\lambda
\mu }\right) S^{\mu }  \notag \\
&&-\tfrac{1}{4}\left( \tilde{U}_{4}\right) ^{\lambda }\left[ D^{\mu \nu \rho
}\left( \tilde{D}_{\nu \alpha }\tilde{D}_{\rho \beta }\sigma ^{\alpha \beta }%
\mathcal{C}_{\lambda \mu }-2\tilde{D}_{\nu \alpha }\tilde{F}_{\rho \beta
|\lambda }\sigma ^{\alpha \beta }S_{\mu }\right) \right.  \notag \\
&&\left. -F^{\mu \nu \rho |\gamma }\tilde{D}_{\nu \alpha }\tilde{D}_{\rho
\beta }S_{\mu }\sigma ^{\alpha \beta }\sigma _{\gamma \lambda }\right] +\bar{%
a}_{4}^{\mathrm{int}}.  \label{ai4}
\end{eqnarray}%
In (\ref{ai4}) $\left( \tilde{U}\right) ^{\lambda }$ are dual to (\ref{3.14}%
), with $W\left( \varphi \right) \rightarrow U\left( \varphi \right) $. In
addition, $\mathcal{C}_{\mu \rho }$ is implicitly defined by formula (\ref%
{relgamex}) so it is a ghost field of pure ghost number one without definite
symmetry/antisymmetry property, $\mathcal{C}^{\ast \nu \lambda }$ is its
associated antifield, defined such that the antibracket $\left( \mathcal{C}%
_{\mu \rho },\mathcal{C}^{\ast \nu \lambda }\right) $ is equal to the `unit'
$\delta _{\mu }^{\nu }\delta _{\rho }^{\lambda }$%
\begin{equation}
\mathcal{C}^{\ast \nu \lambda }\equiv 3S^{\ast \nu \lambda }+A^{\ast \nu
\lambda }.  \label{calcstar}
\end{equation}%
The nonintegrated density $\bar{a}_{4}^{\mathrm{int}}$ stands for the
solution to the homogeneous equation (\ref{3.6}) for $I=4$, showing that $%
\bar{a}_{4}^{\mathrm{int}}$ can be taken as a\ nontrivial element of $%
H\left( \gamma \right) $ in pure ghost number equal to four.

At this stage it is useful to decompose $\bar{a}_{4}^{\mathrm{int}}$ as a
sum between two components
\begin{equation}
\bar{a}_{4}^{\mathrm{int}}=\hat{a}_{4}^{\mathrm{int}}+\check{a}_{4}^{\mathrm{%
int}},  \label{cai4a}
\end{equation}%
where $\hat{a}_{4}^{\mathrm{int}}$ is the solution to (\ref{3.6}) for $I=4$
which is explicitly required by the consistency of $a_{4}^{\mathrm{int}}$ in
antighost number three (ensures that (\ref{3.5}) possesses solutions for $i=4
$ with respect to the terms from (\ref{ai4}) containing the functions of the
type $U$) and $\check{a}_{4}^{\mathrm{int}}$ signifies the part of the
solution to (\ref{3.6}) for $I=4$ that is independently consistent in
antighost number three%
\begin{equation}
\delta \check{a}_{4}^{\mathrm{int}}=-\gamma \check{c}_{3}+\partial _{\mu }%
\check{m}_{3}^{\mu }.  \label{cai4b}
\end{equation}%
Using definitions (\ref{f12a})--(\ref{f13i}) and decomposition (\ref{cai4a}%
), by direct computation we obtain that%
\begin{eqnarray}
\delta a_{4}^{\mathrm{int}} &=&\delta \left[ \hat{a}_{4}^{\mathrm{int}}-%
\tfrac{1}{2}S^{\alpha }S_{\alpha }\left( \left( \tilde{U}_{3}\right) ^{\mu
\nu }B_{\mu \nu }^{\ast }+\tfrac{1}{3}\left( \tilde{U}_{3}\right) ^{\mu \nu
\rho }\eta _{\mu \nu \rho }^{\ast }+\tfrac{1}{12}\left( \tilde{U}_{3}\right)
^{\mu \nu \rho \lambda }\eta _{\mu \nu \rho \lambda }^{\ast }\right. \right.
\notag \\
&&\left. \left. +\tfrac{1}{60}\left( \tilde{U}_{3}\right) ^{\mu \nu \rho
\lambda \sigma }\eta _{\mu \nu \rho \lambda \sigma }^{\ast }\right) \right]
+\gamma c_{3}+\partial _{\mu }j_{3}^{\mu }+\chi _{3},  \label{cai4c}
\end{eqnarray}%
where we made the notations%
\begin{eqnarray}
c_{3} &=&-\check{c}_{3}+\tfrac{1}{12}\left( \tilde{U}_{1}\right) ^{\lambda
\sigma }\tilde{D}_{\mu \rho }\sigma ^{\mu \nu }\left[ \tilde{D}^{\rho \alpha
}\left( \phi _{\lambda \sigma }\tilde{D}_{\alpha \nu }-3C_{\lambda }\tilde{F}%
_{\alpha \nu |\sigma }\right) +\tfrac{3}{2}C\tilde{F}_{\quad |\lambda
}^{\rho \alpha }\tilde{F}_{\alpha \nu |\sigma }\right]   \notag \\
&&-\tfrac{1}{240}\left( \tilde{U}_{2}\right) ^{\lambda \sigma }\tilde{D}%
_{\mu \rho }\sigma ^{\mu \nu }\left[ \tilde{D}^{\rho \alpha }\left( \tilde{K}%
_{\lambda \sigma }\tilde{D}_{\alpha \nu }-12\mathcal{\tilde{G}}_{\lambda }%
\tilde{F}_{\alpha \nu |\sigma }\right) -30\mathcal{\tilde{G}}\tilde{F}%
_{\quad |\lambda }^{\rho \alpha }\tilde{F}_{\alpha \nu |\sigma }\right]
\notag \\
&&-\tfrac{1}{12}\left( \tilde{U}_{3}\right) ^{\lambda \sigma }\left[ S^{\mu
}\left( 6V_{\lambda }\mathcal{C}_{\sigma \mu }-\eta t_{\lambda \sigma |\mu
}\right) +\tfrac{3}{2}\eta \mathcal{C}_{\lambda \rho }\mathcal{C}_{\sigma
\mu }\sigma ^{\rho \mu }\right]   \notag \\
&&-\tfrac{1}{2}\left( \left( \tilde{U}_{3}\right) ^{\mu \nu \sigma }B_{\mu
\nu }^{\ast }+\tfrac{1}{3}\left( \tilde{U}_{3}\right) ^{\mu \nu \rho \sigma
}\eta _{\mu \nu \rho }^{\ast }+\tfrac{1}{12}\left( \tilde{U}_{3}\right)
^{\mu \nu \rho \lambda \sigma }\eta _{\mu \nu \rho \lambda }^{\ast }\right)
S^{\alpha }\mathcal{C}_{\sigma \alpha }  \notag \\
&&-\tfrac{1}{24}\left( \tilde{U}_{4}\right) ^{\lambda \sigma }\sigma
^{\alpha \beta }\left[ D^{\mu \nu \rho }\left( 6\tilde{D}_{\nu \alpha }%
\tilde{F}_{\rho \beta |\lambda }\mathcal{C}_{\sigma \mu }+3\tilde{F}_{\nu
\alpha |\lambda }\tilde{F}_{\rho \beta |\sigma }S_{\mu }+\tilde{D}_{\nu
\alpha }\tilde{D}_{\rho \beta }t_{\lambda \sigma |\mu }\right) \right.
\notag \\
&&\left. +3F_{\qquad |\lambda }^{\mu \nu \rho }\tilde{D}_{\nu \alpha }\left(
\tilde{D}_{\rho \beta }\mathcal{C}_{\sigma \mu }-2\tilde{F}_{\rho \beta
|\sigma }S_{\mu }\right) \right] ,  \label{cai41}
\end{eqnarray}%
\begin{eqnarray}
\chi _{3} &=&-\tfrac{1}{4}\left( \left( \tilde{U}_{1}\right) ^{\lambda
\sigma }C+\left( \tilde{U}_{2}\right) ^{\lambda \sigma }\mathcal{\tilde{G}}%
\right) \sigma ^{\mu \nu }\tilde{D}_{\mu \alpha }\tilde{D}^{\alpha \beta }%
\tilde{R}_{\beta \nu |\lambda \sigma }  \notag \\
&&+\tfrac{1}{6}\left( \tilde{U}_{3}\right) ^{\mu \nu }\eta S^{\rho }D_{\mu
\nu \rho }-\tfrac{1}{12}\left( \tilde{U}_{4}\right) ^{\lambda \sigma }\sigma
^{\alpha \beta }\left[ -3R_{\qquad |\lambda \sigma }^{\mu \nu \rho }\tilde{D}%
_{\nu \alpha }\tilde{D}_{\rho \beta }S_{\mu }\right.   \notag \\
&&\left. +D^{\mu \nu \rho }\tilde{D}_{\nu \alpha }\left( \tilde{D}_{\rho
\beta }D_{\lambda \sigma \mu }-6\tilde{R}_{\rho \beta |\lambda \sigma
}S_{\mu }\right) \right] ,  \label{cai43}
\end{eqnarray}%
and $j_{3}^{\mu }$ are some local currents. In (\ref{cai4c})--(\ref{cai43}) $%
\left( \tilde{U}\right) ^{\mu \nu }$ and $\left( \tilde{U}\right) ^{\mu \nu
\rho }$ denote the duals of (\ref{3.15}) and (\ref{3.16}) with $W\left(
\varphi \right) \rightarrow U\left( \varphi \right) $. In addition, $\left(
\tilde{U}\right) ^{\mu \nu \rho \lambda }$ represents the dual of $\left(
U\right) _{\mu }=\frac{dU}{d\varphi }H_{\mu }^{\ast }$ and $\left( \tilde{U}%
\right) ^{\mu \nu \rho \lambda \sigma }$ the dual of $U\left( \varphi
\right) $. Inspecting (\ref{cai4c}), it follows that the consistency of $%
a_{4}^{\mathrm{int}}$ in antighost number three, namely the existence of $%
a_{3}^{\mathrm{int}}$ as solution to (\ref{3.5}) for $i=4$, requires the
conditions
\begin{equation}
\chi _{3}=\gamma \hat{c}_{3}+\partial _{\mu }\hat{\jmath}_{3}^{\mu }
\label{cai44}
\end{equation}%
and%
\begin{eqnarray}
\hat{a}_{4}^{\mathrm{int}} &=&\tfrac{1}{2}S^{\alpha }S_{\alpha }\left(
\left( \tilde{U}_{3}\right) ^{\mu \nu }B_{\mu \nu }^{\ast }+\tfrac{1}{3}%
\left( \tilde{U}_{3}\right) ^{\mu \nu \rho }\eta _{\mu \nu \rho }^{\ast
}\right.   \notag \\
&&\left. +\tfrac{1}{12}\left( \tilde{U}_{3}\right) ^{\mu \nu \rho \lambda
}\eta _{\mu \nu \rho \lambda }^{\ast }+\tfrac{1}{60}\left( \tilde{U}%
_{3}\right) ^{\mu \nu \rho \lambda \sigma }\eta _{\mu \nu \rho \lambda
\sigma }^{\ast }\right) ,  \label{cai45}
\end{eqnarray}%
where we made the notations $\hat{c}_{3}=-\left( a_{3}^{\mathrm{int}%
}+c_{3}\right) $ and $\hat{\jmath}_{3}^{\mu }=\overset{(3)}{m}^{\mathrm{int~}%
\mu }-j_{3}^{\mu }$. Nevertheless, from (\ref{cai43}) it is obvious that $%
\chi _{3}$ is a nontrivial element from $H\left( \gamma \right) $ in pure
ghost number four, which does not reduce to a full divergence, and therefore
(\ref{cai44}) requires that $\chi _{3}=0$, which further imply that all the
functions of the type $U$ must be some real constants%
\begin{equation}
U_{1}\left( \varphi \right) =u_{1},\qquad U_{2}\left( \varphi \right)
=u_{2},\qquad U_{3}\left( \varphi \right) =u_{3},\qquad U_{4}\left( \varphi
\right) =u_{4}.  \label{umconst}
\end{equation}%
Based on (\ref{umconst}), it is clear that $a_{5}^{\mathrm{int}}$ given by (%
\ref{ai5}) vanishes, and hence we can assume, without loss of nontrivial
terms, that
\begin{equation}
a_{5}^{\mathrm{int}}=0  \label{ai50}
\end{equation}%
in (\ref{fo5}).

\section{No-go result for $I=4$ in $a^{\mathrm{int}}$\label{I=4}}

We have seen in Appendix \ref{I=5} that we can always take (\ref{ai50}) in (%
\ref{fo5}). Consequently, the first-order deformation of the solution to the
master equation in the interacting case stops at antighost number four%
\begin{equation}
a^{\mathrm{int}}=a_{0}^{\mathrm{int}}+a_{1}^{\mathrm{int}}+a_{2}^{\mathrm{int%
}}+a_{3}^{\mathrm{int}}+a_{4}^{\mathrm{int}},  \label{fo8}
\end{equation}%
where the components on the right-hand side of (\ref{fo8}) are subject to
the equations (\ref{3.6}) and (\ref{3.4})--(\ref{3.5}) for $I=4$.

The piece $a_{4}^{\mathrm{int}}$ as solution to equation (\ref{3.6}) for $%
I=4 $ has the general form expressed by (\ref{3.8}) for $I=4$, with $\alpha
_{4}$ from $H_{4}^{\mathrm{inv}}(\delta |d)$. According to (\ref{hinvdelta})
at antighost number four, it follows that $H_{4}^{\mathrm{inv}}(\delta |d)$
is spanned by some representatives involving only BF generators. Since $%
a_{4}^{\mathrm{int}}$ should again mix the BF and the $(2,1)$ tensor field
sectors, it follows that one should retain from the basis elements $%
e^{4}\left( \eta ^{\bar{\Upsilon}}\right) $ only the objects containing at
least one ghost from the $(2,1)$ tensor field sector, namely $D_{\mu \nu
\rho }$ or $S_{\mu } $. The general solution to (\ref{3.6}) for $I=4$ reads
as%
\begin{eqnarray}
a_{4}^{\mathrm{int}} &=&\tfrac{1}{2}\tilde{\eta}^{\ast }\left( q_{1}S_{\mu
}S^{\mu }+\tfrac{q_{2}}{3}\sigma ^{\mu \nu }\tilde{D}_{\mu \alpha }\tilde{D}%
^{\alpha \beta }\tilde{D}_{\beta \nu }\eta \right) +\left( \tilde{U}%
_{5}\right) ^{\mu }\eta \tilde{D}_{\mu \nu }S^{\nu }  \notag \\
&&+\left( \left( \tilde{U}_{6}\right) ^{\mu }C+\left( \tilde{U}_{7}\right)
^{\mu }\mathcal{\tilde{G}}\right) S_{\mu }-\tfrac{1}{4}\left( U_{8}\right)
_{\mu \nu \rho \lambda }\tilde{D}^{\mu \alpha }\tilde{D}^{\nu \beta }\tilde{D%
}^{\rho \gamma }\tilde{D}^{\lambda \delta }\sigma _{\alpha (\gamma }\sigma
_{\delta )\beta }  \notag \\
&&-\tfrac{1}{2}\left( \tilde{U}_{9}\right) ^{\mu }D_{\mu \nu \rho }\tilde{D}%
^{\nu \alpha }\tilde{D}^{\rho \beta }\eta \sigma _{\alpha \beta },
\label{ai4a}
\end{eqnarray}%
where each element generically denoted by $\left( \tilde{U}\right) ^{\mu }$
is the Hodge dual of an object similar to (\ref{3.14}), but with $W$
replaced by the arbitrary, smooth function $U$, depending on the
undifferentiated scalar field, $\left( U_{8}\right) _{\mu \nu \rho \lambda }$
reads as in (\ref{3.14}) with $W\left( \varphi \right) \rightarrow
U_{8}\left( \varphi \right) $, and $q_{1,2}$ are two arbitrary, real
constants.

Introducing (\ref{ai4a}) in equation (\ref{3.4}) for $I=4$ and using
definitions (\ref{f12a})--(\ref{f13i}), we determine the component of
antighost number three from $a^{\mathrm{int}}$ in the form%
\begin{eqnarray}
a_{3}^{\mathrm{int}} &=&\tfrac{1}{2}q_{1}\tilde{\eta}^{\ast \mu }S^{\nu }%
\mathcal{C}_{\mu \nu }+\tfrac{1}{6}q_{2}\tilde{\eta}^{\ast \lambda }\sigma
^{\mu \nu }\tilde{D}_{\mu \alpha }\tilde{D}^{\alpha \beta }\left( \tilde{D}%
_{\beta \nu }V_{\lambda }+\tfrac{3}{2}\tilde{F}_{\beta \nu |\lambda }\eta
\right)  \notag \\
&&+\tfrac{1}{2}\left( \tilde{U}_{5}\right) ^{\mu \nu }\left[ \left( 2V_{\mu }%
\tilde{D}_{\nu \rho }-\eta \tilde{F}_{\mu \rho |\nu }\right) S^{\rho
}+\sigma ^{\rho \lambda }\eta \tilde{D}_{\mu \rho }\mathcal{C}_{\nu \lambda }%
\right]  \notag \\
&&-\tfrac{1}{2}\left( \tilde{U}_{6}\right) ^{\mu \nu }\left( A_{\mu \nu
}C-2S_{\mu }C_{\nu }\right) -\tfrac{1}{2}\left( \tilde{U}_{7}\right) ^{\mu
\nu }\left( A_{\mu \nu }\mathcal{\tilde{G}}+\tfrac{2}{5}S_{\mu }\mathcal{%
\tilde{G}}_{\nu }\right)  \notag \\
&&-\tfrac{1}{2}\left( \tilde{U}_{9}\right) ^{\mu \nu }\sigma _{\alpha \beta }%
\left[ D_{\mu \lambda \rho }\tilde{D}^{\lambda \alpha }\left( \tilde{D}%
^{\rho \beta }V_{\nu }+\tilde{F}_{\quad |\nu }^{\rho \beta }\eta \right) +%
\tfrac{1}{2}F_{\mu \lambda \rho |\nu }\tilde{D}^{\lambda \alpha }\tilde{D}%
^{\rho \beta }\eta \right]  \notag \\
&&-\tfrac{1}{2}\left( \tilde{U}_{8}\right) ^{\mu \tau }\varepsilon _{\mu \nu
\rho \lambda \sigma }\tilde{D}^{\nu \alpha }\tilde{D}^{\rho \beta }\tilde{D}%
^{\lambda \gamma }\tilde{F}_{\quad |\tau }^{\sigma \delta }\sigma _{\alpha
(\gamma }\sigma _{\delta )\beta }+\bar{a}_{3}^{\mathrm{int}},  \label{ai3}
\end{eqnarray}%
where each $\left( \tilde{U}\right) ^{\mu \nu }$ is the Hodge dual of an
object of the type (\ref{3.15}), with $W$ replaced by the corresponding
function of the type $U$. Here, $\bar{a}_{3}^{\mathrm{int}}$ is the general
solution to the homogeneous equation (\ref{3.6}) for $I=3$, showing that $%
\bar{a}_{3}^{\mathrm{int}}$ is a nontrivial object from $H\left( \gamma
\right) $ in pure ghost number three.

At this point we decompose $\bar{a}_{3}^{\mathrm{int}}$ in a manner similar
to (\ref{cai4a})
\begin{equation}
\bar{a}_{3}^{\mathrm{int}}=\hat{a}_{3}^{\mathrm{int}}+\check{a}_{3}^{\mathrm{%
int}},  \label{cai3a}
\end{equation}%
where $\hat{a}_{3}^{\mathrm{int}}$ is the solution to (\ref{3.6}) for $I=3$
that ensures the consistency of $a_{3}^{\mathrm{int}}$ in antighost number
two, namely the existence of $a_{2}^{\mathrm{int}}$ as solution to (\ref{3.5}%
) for $i=3$ with respect to the terms from $a_{3}^{\mathrm{int}}$ containing
the functions of the type $U$ or the constants $q_{1}$ or $q_{2}$, while $%
\check{a}_{3}^{\mathrm{int}}$ is the solution to (\ref{3.6}) for $I=3$ which
is independently consistent in antighost number two%
\begin{equation}
\delta \check{a}_{3}^{\mathrm{int}}=-\gamma \check{c}_{2}+\partial _{\mu }%
\check{m}_{2}^{\mu }.  \label{cai3b}
\end{equation}%
Based on definitions (\ref{f12a})--(\ref{f13i}) and taking into account
decomposition (\ref{cai3a}), we get by direct computation%
\begin{eqnarray}
\delta a_{3}^{\mathrm{int}} &=&\delta \left[ \hat{a}_{3}^{\mathrm{int}}-%
\tfrac{q_{2}}{6}\tilde{\eta}^{\ast \lambda \sigma }\sigma ^{\mu \nu }\tilde{D%
}_{\mu \alpha }\tilde{D}^{\alpha \beta }\tilde{D}_{\beta \nu }B_{\lambda
\sigma }^{\ast }-\left( \left( \tilde{U}_{5}\right) ^{\mu \nu \alpha }B_{\mu
\nu }^{\ast }\right. \right.  \notag \\
&&\left. +\tfrac{1}{3}\left( \tilde{U}_{5}\right) ^{\mu \nu \rho \alpha
}\eta _{\mu \nu \rho }^{\ast }+\tfrac{1}{12}\left( \tilde{U}_{5}\right)
^{\mu \nu \rho \lambda \alpha }\eta _{\mu \nu \rho \lambda }^{\ast }\right)
\tilde{D}_{\alpha \beta }S^{\beta }  \notag \\
&&+\tfrac{1}{2}\sigma _{\alpha \beta }D_{\mu \nu \rho }\tilde{D}^{\nu \alpha
}\tilde{D}^{\rho \beta }\left( \left( \tilde{U}_{9}\right) ^{\mu \lambda
\sigma }B_{\lambda \sigma }^{\ast }-\tfrac{1}{3}\left( \tilde{U}_{9}\right)
^{\mu \lambda \sigma \gamma }\eta _{\lambda \sigma \gamma }^{\ast }\right.
\notag \\
&&\left. \left. +\tfrac{1}{12}\left( \tilde{U}_{9}\right) ^{\mu \lambda
\sigma \gamma \delta }\eta _{\lambda \sigma \gamma \delta }^{\ast }\right) %
\right] +\gamma c_{2}+\partial _{\lambda }j_{2}^{\lambda }+\chi _{2},
\label{cai3c}
\end{eqnarray}%
where%
\begin{eqnarray}
c_{2} &=&-\check{c}_{2}+\tfrac{q_{1}}{12}\tilde{\eta}^{\ast \mu \nu }\left(
S^{\rho }t_{\mu \nu |\rho }-\tfrac{3}{2}\sigma ^{\rho \lambda }\mathcal{C}%
_{\mu \rho }\mathcal{C}_{\nu \lambda }\right)  \notag \\
&&+\tfrac{q_{2}}{4}\tilde{\eta}^{\ast \lambda \sigma }\sigma ^{\mu \nu }%
\tilde{D}_{\mu \alpha }\left( \tilde{D}^{\alpha \beta }V_{\lambda }+\tfrac{1%
}{2}\tilde{F}_{\quad |\lambda }^{\alpha \beta }\eta \right) \tilde{F}_{\beta
\nu |\sigma }  \notag \\
&&+\tfrac{1}{2}\left( \tilde{U}_{5}\right) ^{\mu \nu \rho }\left[ V_{\mu
}\left( \tilde{F}_{\nu \lambda |\rho }S^{\lambda }-\tilde{D}_{\nu \lambda }%
\mathcal{C}_{\rho }^{\quad \lambda }\right) \right.  \notag \\
&&\left. +\tfrac{1}{2}\eta \left( \tilde{F}_{\mu \lambda |\nu }\mathcal{C}%
_{\rho }^{\quad \lambda }+\tfrac{1}{3}\tilde{D}_{\mu }^{\quad \alpha }t_{\nu
\rho |\alpha }\right) \right]  \notag \\
&&+\tfrac{1}{2}\left( \left( \tilde{U}_{5}\right) ^{\mu \nu \lambda \sigma
}B_{\mu \nu }^{\ast }+\tfrac{1}{3}\left( \tilde{U}_{5}\right) ^{\mu \nu \rho
\lambda \sigma }\eta _{\mu \nu \rho }^{\ast }\right) \left( \tilde{F}%
_{\lambda \alpha |\sigma }S^{\alpha }-\tilde{D}_{\lambda \alpha }\mathcal{C}%
_{\sigma }^{\quad \alpha }\right)  \notag \\
&&+\tfrac{1}{2}\left( \tilde{U}_{6}\right) ^{\mu \nu \rho }\left( A_{\mu \nu
}C_{\rho }+S_{\mu }\phi _{\nu \rho }\right) -\tfrac{1}{10}\left( \tilde{U}%
_{7}\right) ^{\mu \nu \rho }\left( A_{\mu \nu }\mathcal{\tilde{G}}_{\rho }+%
\tfrac{1}{4}S_{\mu }\tilde{K}_{\nu \rho }\right)  \notag \\
&&+\tfrac{1}{8}\left( \tilde{U}_{8}\right) ^{\mu \varepsilon \pi
}\varepsilon _{\mu \nu \rho \lambda \sigma }\tilde{D}^{\nu \alpha }\left(
\tilde{D}^{\rho \beta }\tilde{F}_{\quad |\varepsilon }^{\lambda \gamma }+2%
\tilde{F}_{\quad |\varepsilon }^{\rho \beta }\tilde{D}^{\lambda \gamma
}\right) \tilde{F}_{\quad |\pi }^{\sigma \delta }\sigma _{\alpha (\gamma
}\sigma _{\delta )\beta }  \notag \\
&&-\tfrac{1}{8}\left( \tilde{U}_{9}\right) ^{\mu \lambda \sigma }\sigma
_{\alpha \beta }\left[ D_{\mu \nu \rho }\left( 4\tilde{D}^{\nu \alpha }%
\tilde{F}_{\quad |\sigma }^{\rho \beta }V_{\lambda }+\tilde{F}_{\quad
|\lambda }^{\nu \alpha }\tilde{F}_{\quad |\sigma }^{\rho \beta }\eta \right)
\right.  \notag \\
&&\left. +2F_{\mu \nu \rho |\sigma }\tilde{D}^{\nu \alpha }\left( \tilde{D}%
^{\rho \beta }V_{\lambda }+\tilde{F}_{\quad |\lambda }^{\rho \beta }\eta
\right) \right]  \notag \\
&&-\tfrac{1}{4}\left( \tilde{U}_{9}\right) ^{\mu \lambda \sigma \gamma
}\sigma _{\alpha \beta }\left( 2D_{\mu \nu \rho }\tilde{F}_{\quad |\gamma
}^{\nu \alpha }-F_{\mu \nu \rho |\gamma }\tilde{D}^{\nu \alpha }\right)
\tilde{D}^{\rho \beta }B_{\lambda \sigma }^{\ast }  \notag \\
&&+\tfrac{1}{12}\left( \tilde{U}_{9}\right) ^{\mu \lambda \sigma \gamma
\delta }\sigma _{\alpha \beta }\left( 2D_{\mu \nu \rho }\tilde{F}_{\quad
|\delta }^{\nu \alpha }-F_{\mu \nu \rho |\delta }\tilde{D}^{\nu \alpha
}\right) \tilde{D}^{\rho \beta }\eta _{\lambda \sigma \gamma }^{\ast },
\label{cai3d}
\end{eqnarray}%
\begin{eqnarray}
\chi _{2} &=&\tfrac{q_{1}}{6}\tilde{\eta}^{\ast \mu \nu }S^{\rho }D_{\mu \nu
\rho }+\tfrac{1}{6}\left( \tilde{U}_{5}\right) ^{\mu \nu \rho }\eta \left(
\tilde{D}_{\mu }^{\quad \alpha }D_{\nu \rho \alpha }-3\tilde{R}_{\mu \lambda
|\nu \rho }S^{\lambda }\right)  \notag \\
&&+\tfrac{q_{2}}{6}\sigma ^{\mu \nu }\tilde{D}_{\mu \alpha }\tilde{D}%
^{\alpha \beta }\left[ \left( \partial _{\sigma }\tilde{B}^{\ast \lambda
\rho \sigma }\right) \tilde{D}_{\beta \nu }B_{\lambda \rho }^{\ast }+\tfrac{3%
}{2}\tilde{\eta}^{\ast \lambda \rho }\tilde{R}_{\beta \nu |\lambda \rho
}\eta \right]  \notag \\
&&+\tfrac{1}{6}\left( \tilde{U}_{6}\right) ^{\mu \nu \rho }D_{\mu \nu \rho
}C+\tfrac{1}{6}\left( \tilde{U}_{7}\right) ^{\mu \nu \rho }D_{\mu \nu \rho }%
\mathcal{\tilde{G}}  \notag \\
&&-\tfrac{1}{2}\left( \tilde{U}_{8}\right) ^{\mu \varepsilon \pi
}\varepsilon _{\mu \nu \rho \lambda \sigma }\sigma _{\alpha (\gamma }\sigma
_{\delta )\beta }\tilde{D}^{\nu \alpha }\tilde{D}^{\rho \beta }\tilde{D}%
^{\lambda \gamma }\tilde{R}_{\quad |\varepsilon \pi }^{\sigma \delta }
\notag \\
&&+\tfrac{1}{4}\left( \tilde{U}_{9}\right) ^{\mu \lambda \sigma }\sigma
_{\alpha \beta }\left( 2D_{\mu \nu \rho }\tilde{R}_{\quad |\lambda \sigma
}^{\nu \alpha }-R_{\mu \nu \rho |\lambda \sigma }\tilde{D}^{\nu \alpha
}\right) \tilde{D}^{\rho \beta }\eta ,  \label{cai3g}
\end{eqnarray}%
and $j_{2}^{\mu }$ are some local currents. Reprising an argument similar to
that employed in Appendix \ref{I=5} between equations (\ref{cai44}) and (\ref%
{ai50}), we find that the existence of $a_{2}^{\mathrm{int}}$ as solution to
equation (\ref{3.5}) for $i=3$ finally implies that $\chi _{2}$ expressed by
(\ref{cai3g}) must vanish. This is further equivalent to the fact that all
the functions of the type $U$ must be some real constants and both constants
$q_{1,2}$ must vanish
\begin{gather}
U_{5}\left( \varphi \right) =u_{5},\qquad U_{6}\left( \varphi \right)
=u_{6},\qquad U_{7}\left( \varphi \right) =u_{7},  \label{unq120a} \\
U_{8}\left( \varphi \right) =u_{8},\qquad U_{9}\left( \varphi \right)
=u_{9},\qquad q_{1}=0=q_{2}.  \label{unq120b}
\end{gather}%
Inserting (\ref{unq120a}) and (\ref{unq120b}) in (\ref{ai4a}), we conclude
that we can safely take%
\begin{equation}
a_{4}^{\mathrm{int}}=0  \label{a40}
\end{equation}%
in (\ref{fo8}).

\section{No-go result for $I=3$ in $a^{\mathrm{int}}$\label{I=3}}

We have seen in the previous two Appendixes \ref{I=5} and \ref{I=4} that we
can always take (\ref{ai50}) and (\ref{a40}) in (\ref{fo5}). Consequently,
the first-order deformation of the solution to the master equation in the
interacting case stops at antighost number three%
\begin{equation}
a^{\mathrm{int}}=a_{0}^{\mathrm{int}}+a_{1}^{\mathrm{int}}+a_{2}^{\mathrm{int%
}}+a_{3}^{\mathrm{int}},  \label{fo11}
\end{equation}%
where the components on the right-hand side of (\ref{fo11}) are subject to
the equations (\ref{3.6}) and (\ref{3.4})--(\ref{3.5}) for $I=3$.

The piece $a_{3}^{\mathrm{int}}$ as solution to equation (\ref{3.6}) for $I=3
$ has the general form expressed by (\ref{3.8}) for $I=3$, with $\alpha _{3}$
from $H_{3}^{\mathrm{inv}}(\delta |d)$. Looking at formula (\ref{hgamma})
and also at relation (\ref{hinvdelta}) in antighost number three and
requiring that $a_{3}^{\mathrm{int}}$ mixes BRST generators from the BF and $%
(2,1)$ sectors, we find that the most general solution to (\ref{3.6}) for $%
I=3$ reads as\footnote{%
In principle, one can add to $a_{3}^{\mathrm{int}}$ the term $\left(
M_{1}\right) _{\mu \nu \rho }\tilde{D}^{\mu \nu }S^{\rho }$, where $\left(
M_{1}\right) _{\mu \nu \rho }$ reads as in (\ref{3.15}), with $W\left(
\varphi \right) \rightarrow M_{1}\left( \varphi \right) $. It is possible to
show that such a term outputs only trivial deformations.}%
\begin{eqnarray}
a_{3}^{\mathrm{int}} &=&\tilde{\eta}^{\ast \mu }\left( q_{3}\eta S_{\mu
}+q_{4}S^{\nu }\tilde{D}_{\mu \nu }-\tfrac{1}{2}q_{5}\sigma _{\alpha \beta
}D_{\mu \nu \rho }\tilde{D}^{\nu \alpha }\tilde{D}^{\rho \beta }\right)
\notag \\
&&+q_{6}S^{\ast \mu }\eta S_{\mu }+\tfrac{1}{6}\sigma ^{\mu \nu }\left(
q_{7}C^{\ast }+q_{8}\mathcal{\tilde{G}}^{\ast }\right) \tilde{D}_{\mu \alpha
}\tilde{D}^{\alpha \beta }\tilde{D}_{\beta \nu }  \notag \\
&&+\left( \tilde{U}_{10}\right) ^{\mu \nu }\tilde{D}_{\mu \nu }\mathcal{%
\tilde{G}}+\left( \tilde{U}_{11}\right) ^{\mu \nu }\tilde{D}_{\mu \nu }C+%
\tfrac{1}{2}\left( \tilde{U}_{12}\right) ^{\mu \nu }\sigma ^{\alpha \beta
}\eta \tilde{D}_{\mu \alpha }\tilde{D}_{\nu \beta },  \label{ai3n}
\end{eqnarray}%
where any object denoted by $q$ represents an arbitrary, real constant.
Inserting (\ref{ai3n}) in equation (\ref{3.4}) for $I=3$ and using
definitions (\ref{f12a})--(\ref{f13i}), we can write%
\begin{eqnarray}
a_{2}^{\mathrm{int}} &=&-q_{3}\tilde{\eta}^{\ast \mu \nu }\left( V_{\mu
}S_{\nu }+\tfrac{1}{2}\eta A_{\mu \nu }\right) +\tfrac{q_{4}}{2}\tilde{\eta}%
^{\ast \mu \nu }\left( \mathcal{C}_{\mu \rho }\tilde{D}_{\nu }^{\quad \rho
}+S^{\rho }\tilde{F}_{\rho \mu |\nu }\right)   \notag \\
&&-\tfrac{q_{5}}{4}\tilde{\eta}^{\ast \mu \varepsilon }\sigma _{\alpha \beta
}\left( 2D_{\mu \nu \rho }\tilde{F}_{\quad |\varepsilon }^{\nu \alpha
}-F_{\mu \nu \rho |\varepsilon }\tilde{D}^{\nu \alpha }\right) \tilde{D}%
^{\rho \beta }  \notag \\
&&-q_{6}\mathcal{C}^{\ast \mu \nu }\left( 2V_{\mu }S_{\nu }+\eta \mathcal{C}%
_{\mu \nu }\right) +\tfrac{1}{4}\sigma ^{\mu \nu }\left( q_{7}C^{\ast
\lambda }-q_{8}\mathcal{\tilde{G}}^{\ast \lambda }\right) \tilde{D}_{\mu
\alpha }\tilde{D}^{\alpha \beta }\tilde{F}_{\beta \nu |\lambda }  \notag \\
&&-\tfrac{1}{2}\left( \tilde{U}_{10}\right) ^{\mu \nu \rho }\left( \tilde{F}%
_{\mu \nu |\rho }\mathcal{\tilde{G}}+\tfrac{2}{5}\tilde{D}_{\mu \nu }%
\mathcal{\tilde{G}}_{\rho }\right) +\left( \tilde{U}_{11}\right) ^{\mu \nu
\rho }\left( \tilde{D}_{\mu \nu }C_{\rho }-\tfrac{1}{2}\tilde{F}_{\mu \nu
|\rho }C\right)   \notag \\
&&+\tfrac{1}{2}\left( \tilde{U}_{12}\right) ^{\mu \nu \rho }\sigma ^{\alpha
\beta }\left( V_{\mu }\tilde{D}_{\nu \alpha }+\eta \tilde{F}_{\alpha \mu
|\nu }\right) \tilde{D}_{\rho \beta }+\bar{a}_{2}^{\mathrm{int}}.
\label{ai2i}
\end{eqnarray}%
The component $\bar{a}_{2}^{\mathrm{int}}$ represents the solution to the
homogeneous equation in antighost number two (\ref{3.6}) for $I=2$, so $\bar{%
a}_{2}^{\mathrm{int}}$ is a nontrivial element from $H\left( \gamma \right) $
of pure ghost number two and antighost number two. It is useful to decompose
$\bar{a}_{2}^{\mathrm{int}}$ as a sum between two terms
\begin{equation}
\bar{a}_{2}^{\mathrm{int}}=\hat{a}_{2}^{\mathrm{int}}+\check{a}_{2}^{\mathrm{%
int}},  \label{descom2}
\end{equation}%
with $\hat{a}_{2}^{\mathrm{int}}$ the solution to (\ref{3.6}) for $I=2$ that
ensures the consistency of $a_{2}^{\mathrm{int}}$ in antighost number one,
namely the existence of $a_{1}^{\mathrm{int}}$ as solution to (\ref{3.5})
for $i=2$ with respect to the terms from $a_{2}^{\mathrm{int}}$ containing
the functions of the type $U$ or the constants denoted by $q$, and $\check{a}%
_{2}^{\mathrm{int}}$ the solution to (\ref{3.6}) for $I=2$ that is
independently consistent in antighost number one%
\begin{equation}
\delta \check{a}_{2}^{\mathrm{int}}=-\gamma \check{c}_{1}+\partial _{\mu }%
\check{m}_{1}^{\mu }.  \label{cai2ia}
\end{equation}

Using definitions (\ref{f12a})--(\ref{f13f}) and decomposition (\ref{descom2}%
), by direct computation we obtain that%
\begin{eqnarray}
\delta a_{2}^{\mathrm{int}} &=&\delta \left[ \hat{a}_{2}^{\mathrm{int}}-%
\tfrac{1}{2}\left( \left( \tilde{U}_{12}\right) ^{\mu \nu \lambda \sigma
}B_{\mu \nu }^{\ast }+\tfrac{1}{3}\left( \tilde{U}_{12}\right) ^{\mu \nu
\rho \lambda \sigma }\eta _{\mu \nu \rho }^{\ast }\right) \sigma ^{\alpha
\beta }\tilde{D}_{\lambda \alpha }\tilde{D}_{\sigma \beta }\right]  \notag \\
&&+\gamma c_{1}+\partial _{\lambda }j_{1}^{\lambda }+\chi _{1},
\label{cai2ib}
\end{eqnarray}%
where we used the notations%
\begin{eqnarray}
c_{1} &=&-\check{c}_{1}-\tfrac{1}{2}\tilde{B}^{\ast \mu \nu \rho }\left[
q_{3}V_{\mu }A_{\nu \rho }-\tfrac{q_{4}}{6}\sigma ^{\alpha \beta }\left( 3%
\mathcal{C}_{\mu \alpha }\tilde{F}_{\nu \beta |\rho }+t_{\mu \nu |\alpha }%
\tilde{D}_{\rho \beta }\right) \right.  \notag \\
&&\left. +\tfrac{q_{5}}{4}\sigma _{\lambda \sigma }\left( D_{\mu \alpha
\beta }\tilde{F}_{\quad |\nu }^{\alpha \lambda }-2F_{\mu \alpha \beta |\nu }%
\tilde{D}^{\alpha \lambda }\right) \tilde{F}_{\quad |\rho }^{\beta \sigma }%
\right]  \notag \\
&&+q_{6}t^{\ast \mu \nu |\rho }\left( 6V_{\mu }\mathcal{C}_{\nu \rho }-\eta
t_{\mu \nu |\rho }\right) +\tfrac{1}{2}\left( \tilde{U}_{11}\right) ^{\mu
\nu \rho \lambda }\left( \tilde{F}_{\mu \nu |\rho }C_{\lambda }+\tilde{D}%
_{\mu \nu }\phi _{\rho \lambda }\right)  \notag \\
&&-\tfrac{1}{4}\sigma ^{\mu \nu }\left( q_{7}\phi ^{\ast \lambda \rho }-%
\tfrac{q_{8}}{2}\tilde{K}^{\ast \lambda \rho }\right) \tilde{D}_{\mu \alpha }%
\tilde{F}_{\quad |\lambda }^{\alpha \beta }\tilde{F}_{\beta \nu |\rho }
\notag \\
&&-\tfrac{1}{10}\left( \tilde{U}_{10}\right) ^{\mu \nu \rho \lambda }\left(
\tilde{F}_{\mu \nu |\rho }\mathcal{\tilde{G}}_{\lambda }+\tfrac{1}{4}\tilde{D%
}_{\mu \nu }\tilde{K}_{\rho \lambda }\right)  \notag \\
&&+\tfrac{1}{2}\left( \tilde{U}_{12}\right) ^{\mu \nu \rho \lambda }\sigma
^{\alpha \beta }\left( V_{\mu }\tilde{D}_{\nu \alpha }\tilde{F}_{\rho \beta
|\lambda }-\tfrac{1}{4}\eta \tilde{F}_{\mu \alpha |\nu }\tilde{F}_{\rho
\beta |\lambda }\right)  \notag \\
&&+\tfrac{1}{2}\left( \tilde{U}_{12}\right) ^{\mu \nu \rho \lambda \sigma
}\sigma ^{\alpha \beta }B_{\mu \nu }^{\ast }\tilde{D}_{\sigma \alpha }\tilde{%
F}_{\rho \beta |\lambda },  \label{cai2ic}
\end{eqnarray}%
\begin{eqnarray}
\chi _{1} &=&\tfrac{1}{6}\tilde{B}^{\ast \mu \nu \rho }\left[ q_{3}\left(
\eta D_{\mu \nu \rho }+3S_{\rho }\partial _{\lbrack \mu }V_{\nu ]}\right)
-q_{4}\sigma ^{\alpha \beta }\left( D_{\mu \nu \alpha }\tilde{D}_{\rho \beta
}+3\tilde{R}_{\mu \alpha |\nu \rho }S_{\beta }\right) \right.  \notag \\
&&\left. +\tfrac{3q_{5}}{2}\sigma _{\lambda \sigma }\left( R_{\mu \alpha
\beta |\nu \rho }\tilde{D}^{\alpha \lambda }-2D_{\mu \alpha \beta }\tilde{R}%
_{\quad |\nu \rho }^{\alpha \lambda }\right) \tilde{D}^{\beta \sigma }\right]
-6q_{6}t^{\ast \mu \nu |\rho }\left( \partial _{\lbrack \mu }V_{\nu
]}\right) S_{\rho }  \notag \\
&&+\tfrac{1}{2}\sigma ^{\mu \nu }\left( q_{7}\phi ^{\ast \rho \lambda }-%
\tfrac{q_{8}}{2}\tilde{K}^{\ast \rho \lambda }\right) \tilde{D}_{\mu \alpha }%
\tilde{D}^{\alpha \beta }\tilde{R}_{\beta \nu |\rho \lambda }-\tfrac{1}{2}%
\left( \tilde{U}_{10}\right) ^{\mu \nu \rho \lambda }\tilde{R}_{\mu \nu
|\rho \lambda }\mathcal{\tilde{G}}  \notag \\
&&-\tfrac{1}{2}\left( \tilde{U}_{11}\right) ^{\mu \nu \rho \lambda }\tilde{R}%
_{\mu \nu |\rho \lambda }C-\tfrac{1}{2}\left( \tilde{U}_{12}\right) ^{\mu
\nu \rho \lambda }\sigma ^{\alpha \beta }\eta \tilde{D}_{\mu \alpha }\tilde{R%
}_{\nu \beta |\rho \lambda },  \label{cai2ie}
\end{eqnarray}%
and $j_{1}^{\mu }$ are some local currents. It is easy to see that $\chi
_{1} $ given in (\ref{cai2ie}) is a nontrivial object from $H\left( \gamma
\right) $ in pure ghost number two, which obviously does not reduce to a
full divergence. Then, since (\ref{cai2ib}) requires that it is $\gamma $%
-exact modulo $d$, it must vanish, which further implies that all the
functions of the type $U\left( \varphi \right) $ are some real constants and
all the constants denoted by $q$ vanish%
\begin{gather}
U_{10}\left( \varphi \right) =u_{10},\qquad U_{11}\left( \varphi \right)
=u_{11},\qquad U_{12}\left( \varphi \right) =u_{12},  \label{conda31} \\
q_{3}=q_{4}=q_{5}=q_{6}=q_{7}=q_{8}=0.  \label{conda32}
\end{gather}

Inserting conditions (\ref{conda31}) and (\ref{conda32}) into (\ref{ai3n}),
we conclude that we conclude that we can safely take%
\begin{equation}
a_{3}^{\mathrm{int}}=0  \label{ai3good}
\end{equation}%
in (\ref{fo11}).

\section{No-go result for $I=0$ in $a^{\mathrm{int}}$\label{I=0}}

The solution to the `homogeneous' equation (\ref{om0}) can be represented as
\begin{equation}
\bar{a}_{0}^{\mathrm{int}}=\bar{a}_{0}^{\prime \mathrm{int}}+\bar{a}%
_{0}^{\prime \prime \mathrm{int}},  \label{om01}
\end{equation}%
where
\begin{eqnarray}
\gamma \bar{a}_{0}^{\prime \mathrm{int}} &=&0,  \label{om02a} \\
\gamma \bar{a}_{0}^{\prime \prime \mathrm{int}} &=&\partial _{\mu }\bar{m}%
_{0}^{\mu }  \label{om02b}
\end{eqnarray}%
and $\bar{m}_{0}^{\mu }$ is a nonvanishing, local current.

According to the general result expressed by (\ref{3.8}) in both antighost
and pure ghost numbers equal to zero, equation (\ref{om02a}) implies%
\begin{equation}
\bar{a}_{0}^{\prime \mathrm{int}}=\bar{a}_{0}^{\prime \mathrm{int}}\left( %
\left[ F_{\bar{A}}\right] \right) ,  \label{om03}
\end{equation}%
where $F_{\bar{A}}$ are listed in (\ref{3.8}). Solution (\ref{om03}) is
assumed to provide a cross-coupling Lagrangian. Therefore, since $R_{\mu \nu
\rho |\alpha \beta }$ is the most general gauge-invariant quantity depending
on the field $t_{\mu \nu |\alpha }$, it follows that each interaction vertex
from $\bar{a}_{0}^{\prime \mathrm{int}}$ is required to be at least linear
in $R_{\mu \nu \rho |\alpha \beta }$ and to depend at least on a BF field.
But $R_{\mu \nu \rho |\alpha \beta }$ contains two spacetime derivatives, so
the emerging interacting field equations would exhibit at least two
spacetime derivatives acting on the BF field(s) from the interaction
vertices. Nevertheless, this contradicts the general assumption on the
preservation of the differential order of each field equation with respect
to the free theory (see assumption ii) from the beginning of section \ref%
{stand}), so we must set%
\begin{equation}
\bar{a}_{0}^{\prime \mathrm{int}}=0.  \label{om04}
\end{equation}

Next, we solve equation (\ref{om02b}). In view of this, we decompose $\bar{a}%
_{0}^{\prime \prime \mathrm{int}}$ with respect to the number of derivatives
acting on the fields as%
\begin{equation}
\bar{a}_{0}^{\prime \prime \mathrm{int}}=\overset{(0)}{\pi }+\overset{(1)}{%
\pi }+\overset{(2)}{\pi },  \label{om05}
\end{equation}%
where each $\overset{(i)}{\pi }$ contains precisely $i$ spacetime
derivatives. Of course, each $\overset{(i)}{\pi }$ is required to mix the BF
and $(2,1)$ field sectors in order to produce cross-interactions. In
agreement with (\ref{om05}), equation (\ref{om02b}) is equivalent to
\begin{eqnarray}
\gamma \overset{(0)}{\pi } &=&\partial _{\mu }\overset{(0)}{m}_{0}^{\mu },
\label{om06a} \\
\gamma \overset{(1)}{\pi } &=&\partial _{\mu }\overset{(1)}{m}_{0}^{\mu },
\label{om06b} \\
\gamma \overset{(2)}{\pi } &=&\partial _{\mu }\overset{(2)}{m}_{0}^{\mu }.
\label{om06c}
\end{eqnarray}%
Using definitions (\ref{f13b})--(\ref{f13d}) and an integration by parts it is
possible to show that%
\begin{eqnarray}
\gamma \overset{(0)}{\pi } &=&\partial _{\mu }\overset{(0)}{m}_{0}^{\mu
}-\left( \partial _{\mu }\frac{\partial \overset{(0)}{\pi }}{\partial t_{\mu
(\nu |\alpha )}}\right) S_{\nu \alpha }+\left( 2\partial _{\mu }\frac{%
\partial \overset{(0)}{\pi }}{\partial t_{\alpha \beta |\mu }}-\partial
_{\mu }\frac{\partial \overset{(0)}{\pi }}{\partial t_{\mu \lbrack \alpha
|\beta ]}}\right) A_{\alpha \beta }  \notag \\
&&+\left( \partial _{\lbrack \mu }\frac{\partial \overset{(0)}{\pi }}{%
\partial H^{\nu ]}}\right) C^{\mu \nu }-\left( \partial _{\mu }\frac{%
\partial \overset{(0)}{\pi }}{\partial V_{\mu }}\right) \eta +\left(
\partial _{\lbrack \mu }\frac{\partial \overset{(0)}{\pi }}{\partial B^{\nu
\rho ]}}\right) \eta ^{\mu \nu \rho }  \notag \\
&&-2\left( \partial _{\mu }\frac{\partial \overset{(0)}{\pi }}{\partial \phi
_{\mu \nu }}\right) C_{\nu }+\left( \partial _{\lbrack \mu }\frac{\partial
\overset{(0)}{\pi }}{\partial K^{\nu \rho \lambda ]}}\right) \mathcal{G}%
^{\mu \nu \rho \lambda }.  \label{om07}
\end{eqnarray}%
From (\ref{om07}) we observe that $\overset{(0)}{\pi }$ is solution to (\ref%
{om06a}) if and only if the following conditions are satisfied
simultaneously
\begin{gather}
\partial _{\mu }\frac{\partial \overset{(0)}{\pi }}{\partial t_{\mu (\nu
|\alpha )}}=0,\qquad \partial _{\mu }\frac{\partial \overset{(0)}{\pi }}{%
\partial t_{\alpha \beta |\mu }}=0,\qquad \partial _{\lbrack \mu }\frac{%
\partial \overset{(0)}{\pi }}{\partial H^{\nu ]}}=0,  \label{om08a} \\
\partial _{\mu }\frac{\partial \overset{(0)}{\pi }}{\partial V_{\mu }}%
=0,\qquad \partial _{\lbrack \mu }\frac{\partial \overset{(0)}{\pi }}{%
\partial B^{\nu \rho ]}}=0,\qquad \partial _{\mu }\frac{\partial \overset{(0)%
}{\pi }}{\partial \phi _{\mu \nu }}=0,\qquad \partial _{\lbrack \mu }\frac{%
\partial \overset{(0)}{\pi }}{\partial K^{\nu \rho \lambda ]}}=0.
\label{om08b}
\end{gather}%
Because $\overset{(0)}{\pi }$ is derivative-free, the solutions to equations
(\ref{om08a})--(\ref{om08b}) read as
\begin{align}
\frac{\partial \overset{(0)}{\pi }}{\partial t_{\mu \nu |\alpha }}& =\tau
^{\mu \nu |\alpha }, & \frac{\partial \overset{(0)}{\pi }}{\partial H^{\mu }}%
& =h_{\mu }, & \frac{\partial \overset{(0)}{\pi }}{\partial V_{\mu }}&
=v^{\mu },  \label{om09a} \\
\frac{\partial \overset{(0)}{\pi }}{\partial B^{\mu \nu }}& =b_{\mu \nu }, &
\frac{\partial \overset{(0)}{\pi }}{\partial \phi _{\mu \nu }}& =f_{\mu \nu
}, & \frac{\partial \overset{(0)}{\pi }}{\partial K^{\mu \nu \rho }}&
=k_{\mu \nu \rho },  \label{om09b}
\end{align}%
where $\tau ^{\mu \nu |\alpha }$, $h_{\mu }$, $v^{\mu }$, $b_{\mu \nu }$, $%
f_{\mu \nu }$, and $k_{\mu \nu \rho }$ are some real, constant tensors. In
addition, $\tau ^{\mu \nu |\alpha }$ display the same mixed symmetry
properties like the tensor field $t^{\mu \nu |\alpha }$ and $b_{\mu \nu }$, $%
f_{\mu \nu }$, and $k_{\mu \nu \rho }$ are completely antisymmetric. Because
there are no such constant tensors in $D=5$, we conclude that (\ref{om08a}%
)--(\ref{om08b}) possess only the trivial solution, which further implies
that
\begin{equation}
\overset{(0)}{\pi }=0.  \label{om010}
\end{equation}

Related to equation (\ref{om06b}), we use again definitions (\ref{f13b})--(%
\ref{f13d}) and integrate twice by parts, obtaining
\begin{eqnarray}
\gamma \overset{(1)}{\pi } &=&\partial _{\mu }\overset{(1)}{m}_{0}^{\mu
}-\left( \partial _{\mu }\frac{\delta \overset{(1)}{\pi }}{\delta t_{\mu
(\alpha |\beta )}}\right) S_{\alpha \beta }-\left( \partial _{\mu }\frac{%
\delta \overset{(1)}{\pi }}{\delta t_{\mu \lbrack \alpha |\beta ]}}%
-2\partial _{\mu }\frac{\delta \overset{(1)}{\pi }}{\delta t_{\alpha \beta
|\mu }}\right) A_{\alpha \beta }  \notag \\
&&+\left( \partial _{\lbrack \mu }\frac{\delta \overset{(1)}{\pi }}{\delta
H^{\nu ]}}\right) C^{\mu \nu }-\left( \partial _{\mu }\frac{\delta \overset{%
(1)}{\pi }}{\delta V_{\mu }}\right) \eta +\left( \partial _{\lbrack \mu }%
\frac{\delta \overset{(1)}{\pi }}{\delta B^{\nu \rho ]}}\right) \eta ^{\mu
\nu \rho }  \notag \\
&&-2\left( \partial _{\mu }\frac{\delta \overset{(1)}{\pi }}{\delta \phi
_{\mu \nu }}\right) C_{\nu }+\left( \partial _{\lbrack \mu }\frac{\delta
\overset{(1)}{\pi }}{\delta K^{\nu \rho \lambda ]}}\right) \mathcal{G}^{\mu
\nu \rho \lambda }.  \label{om11}
\end{eqnarray}%
Inspecting (\ref{om11}), we observe that $\overset{(1)}{\pi }$ satisfies
equation (\ref{om06b}) if and only if the following relations take place
simultaneously
\begin{gather}
\partial _{\mu }\frac{\delta \overset{(1)}{\pi }}{\delta t_{\mu (\alpha
|\beta )}}=0,\qquad \partial _{\mu }\frac{\delta \overset{(1)}{\pi }}{\delta
t_{\alpha \beta |\mu }}=0,\qquad \partial _{\lbrack \mu }\frac{\delta
\overset{(1)}{\pi }}{\delta H^{\nu ]}}=0,  \label{om012a} \\
\partial _{\mu }\frac{\delta \overset{(1)}{\pi }}{\delta V_{\mu }}=0,\qquad
\partial _{\lbrack \mu }\frac{\delta \overset{(1)}{\pi }}{\delta B^{\nu \rho
]}}=0,\qquad \partial _{\mu }\frac{\delta \overset{(1)}{\pi }}{\delta \phi
_{\mu \nu }}=0,\qquad \partial _{\lbrack \mu }\frac{\delta \overset{(1)}{\pi
}}{\delta K^{\nu \rho \lambda ]}}=0.  \label{om012b}
\end{gather}%
The solutions to equations (\ref{om012a})--(\ref{om012b}) are expressed by%
\begin{gather}
\frac{\delta \overset{(1)}{\pi }}{\delta t_{\mu (\alpha |\beta )}}=\partial
_{\nu }s^{\mu \nu \alpha \beta },\qquad \frac{\delta \overset{(1)}{\pi }}{%
\delta t_{\alpha \beta |\mu }}=\partial _{\nu }\tau ^{\alpha \beta \mu \nu },
\label{om013a} \\
\frac{\delta \overset{(1)}{\pi }}{\delta H^{\mu }}=\partial _{\mu }h,\qquad
\frac{\delta \overset{(1)}{\pi }}{\delta V_{\mu }}=\partial _{\nu }v^{\mu
\nu },\qquad \frac{\delta \overset{(1)}{\pi }}{\delta B^{\mu \nu }}=\partial
_{\lbrack \mu }b_{\nu ]},  \label{om013b} \\
\frac{\delta \overset{(1)}{\pi }}{\delta \phi _{\mu \nu }}=\partial _{\rho
}f^{\mu \nu \rho },\qquad \frac{\delta \overset{(1)}{\pi }}{\delta K^{\mu
\nu \rho }}=\partial _{\lbrack \mu }k_{\nu \rho ]},  \label{om013c}
\end{gather}%
where the quantities $s^{\mu \nu \alpha \beta }$, $\tau ^{\alpha \beta \mu
\nu }$, $h$, $v^{\mu \nu }$, $b_{\mu }$, $f^{\mu \nu \rho }$, and $k_{\mu
\nu }$ are some tensors depending at most on the undifferentiated fields $%
\Phi ^{\alpha _{0}}$ from (\ref{f10a}). In addition, they display the
symmetry/antisymmetry properties%
\begin{gather}
s^{\mu \nu \alpha \beta }=-s^{\nu \mu \alpha \beta }=s^{\mu \nu \beta \alpha
},  \label{symm1} \\
\tau ^{\alpha \beta \mu \nu }=-\tau ^{\beta \alpha \mu \nu }=-\tau ^{\alpha
\beta \nu \mu },  \label{symm2} \\
\tau ^{\lbrack \alpha \beta \mu ]\nu }=0,  \label{symm3}
\end{gather}%
and $v^{\mu \nu }$, $f^{\mu \nu \rho }$, and $k_{\mu \nu }$ are completely
antisymmetric. Because both tensors $s^{\mu \nu \alpha \beta }$ and $\tau
^{\alpha \beta \mu \nu }$ are derivative-free, their are related through%
\begin{equation}
s^{\mu \nu \alpha \beta }=\tau ^{\mu (\alpha \beta )\nu }.  \label{legs_a}
\end{equation}%
Using successively properties (\ref{symm1})--(\ref{symm3}) and formula (\ref%
{legs_a}), it can be shown that $\tau ^{\alpha \beta \mu \nu }$ is
completely antisymmetric. This last property together with (\ref{symm3})
leads to
\begin{equation*}
\tau ^{\alpha \beta \mu \nu }=0,
\end{equation*}%
which replaced in the latter equality from (\ref{om013a}) produces
\begin{equation*}
\frac{\delta \overset{(1)}{\pi }}{\delta t_{\alpha \beta |\mu }}=0.
\end{equation*}%
This means that the entire dependence of $\overset{(1)}{\pi }$ on $t_{\alpha
\beta |\mu }$ is trivial (reduces to a full divergence), and therefore $%
\overset{(1)}{\pi }$ can at most describe self-interactions in the BF
sector. Since there is no nontrivial solution to (\ref{om06b}) that mixes
the BF and $(2,1)$ field sectors, we can safely take%
\begin{equation}
\overset{(1)}{\pi }=0.  \label{pi1}
\end{equation}

In the end of this section we analyze equation (\ref{om06c}). Taking one
more time into account definitions (\ref{f13b})--(\ref{f13d}), it is easy to
see that (\ref{om06c}) implies that the EL derivatives of $\overset{(2)}{\pi
}$ are subject to the equations
\begin{gather}
\partial _{\mu }\frac{\delta \overset{(2)}{\pi }}{\delta t_{\mu (\alpha
|\beta )}}=0,\qquad \partial _{\mu }\frac{\delta \overset{(2)}{\pi }}{\delta
t_{\alpha \beta |\mu }}=0,  \label{om014a} \\
\partial _{\lbrack \mu }\frac{\delta \overset{(2)}{\pi }}{\delta H^{\nu ]}}%
=0,\qquad \partial _{\mu }\frac{\delta \overset{(2)}{\pi }}{\delta V_{\mu }}%
=0,\qquad \partial _{\lbrack \mu }\frac{\delta \overset{(2)}{\pi }}{\delta
B^{\nu \rho ]}}=0,  \label{om014b} \\
\partial _{\mu }\frac{\delta \overset{(2)}{\pi }}{\delta \phi _{\mu \nu }}%
=0,\qquad \partial _{\lbrack \mu }\frac{\delta \overset{(2)}{\pi }}{\delta
K^{\nu \rho \lambda ]}}=0.  \label{om014c}
\end{gather}%
Because $\overset{(2)}{\pi }$ (and also its EL derivatives) contains two
spacetime derivatives, the solution to both equations from (\ref{om014a}) is
of the type%
\begin{equation}
\frac{\delta \overset{(2)}{\pi }}{\delta t_{\mu \nu |\alpha }}=\partial
_{\rho }\partial _{\beta }\bar{\tau}^{\mu \nu \rho |\alpha \beta },
\label{som014a}
\end{equation}%
where $\bar{\tau}^{\mu \nu \rho |\alpha \beta }$ depends only on the
undifferentiated fields $\Phi ^{\alpha _{0}}$ and exhibits the mixed
symmetry $\left( 3,2\right) $. This means that $\bar{\tau}^{\mu \nu \rho
|\alpha \beta }$ is simultaneously antisymmetric in its first three and
respectively last two indices and satisfies the identity $\bar{\tau}^{[\mu
\nu \rho |\alpha ]\beta }=0$. The solutions to the remaining equations, (\ref%
{om014b}) and (\ref{som14c}), can be represented as%
\begin{gather}
\frac{\delta \overset{(2)}{\pi }}{\delta H^{\mu }}=\partial _{\mu }\bar{h}%
,\qquad \frac{\delta \overset{(2)}{\pi }}{\delta V_{\mu }}=\partial _{\nu }%
\bar{v}^{\mu \nu },\qquad \frac{\delta \overset{(2)}{\pi }}{\delta B^{\mu
\nu }}=\partial _{\lbrack \mu }\bar{b}_{\nu ]},  \label{som14b} \\
\frac{\delta \overset{(2)}{\pi }}{\delta \phi _{\mu \nu }}=\partial _{\rho }%
\bar{f}^{\mu \nu \rho },\qquad \frac{\delta \overset{(2)}{\pi }}{\delta
K^{\mu \nu \rho }}=\partial _{\lbrack \mu }\bar{k}_{\nu \rho ]},
\label{som14c}
\end{gather}%
where the functions $\bar{v}^{\mu \nu }$, $\bar{f}^{\mu \nu \rho }$, and $%
\bar{k}_{\mu \nu }$ are completely antisymmetric and contain a single
spacetime derivative.

Let $N$ be a derivation in the algebra of the fields $t_{\mu \nu |\alpha }$,
$H^{\mu }$, $V_{\mu }$, $B^{\mu \nu }$, $\phi _{\mu \nu }$, $K^{\mu \nu \rho
}$, and of their derivatives, which counts the powers of these fields and of
their derivatives
\begin{eqnarray}
N &=&\sum\limits_{n\geq 0}\left( \left( \partial _{\mu _{1}\cdots \mu
_{n}}t_{\mu \nu |\alpha }\right) \frac{\partial }{\partial \left( \partial
_{\mu _{1}\cdots \mu _{n}}t_{\mu \nu |\alpha }\right) }+\left( \partial
_{\mu _{1}\cdots \mu _{n}}H^{\mu }\right) \frac{\partial }{\partial \left(
\partial _{\mu _{1}\cdots \mu _{n}}H^{\mu }\right) }\right.  \notag \\
&&+\left( \partial _{\mu _{1}\cdots \mu _{n}}V_{\mu }\right) \frac{\partial
}{\partial \left( \partial _{\mu _{1}\cdots \mu _{n}}V_{\mu }\right) }%
+\left( \partial _{\mu _{1}\cdots \mu _{n}}B^{\mu \nu }\right) \frac{%
\partial }{\partial \left( \partial _{\mu _{1}\cdots \mu _{n}}B^{\mu \nu
}\right) }  \notag \\
&&\left. +\left( \partial _{\mu _{1}\cdots \mu _{n}}\phi _{\mu \nu }\right)
\frac{\partial }{\partial \left( \partial _{\mu _{1}\cdots \mu _{n}}\phi
_{\mu \nu }\right) }+\left( \partial _{\mu _{1}\cdots \mu _{n}}K^{\mu \nu
\rho }\right) \frac{\partial }{\partial \left( \partial _{\mu _{1}\cdots \mu
_{n}}K^{\mu \nu \rho }\right) }\right) .  \label{om015}
\end{eqnarray}%
We emphasize that $N$ does not `see' either the scalar field $\varphi $ or
its spacetime derivatives. It is easy to check that for every nonintegrated
density $\Psi $ we have%
\begin{eqnarray}
N\Psi &=&\frac{\delta \Psi }{\delta t_{\mu \nu |\alpha }}t_{\mu \nu |\alpha
}+\frac{\delta \Psi }{\delta H^{\mu }}H^{\mu }+\frac{\delta \Psi }{\delta
V_{\mu }}V_{\mu }+\frac{\delta \Psi }{\delta B^{\mu \nu }}B^{\mu \nu }
\notag \\
&&+\frac{\delta \Psi }{\delta \phi _{\mu \nu }}\phi _{\mu \nu }+\frac{\delta
\Psi }{\delta K^{\mu \nu \rho }}K^{\mu \nu \rho }+\partial _{\mu }s^{\mu }.
\label{om016}
\end{eqnarray}%
If $\Psi ^{(n)}$ is a homogeneous polynomial of degree $n$ in the fields $%
t_{\mu \nu |\alpha }$, $H^{\mu }$, $V_{\mu }$, $B^{\mu \nu }$, $\phi _{\mu
\nu }$, $K^{\mu \nu \rho }$ and their derivatives (such a polynomial may
depend also on $\varphi $ and its spacetime derivatives, but the homogeneity
does not take them into consideration since $\Psi $ is allowed to be a
series in $\varphi $), then
\begin{equation*}
N\Psi ^{(n)}=n\Psi ^{(n)}.
\end{equation*}%
Based on results (\ref{som014a})--(\ref{som14c}), we can write%
\begin{eqnarray}
N\overset{(2)}{\pi } &=&-\tfrac{1}{3}\bar{\tau}^{\mu \nu \rho |\alpha \beta
}R_{\mu \nu \rho |\alpha \beta }-\bar{h}\partial _{\mu }H^{\mu }+\tfrac{1}{2}%
\bar{v}^{\mu \nu }\partial _{\lbrack \mu }V_{\nu ]}+2\bar{b}_{\mu }\partial
_{\nu }B^{\mu \nu }  \notag \\
&&-\tfrac{1}{3}\bar{f}^{\mu \nu \rho }\partial _{\lbrack \mu }\phi _{\nu
\rho ]}-3\bar{k}_{\mu \nu }\partial _{\rho }K^{\mu \nu \rho }+\partial _{\mu
}m^{\mu }.  \label{om017}
\end{eqnarray}%
We decompose $\overset{(2)}{\pi }$ along the degree $n$ as%
\begin{equation}
\overset{(2)}{\pi }=\sum\limits_{n\geq 2}\overset{(2)}{\pi }^{(n)},
\label{om018}
\end{equation}%
where $N\overset{(2)}{\pi }^{(n)}=n\overset{(2)}{\pi }^{(n)}$ ($n\geq 2$ in (%
\ref{om018}) because $\overset{(2)}{\pi }$, and hence every $\overset{(2)}{%
\pi }^{(n)}$, is assumed to describe cross-interactions between the BF model
and the tensor field with the mixed symmetry $(2,1)$), and find that
\begin{equation}
N\overset{(2)}{\pi }=\sum\limits_{n\geq 2}n\overset{(2)}{\pi }^{(n)}.
\label{om019}
\end{equation}%
Comparing (\ref{om019}) with (\ref{om017}), it follows that decomposition (%
\ref{om018}) induces a similar one with respect to each function $\bar{\tau}%
^{\mu \nu \rho |\alpha \beta }$, $\bar{h}$, $\bar{v}^{\mu \nu }$, $\bar{b}%
_{\mu }$, $\bar{f}^{\mu \nu \rho }$, and $\bar{k}_{\mu \nu }$
\begin{align}
\bar{\tau}^{\mu \nu \rho |\alpha \beta }& =\sum\limits_{n\geq 2}\bar{\tau}%
_{(n-1)}^{\mu \nu \rho |\alpha \beta }, & \bar{h}& =\sum\limits_{n\geq 2}%
\bar{h}_{(n-1)}, & \bar{v}^{\mu \nu }& =\sum\limits_{n\geq 2}\bar{v}%
_{(n-1)}^{\mu \nu },  \label{om020a} \\
\bar{b}^{\mu }& =\sum\limits_{n\geq 2}\bar{b}_{(n-1)}^{\mu }, & \bar{f}^{\mu
\nu \rho }& =\sum\limits_{n\geq 2}\bar{f}_{(n-1)}^{\mu \nu \rho }, & \bar{k}%
^{\mu \nu }& =\sum\limits_{n\geq 2}\bar{k}_{(n-1)}^{\mu \nu }.
\label{om020b}
\end{align}%
Inserting (\ref{om020a}) and (\ref{om020b}) in (\ref{om017}) and comparing
the resulting expression with (\ref{om019}), we get
\begin{eqnarray}
\overset{(2)}{\pi }^{(n)} &=&-\tfrac{1}{3n}\bar{\tau}_{(n-1)}^{\mu \nu \rho
|\alpha \beta }R_{\mu \nu \rho |\alpha \beta }-\tfrac{1}{n}\bar{h}%
_{(n-1)}\partial _{\mu }H^{\mu }+\tfrac{1}{2n}\bar{v}_{(n-1)}^{\mu \nu
}\partial _{\lbrack \mu }V_{\nu ]}  \notag \\
&&+\tfrac{2}{n}\bar{b}_{(n-1)}^{\mu }\partial ^{\nu }B_{\mu \nu }-\tfrac{1}{%
3n}\bar{f}_{(n-1)}^{\mu \nu \rho }\partial _{\lbrack \mu }\phi _{\nu \rho ]}-%
\tfrac{3}{n}\bar{k}_{(n-1)}^{\mu \nu }\partial ^{\rho }K_{\mu \nu \rho
}+\partial _{\mu }m_{(n)}^{\mu }.  \label{om021}
\end{eqnarray}%
Replacing the last result, (\ref{om021}), into (\ref{om018}), we further
obtain%
\begin{eqnarray}
\overset{(2)}{\pi } &=&-\tfrac{1}{3}\hat{\tau}^{\mu \nu \rho |\alpha \beta
}R_{\mu \nu \rho |\alpha \beta }-\hat{h}\partial _{\mu }H^{\mu }+\tfrac{1}{2}%
\hat{v}^{\mu \nu }\partial _{\lbrack \mu }V_{\nu ]}+2\hat{b}_{\mu }\partial
_{\nu }B^{\mu \nu }  \notag \\
&&-\tfrac{1}{3}\hat{f}^{\mu \nu \rho }\partial _{\lbrack \mu }\phi _{\nu
\rho ]}-3\hat{k}_{\mu \nu }\partial _{\rho }K^{\mu \nu \rho }+\partial _{\mu
}\hat{m}^{\mu },  \label{om022}
\end{eqnarray}%
where%
\begin{align}
\hat{\tau}^{\mu \nu \rho |\alpha \beta }& =\sum\limits_{n\geq 2}\tfrac{1}{n}%
\bar{\tau}_{(n-1)}^{\mu \nu \rho |\alpha \beta }, & \hat{h}&
=\sum\limits_{n\geq 2}\tfrac{1}{n}\bar{h}_{(n-1)}, & \hat{v}^{\mu \nu }&
=\sum\limits_{n\geq 2}\tfrac{1}{n}\bar{v}_{(n-1)}^{\mu \nu },  \label{om023a}
\\
\hat{b}^{\mu }& =\sum\limits_{n\geq 2}\tfrac{1}{n}\bar{b}_{(n-1)}^{\mu }, &
\hat{f}^{\mu \nu \rho }& =\sum\limits_{n\geq 2}\tfrac{1}{n}\bar{f}%
_{(n-1)}^{\mu \nu \rho }, & \hat{k}^{\mu \nu }& =\sum\limits_{n\geq 2}\tfrac{%
1}{n}\bar{k}_{(n-1)}^{\mu \nu }.  \label{om023b}
\end{align}%
So far, we showed that the solution to (\ref{om06c}) can be put in the form (%
\ref{om022}). By means of definitions (\ref{f12b})--(\ref{f12c}), we can
bring (\ref{om022}) to the expression
\begin{eqnarray}
\overset{(2)}{\pi } &=&-\tfrac{1}{3}\hat{\tau}^{\mu \nu \rho |\alpha \beta
}R_{\mu \nu \rho |\alpha \beta }+\partial _{\mu }\hat{m}^{\mu }  \notag \\
&&+\delta \left( -\varphi ^{\ast }\hat{h}-B_{\mu \nu }^{\ast }\hat{v}^{\mu
\nu }-2V_{\mu }^{\ast }\hat{b}^{\mu }+K_{\mu \nu \rho }^{\ast }\hat{f}^{\mu
\nu \rho }-3\phi ^{\ast \mu \nu }\hat{k}_{\mu \nu }\right) .  \label{om024}
\end{eqnarray}%
The $\delta $-exact modulo $d$ terms in the right-hand side of (\ref{om024})
produce purely trivial interactions, which can be eliminated via field
redefinitions. This is due to the isomorphism $H^{i}\left( s|d\right) \simeq
H^{i}\left( \gamma |d,H_{0}\left( \delta \right) \right) $ in all positive
values of the ghost number and respectively of the pure ghost number~\cite%
{gen1}, which at $i=0$ allows one to state that any solution of equation (%
\ref{om06c}) that is $\delta $-exact modulo $d$ is in fact a trivial cocycle
from $H^{0}\left( s|d\right) $. In conclusion, the only nontrivial solution
to (\ref{om06c}) can be written as
\begin{equation}
\overset{(2)}{\pi }=-\tfrac{1}{3}\hat{\tau}^{\mu \nu \rho |\alpha \beta
}R_{\mu \nu \rho |\alpha \beta },  \label{om025}
\end{equation}%
where $\hat{\tau}^{\mu \nu \rho |\alpha \beta }$ displays the mixed symmetry
$\left( 3,2\right) $, is derivative-free, and is required to depend at least
on one field from the BF sector. But $R_{\mu \nu \rho |\alpha \beta }$
already contains two spacetime derivatives, so such a $\overset{(2)}{\pi }$
disagrees with the hypothesis on the differential order of the interacting
field equations (see also the discussion following formula (\ref{om03})),
which means that we must set
\begin{equation}
\overset{(2)}{\pi }=0.  \label{om033}
\end{equation}

Substituting results (\ref{om010}), (\ref{pi1}), and (\ref{om033}) into
decomposition (\ref{om05}), we obtain%
\begin{equation}
\bar{a}_{0}^{\prime \prime \mathrm{int}}=0,  \label{om034}
\end{equation}%
which combined with (\ref{om04}) proves that indeed there is no nontrivial
solution to the `homogeneous' equation (\ref{om0}) that complies with all
the working hypotheses
\begin{equation}
\bar{a}_{0}^{\mathrm{int}}=0.  \label{om035}
\end{equation}

\section{Notations from section \protect\ref{higher}\label{struct}}

In this Appendix we list the concrete form of the various notations made in
section \ref{higher}.

The polynomials denoted by $\bar{X}_{p}^{\left( i\right) }$ that enter $\bar{%
\Delta}^{\mathrm{int}}$ given in (\ref{delbar_int}) read as%
\begin{eqnarray}
\bar{X}_{0}^{\left( 1\right) } &=&6S^{\ast }\eta C+12t^{\ast \mu }\left(
V_{\mu }C+\eta C_{\mu }\right) +6\left( 2B_{\mu \nu }^{\ast }C+V_{[\mu
}C_{\nu ]}-\phi _{\mu \nu }\eta \right) F^{\mu \nu }  \notag \\
&&-2\left( 2\eta _{\mu \nu \rho }^{\ast }C+2B_{[\mu \nu }^{\ast }C_{\rho
]}-3K_{\mu \nu \rho }^{\ast }\eta -\phi _{\lbrack \mu \nu }V_{\rho ]}\right)
\tilde{D}_{\lambda \sigma }\varepsilon ^{\mu \nu \rho \lambda \sigma },
\label{xxt1}
\end{eqnarray}%
\begin{eqnarray}
\bar{X}_{1}^{\left( 1\right) } &=&\left[ \left( -2C_{\mu \nu \rho }^{\ast
}\eta -2C_{[\mu \nu }^{\ast }V_{\rho ]}-4H_{[\mu }^{\ast }B_{\nu \rho
]}^{\ast }\right) C\right.  \notag \\
&&\left. +\left( -2H_{[\mu }^{\ast }V_{\nu }C_{\rho ]}+2H_{[\mu }^{\ast
}\phi _{\nu \rho ]}\eta \right) +2C_{[\mu \nu }^{\ast }C_{\rho ]}\eta \right]
\tilde{D}_{\lambda \sigma }\varepsilon ^{\mu \nu \rho \lambda \sigma }
\notag \\
&&-12H_{\mu }^{\ast }t^{\ast \mu }\eta C+6\left( H_{[\mu }^{\ast }V_{\nu
]}C+2H_{\mu }^{\ast }\eta C_{\nu }+C_{\mu \nu }^{\ast }\eta C\right) F^{\mu
\nu },  \label{xxt2}
\end{eqnarray}%
\begin{eqnarray}
\bar{X}_{2}^{\left( 1\right) } &=&\left[ \left( -2H_{[\mu }^{\ast }C_{\nu
\rho ]}^{\ast }\eta -2H_{[\mu }^{\ast }H_{\nu }^{\ast }V_{\rho ]}\right)
C\right.  \notag \\
&&\left. +2H_{[\mu }^{\ast }H_{\nu }^{\ast }C_{\rho ]}\eta \right] \tilde{D}%
_{\lambda \sigma }\varepsilon ^{\mu \nu \rho \lambda \sigma }+6H_{\mu
}^{\ast }H_{\nu }^{\ast }\eta CF^{\mu \nu },  \label{xxt3}
\end{eqnarray}%
\begin{equation}
\bar{X}_{3}^{\left( 1\right) }=4H_{\mu }^{\ast }H_{\nu }^{\ast }H_{\rho
}^{\ast }\eta CD^{\mu \nu \rho },  \label{xxt4}
\end{equation}%
\begin{eqnarray}
\bar{X}_{0}^{\left( 2\right) } &=&-12\cdot 5!\left( S^{\ast }\eta +2t^{\ast
\mu }V_{\mu }+2B_{\mu \nu }^{\ast }F^{\mu \nu }\right) \mathcal{\tilde{G}}%
-4\cdot 5!\eta _{\mu \nu \rho }^{\ast }\tilde{D}_{\lambda \sigma }\mathcal{G}%
^{\mu \nu \rho \lambda \sigma }  \notag \\
&&+4!\cdot 4!t^{\ast \mu }\eta \mathcal{\tilde{G}}_{\mu }-4!\cdot 4!B_{\mu
\nu }^{\ast }\tilde{D}_{\rho \lambda }\mathcal{G}^{\mu \nu \rho \lambda
}+6\cdot 4!\left( \phi ^{\ast \mu \nu }\eta \right.  \notag \\
&&\left. -K^{\mu \nu \rho }V_{\rho }\right) \tilde{D}_{\mu \nu }-3\cdot
4!\left( \tilde{K}_{\mu \nu }\eta -4V_{[\mu }\mathcal{\tilde{G}}_{\nu
]}\right) F^{\mu \nu },  \label{xxt5}
\end{eqnarray}%
\begin{eqnarray}
\bar{X}_{1}^{\left( 2\right) } &=&-4\cdot 5!\left( C_{\mu \nu \rho }^{\ast
}\eta +C_{[\mu \nu }^{\ast }V_{\rho ]}+2H_{[\mu }^{\ast }B_{\nu \rho
]}^{\ast }\right) \tilde{D}_{\lambda \sigma }\mathcal{G}^{\mu \nu \rho
\lambda \sigma }  \notag \\
&&-12\cdot 5!\left( C_{\mu \nu }^{\ast }F^{\mu \nu }\eta -2H_{\mu }^{\ast
}t^{\ast \mu }\eta +H_{[\mu }^{\ast }V_{\nu ]}F^{\mu \nu }\right) \tilde{%
\mathcal{G}}-12\cdot 4!H_{[\mu }^{\ast }\mathcal{\tilde{G}}_{\nu ]}\eta
F^{\mu \nu }  \notag \\
&&-12\cdot 4!\left( C_{\mu \nu }^{\ast }\eta +H_{[\mu }^{\ast }V_{\nu
]}\right) \tilde{D}_{\rho \lambda }\mathcal{G}^{\mu \nu \rho \lambda
}-6\cdot 4!H_{\mu }^{\ast }K^{\mu \nu \rho }\eta \tilde{D}_{\nu \rho },
\label{xxt6}
\end{eqnarray}%
\begin{eqnarray}
\bar{X}_{2}^{\left( 2\right) } &=&-4\cdot 5!\left( H_{[\mu }^{\ast }C_{\nu
\rho ]}^{\ast }\eta +H_{[\mu }^{\ast }H_{\nu }^{\ast }V_{\rho ]}\right)
\tilde{D}_{\lambda \sigma }\mathcal{G}^{\mu \nu \rho \lambda \sigma }  \notag
\\
&&-12\cdot 4!H_{\mu }^{\ast }H_{\nu }^{\ast }\eta \tilde{D}_{\rho \lambda }%
\mathcal{G}^{\mu \nu \rho \lambda }-12\cdot 5!H_{\mu }^{\ast }H_{\nu }^{\ast
}\eta F^{\mu \nu }\mathcal{\tilde{G}},  \label{xxt7}
\end{eqnarray}%
\begin{equation}
\bar{X}_{3}^{\left( 2\right) }=-4\cdot 5!H_{\mu }^{\ast }H_{\nu }^{\ast
}H_{\rho }^{\ast }\eta \tilde{D}_{\lambda \sigma }\mathcal{G}^{\mu \nu \rho
\lambda \sigma },  \label{xxt8}
\end{equation}%
\begin{equation}
\bar{X}_{0}^{\left( 3\right) }=-6\cdot 5!S^{\ast }\tilde{\eta}+12\cdot
4!t_{\mu }^{\ast }\tilde{\eta}^{\mu }+4!B^{\mu \nu }\tilde{D}_{\mu \nu }-36%
\tilde{\eta}_{\mu \nu }F^{\mu \nu },  \label{xxt9}
\end{equation}%
\begin{eqnarray}
\bar{X}_{1}^{\left( 3\right) } &=&2\cdot 5!C_{\mu \nu \rho }^{\ast }\tilde{D}%
_{\lambda \sigma }\eta ^{\mu \nu \rho \lambda \sigma }+6\cdot 5!\left(
2H_{\mu }^{\ast }t^{\ast \mu }-C_{\mu \nu }^{\ast }F^{\mu \nu }\right)
\tilde{\eta}  \notag \\
&&+6\cdot 4!C_{\mu \nu }^{\ast }\tilde{D}_{\rho \lambda }\eta ^{\mu \nu \rho
\lambda }+3\cdot 4!H_{\mu }^{\ast }\tilde{D}_{\nu \rho }\eta ^{\mu \nu \rho
}+6\cdot 4!H_{[\mu }^{\ast }\tilde{\eta}_{\nu ]}F^{\mu \nu },  \label{xxt10}
\end{eqnarray}%
\begin{equation}
\bar{X}_{2}^{\left( 3\right) }=2\cdot 5!H_{[\mu }^{\ast }C_{\nu \rho
]}^{\ast }\tilde{D}_{\lambda \sigma }\eta ^{\mu \nu \rho \lambda \sigma
}+6\cdot 4!H_{\mu }^{\ast }H_{\nu }^{\ast }\left( \tilde{D}_{\rho \lambda
}\eta ^{\mu \nu \rho \lambda }-5F^{\mu \nu }\tilde{\eta}\right) ,
\label{xxt11}
\end{equation}%
\begin{equation}
\bar{X}_{3}^{\left( 3\right) }=2\cdot 5!H_{\mu }^{\ast }H_{\nu }^{\ast
}H_{\rho }^{\ast }\tilde{D}_{\lambda \sigma }\eta ^{\mu \nu \rho \lambda
\sigma }.  \label{xxt25}
\end{equation}

The functions appearing in (\ref{omega}) and denoted by $U_{p}^{\left(
i\right) }$ are of the form
\begin{equation}
U_{0}^{\left( 1\right) }=-9\left( 2k_{1}\phi ^{\mu \nu }-\tfrac{k_{2}}{10}%
\tilde{K}^{\mu \nu }\right) \left( 2B_{\mu \nu }^{\ast }C+V_{[\mu }C_{\nu
]}-\phi _{\mu \nu }\eta \right) ,  \label{u01}
\end{equation}%
\begin{equation}
U_{1}^{\left( 1\right) }=-9\left( k_{1}\phi ^{\mu \nu }-\tfrac{k_{2}}{20}%
\tilde{K}^{\mu \nu }\right) \left[ C_{\mu \nu }^{\ast }\eta C-H_{\mu }^{\ast
}\left( V_{\nu }C+\eta C_{\nu }\right) \right] ,  \label{u11}
\end{equation}%
\begin{equation}
U_{2}^{\left( 1\right) }=-9\left( k_{1}\phi ^{\mu \nu }-\tfrac{k_{2}}{20}%
\tilde{K}^{\mu \nu }\right) H_{\mu }^{\ast }H_{\nu }^{\ast }\eta C,
\label{u21}
\end{equation}

\begin{equation}
U_{0}^{\left( 2\right) }=108\left( k_{1}\phi ^{\mu \nu }-\tfrac{k_{2}}{20}%
\tilde{K}^{\mu \nu }\right) \left( 40B_{\mu \nu }^{\ast }\mathcal{\tilde{G}}%
+\eta \tilde{K}_{\mu \nu }-8V_{\mu }\mathcal{\tilde{G}}_{\nu }\right) ,
\label{u02}
\end{equation}%
\begin{eqnarray}
U_{1}^{\left( 2\right) } &=&18\varepsilon ^{\alpha \beta \gamma \delta
\varepsilon }\left( C_{\mu \nu }^{\ast }\eta +H_{[\mu }^{\ast }V_{\nu
]}\right) \left( k_{1}\phi ^{\mu \nu }-\tfrac{k_{2}}{20}\tilde{K}^{\mu \nu
}\right) \mathcal{G}_{\alpha \beta \gamma \delta \varepsilon }  \notag \\
&&-36\varepsilon _{\rho \alpha \beta \gamma \delta }H_{\mu }^{\ast }\left(
k_{1}\phi ^{\mu \rho }-\tfrac{k_{2}}{20}\tilde{K}^{\mu \rho }\right) \eta
\mathcal{G}^{\alpha \beta \gamma \delta },  \label{u12}
\end{eqnarray}%
\begin{equation}
U_{2}^{\left( 2\right) }=18\varepsilon _{\alpha \beta \gamma \delta
\varepsilon }\left( k_{1}\phi ^{\mu \nu }-\tfrac{k_{2}}{20}\tilde{K}^{\mu
\nu }\right) H_{\mu }^{\ast }H_{\nu }^{\ast }\eta \mathcal{G}^{\alpha \beta
\gamma \delta \varepsilon },  \label{u22}
\end{equation}

\begin{equation}
U_{0}^{\left( 3\right) }=9\varepsilon _{\nu \rho \alpha \beta \gamma }\left(
k_{1}\phi ^{\nu \rho }-\tfrac{k_{2}}{20}\tilde{K}^{\nu \rho }\right) \eta
^{\alpha \beta \gamma },  \label{u03}
\end{equation}%
\begin{eqnarray}
U_{1}^{\left( 3\right) } &=&\tfrac{9}{4}\varepsilon _{\alpha \beta \gamma
\delta \varepsilon }C_{\mu \nu }^{\ast }\left( k_{1}\phi ^{\mu \nu }-\tfrac{%
k_{2}}{20}\tilde{K}^{\mu \nu }\right) \eta ^{\alpha \beta \gamma \delta
\varepsilon }  \notag \\
&&-18\varepsilon _{\rho \beta \gamma \delta \varepsilon }H_{\mu }^{\ast
}\left( k_{1}\phi ^{\mu \rho }-\tfrac{k_{2}}{20}\tilde{K}^{\mu \rho }\right)
\eta ^{\beta \gamma \delta \varepsilon },  \label{u13}
\end{eqnarray}%
\begin{equation}
U_{2}^{\left( 3\right) }=9\varepsilon _{\alpha \beta \gamma \delta
\varepsilon }H_{\mu }^{\ast }H_{\nu }^{\ast }\left( k_{1}\phi ^{\mu \nu }-%
\tfrac{k_{2}}{20}\tilde{K}^{\mu \nu }\right) \eta ^{\alpha \beta \gamma
\delta \varepsilon }.  \label{u24}
\end{equation}

\section{Deformed gauge structure\label{gauge}}

If we denote by $\Omega _{1}^{\alpha _{1}}$ and $\Omega _{2}^{\alpha _{1}}$
two independent sets of gauge parameters,%
\begin{eqnarray}
\Omega _{1}^{\alpha _{1}} &\equiv &\left( \epsilon ^{(1)\mu \nu },\epsilon
^{(1)},\epsilon ^{(1)\mu \nu \rho },\xi _{\mu }^{(1)},\xi ^{(1)\mu \nu \rho
\lambda },\theta _{\mu \nu }^{(1)},\chi _{\mu \nu }^{(1)}\right) ,
\label{g1} \\
\Omega _{2}^{\alpha _{1}} &\equiv &\left( \epsilon ^{(2)\mu \nu },\epsilon
^{(2)},\epsilon ^{(2)\mu \nu \rho },\xi _{\mu }^{(2)},\xi ^{(2)\mu \nu \rho
\lambda },\theta _{\mu \nu }^{(2)},\chi _{\mu \nu }^{(2)}\right) ,
\label{g2}
\end{eqnarray}%
then the concrete form of the commutators among the deformed gauge
transformations of the fields associated with (\ref{g1}) and (\ref{g2}) (and
generically written as in (\ref{defalgebra})) read as
\begin{equation}
\left[ \bar{\delta}_{\Omega _{1}},\bar{\delta}_{\Omega _{2}}\right] \varphi
=0,  \label{com1}
\end{equation}%
\begin{eqnarray}
\left[ \bar{\delta}_{\Omega _{1}},\bar{\delta}_{\Omega _{2}}\right] H^{\mu }
&=&\bar{\delta}_{\Omega }H^{\mu }-2\frac{\delta S^{\mathrm{L}}}{\delta
H^{\nu }}\frac{d\epsilon ^{\mu \nu }}{d\varphi }-3\frac{\delta S^{\mathrm{L}}%
}{\delta B^{\nu \rho }}\frac{d\epsilon ^{\mu \nu \rho }}{d\varphi }  \notag
\\
&&+2\frac{\delta S^{\mathrm{L}}}{\delta \phi _{\mu \nu }}\frac{d\xi _{\nu }}{%
d\varphi }-4\frac{\delta S^{\mathrm{L}}}{\delta K^{\nu \rho \lambda }}\frac{%
d\xi ^{\mu \nu \rho \lambda }}{d\varphi },  \label{com2}
\end{eqnarray}%
\begin{equation}
\left[ \bar{\delta}_{\Omega _{1}},\bar{\delta}_{\Omega _{2}}\right] V_{\mu }=%
\bar{\delta}_{\Omega }V_{\mu },  \label{com3}
\end{equation}%
\begin{equation}
\left[ \bar{\delta}_{\Omega _{1}},\bar{\delta}_{\Omega _{2}}\right] B^{\mu
\nu }=\bar{\delta}_{\Omega }B^{\mu \nu }+3\frac{\delta S^{\mathrm{L}}}{%
\delta H^{\rho }}\frac{d\epsilon ^{\mu \nu \rho }}{d\varphi },  \label{com4}
\end{equation}%
\begin{equation}
\left[ \bar{\delta}_{\Omega _{1}},\bar{\delta}_{\Omega _{2}}\right] \phi
_{\mu \nu }=\bar{\delta}_{\Omega }\phi _{\mu \nu }-\frac{\delta S^{\mathrm{L}%
}}{\delta H^{[\mu }}\frac{d\xi _{\nu ]}}{d\varphi },  \label{comm5}
\end{equation}%
\begin{equation}
\left[ \bar{\delta}_{\Omega _{1}},\bar{\delta}_{\Omega _{2}}\right] K^{\mu
\nu \rho }=\bar{\delta}_{\Omega }K^{\mu \nu \rho }-4\frac{\delta S^{\mathrm{L%
}}}{\delta H^{\lambda }}\frac{d\xi ^{\mu \nu \rho \lambda }}{d\varphi },
\label{comm6}
\end{equation}%
\begin{equation}
\left[ \bar{\delta}_{\Omega _{1}},\bar{\delta}_{\Omega _{2}}\right] t_{\mu
\nu |\alpha }=0.  \label{com7}
\end{equation}%
The gauge parameters from the right-hand side of the above formulas are
defined through
\begin{equation}
\Omega ^{\alpha _{1}}=\left( \epsilon ^{\mu \nu },\epsilon =0,\epsilon ^{\mu
\nu \rho },\xi _{\mu },\xi ^{\mu \nu \rho \lambda },\theta _{\mu \nu
}=0,\chi _{\mu \nu }=0\right) ,  \label{parcom}
\end{equation}%
where%
\begin{eqnarray}
\epsilon ^{\mu \nu } &=&\lambda \left\{ -\frac{dW_{1}}{d\varphi }\left(
\epsilon ^{(1)}\epsilon ^{(2)\mu \nu }-\epsilon ^{(2)}\epsilon ^{(1)\mu \nu
}\right) \right.  \notag \\
&&+6\frac{dW_{3}}{d\varphi }\left[ \phi _{\rho \lambda }\left( \epsilon
^{(1)}\xi ^{(2)\mu \nu \rho \lambda }-\epsilon ^{(2)}\xi ^{(1)\mu \nu \rho
\lambda }\right) \right.  \notag \\
&&\left. +\tfrac{1}{2}K^{\mu \nu \rho }\left( \epsilon ^{(1)}\xi _{\rho
}^{(2)}-\epsilon ^{(2)}\xi _{\rho }^{(1)}\right) -2V_{\rho }\left( \xi
_{\lambda }^{(1)}\xi ^{(2)\mu \nu \rho \lambda }-\xi _{\lambda }^{(2)}\xi
^{(1)\mu \nu \rho \lambda }\right) \right]  \notag \\
&&-3\frac{dW_{2}}{d\varphi }\left( \xi _{\rho }^{(1)}\epsilon ^{(2)\mu \nu
\rho }-\xi _{\rho }^{(2)}\epsilon ^{(1)\mu \nu \rho }\right)  \notag \\
&&+3\frac{dW_{6}}{d\varphi }\varepsilon _{\rho \alpha \beta \gamma \delta
}\left( \epsilon ^{(1)\mu \nu \rho }\xi ^{(2)\alpha \beta \gamma \delta
}-\epsilon ^{(2)\mu \nu \rho }\xi ^{(1)\alpha \beta \gamma \delta }\right)
\notag \\
&&+6\frac{dW_{4}}{d\varphi }\left[ \varepsilon _{\rho \alpha \beta \gamma
\delta }K^{\mu \nu \rho }\left( \epsilon ^{(1)}\xi ^{(2)\alpha \beta \gamma
\delta }-\epsilon ^{(2)}\xi ^{(1)\alpha \beta \gamma \delta }\right) \right.
\notag \\
&&\left. +\tfrac{1}{6}\varepsilon ^{\mu \nu \rho \lambda \sigma }\varepsilon
_{\lambda \alpha \beta \gamma \delta }\varepsilon _{\sigma \alpha ^{\prime
}\beta ^{\prime }\gamma ^{\prime }\delta ^{\prime }}V_{\rho }\xi ^{(1)\alpha
\beta \gamma \delta }\xi ^{(2)\alpha ^{\prime }\beta ^{\prime }\gamma
^{\prime }\delta ^{\prime }}\right]  \notag \\
&&\left. -\tfrac{1}{2}\varepsilon ^{\mu \nu \rho \lambda \sigma }\frac{dW_{5}%
}{d\varphi }\left[ \phi _{\rho \lambda }\left( \epsilon ^{(1)}\xi _{\sigma
}^{(2)}-\epsilon ^{(2)}\xi _{\sigma }^{(1)}\right) -2V_{\rho }\xi _{\lambda
}^{(1)}\xi _{\sigma }^{(2)}\right] \right\} ,  \label{defpar1}
\end{eqnarray}%
\begin{eqnarray}
\epsilon ^{\mu \nu \rho } &=&-8\lambda \left[ W_{3}\left( \xi _{\lambda
}^{(1)}\xi ^{(2)\mu \nu \rho \lambda }-\xi _{\lambda }^{(2)}\xi ^{(1)\mu \nu
\rho \lambda }\right) \right.  \notag \\
&&-\tfrac{1}{12}\varepsilon ^{\mu \nu \rho \lambda \sigma }\left(
W_{4}\varepsilon _{\lambda \alpha \beta \gamma \delta }\varepsilon _{\sigma
\alpha ^{\prime }\beta ^{\prime }\gamma ^{\prime }\delta ^{\prime }}\xi
^{(1)\alpha \beta \gamma \delta }\xi ^{(2)\alpha ^{\prime }\beta ^{\prime
}\gamma ^{\prime }\delta ^{\prime }}\right.  \notag \\
&&\left. \left. +W_{5}\xi _{\lambda }^{(1)}\xi _{\sigma }^{(2)}\right)
\right] ,  \label{defpar2}
\end{eqnarray}%
\begin{eqnarray}
\xi _{\mu } &=&-3\lambda \left[ W_{3}\left( \epsilon ^{(1)}\xi _{\mu
}^{(2)}-\epsilon ^{(2)}\xi _{\mu }^{(1)}\right) \right.  \notag \\
&&\left. +2W_{4}\varepsilon _{\mu \nu \rho \lambda \sigma }\left( \epsilon
^{(1)}\xi ^{(2)\nu \rho \lambda \sigma }-\epsilon ^{(2)}\xi ^{(1)\nu \rho
\lambda \sigma }\right) \right] ,  \label{defpar3}
\end{eqnarray}%
\begin{eqnarray}
\xi ^{\mu \nu \rho \lambda } &=&3\lambda \left[ W_{3}\left( \epsilon
^{(1)}\xi ^{(2)\mu \nu \rho \lambda }-\epsilon ^{(2)}\xi ^{(1)\mu \nu \rho
\lambda }\right) \right.  \notag \\
&&\left. -\tfrac{1}{12}W_{4}\varepsilon ^{\mu \nu \rho \lambda \sigma
}\left( \epsilon ^{(1)}\xi _{\sigma }^{(2)}-\epsilon ^{(2)}\xi _{\sigma
}^{(1)}\right) \right] .  \label{defpar4}
\end{eqnarray}%
In addition, we made the notations%
\begin{equation}
\theta ^{(i)}=\sigma _{\alpha \beta }\theta ^{(i)\alpha \beta },\qquad i=%
\overline{1,2}.  \label{parcom'}
\end{equation}

Related to the first-order reducibility, the transformations (\ref{red1par})
are given by
\begin{eqnarray}
\epsilon ^{\mu \nu }\left( \bar{\Omega}\right) &=&-3D_{\rho }\bar{\epsilon}%
^{\mu \nu \rho }-\lambda \frac{dW_{2}}{d\varphi }\left( B^{\mu \nu }\bar{\xi}%
-6\phi _{\rho \lambda }\bar{\epsilon}^{\mu \nu \rho \lambda }\right)  \notag
\\
&&+3\lambda \frac{dW_{3}}{d\varphi }V_{\rho }\left( K^{\mu \nu \rho }\bar{\xi%
}-10\phi _{\lambda \sigma }\bar{\xi}^{\mu \nu \rho \lambda \sigma }\right)
\notag \\
&&-6\lambda \varepsilon _{\alpha \beta \gamma \delta \varepsilon }\frac{%
dW_{4}}{d\varphi }K^{\mu \nu \rho }V_{\rho }\bar{\xi}^{\alpha \beta \gamma
\delta \varepsilon }-\tfrac{\lambda }{2}\varepsilon ^{\mu \nu \rho \lambda
\sigma }\frac{dW_{5}}{d\varphi }V_{\rho }\phi _{\lambda \sigma }\bar{\xi}
\notag \\
&&+\lambda \frac{dW_{6}}{d\varphi }\left( \varepsilon _{\alpha \beta \gamma
\delta \varepsilon }B^{\mu \nu }\bar{\xi}^{\alpha \beta \gamma \delta
\varepsilon }+3K^{\mu \nu \rho }\varepsilon _{\rho \alpha \beta \gamma
\delta }\bar{\epsilon}^{\alpha \beta \gamma \delta }\right) ,  \label{tpar2}
\end{eqnarray}%
\begin{equation}
\epsilon \left( \bar{\Omega}\right) =2\lambda \left( W_{2}\bar{\xi}%
-\varepsilon _{\alpha \beta \gamma \delta \varepsilon }W_{6}\bar{\xi}%
^{\alpha \beta \gamma \delta \varepsilon }\right) ,  \label{tpar1}
\end{equation}%
\begin{eqnarray}
\epsilon ^{\mu \nu \rho }\left( \bar{\Omega}\right) &=&4\partial _{\lambda }%
\bar{\epsilon}^{\mu \nu \rho \lambda }+2\lambda W_{1}\bar{\epsilon}^{\mu \nu
\rho }-20\lambda W_{3}\phi _{\lambda \sigma }\bar{\xi}^{\mu \nu \rho \lambda
\sigma }  \notag \\
&&+2\lambda K^{\mu \nu \rho }\left( W_{3}\bar{\xi}-2\varepsilon _{\alpha
\beta \gamma \delta \varepsilon }W_{4}\bar{\xi}^{\alpha \beta \gamma \delta
\varepsilon }\right)  \notag \\
&&-\tfrac{\lambda }{3}\varepsilon ^{\mu \nu \rho \lambda \sigma }W_{5}\phi
_{\lambda \sigma }\bar{\xi},  \label{tpar3}
\end{eqnarray}%
\begin{equation}
\xi _{\mu }\left( \bar{\Omega}\right) =D_{\mu }^{(-)}\bar{\xi}+6\lambda
\varepsilon _{\alpha \beta \gamma \delta \varepsilon }W_{4}V_{\mu }\bar{\xi}%
^{\alpha \beta \gamma \delta \varepsilon }-3\lambda \varepsilon _{\mu \nu
\rho \lambda \sigma }W_{6}\bar{\epsilon}^{\nu \rho \lambda \sigma },
\label{tpar4}
\end{equation}%
\begin{equation}
\xi ^{\mu \nu \rho \lambda }\left( \bar{\Omega}\right) =-5D_{\sigma }^{(+)}%
\bar{\xi}^{\mu \nu \rho \lambda \sigma }+3\lambda W_{2}\bar{\epsilon}^{\mu
\nu \rho \lambda }-\tfrac{\lambda }{4}\varepsilon ^{\mu \nu \rho \lambda
\sigma }W_{5}V_{\sigma }\bar{\xi},  \label{tpar5}
\end{equation}%
\begin{equation}
\theta _{\mu \nu }\left( \bar{\Omega}\right) =3\partial _{(\mu }\bar{\theta}%
_{\nu )}+\lambda \sigma _{\mu \nu }\left( k_{1}\bar{\xi}+\tfrac{k_{2}}{5!}%
\varepsilon _{\alpha \beta \gamma \delta \varepsilon }\bar{\xi}^{\alpha
\beta \gamma \delta \varepsilon }\right) ,  \label{tpar6}
\end{equation}%
\begin{equation}
\chi _{\mu \nu }\left( \bar{\Omega}\right) =\partial _{\lbrack \mu }\bar{%
\theta}_{\nu ]},  \label{tpar7}
\end{equation}%
while the first-order reducibility relations (\ref{red1def}) read as%
\begin{equation}
\bar{\delta}_{\Omega \left( \bar{\Omega}\right) }\varphi =0,  \label{r1def1}
\end{equation}%
\begin{eqnarray}
\bar{\delta}_{\Omega \left( \bar{\Omega}\right) }H^{\mu } &=&\lambda \frac{%
\delta S^{\mathrm{L}}}{\delta H^{\nu }}\left\{ 6V_{\rho }\left[ \frac{%
d^{2}W_{1}}{d\varphi ^{2}}\bar{\epsilon}^{\mu \nu \rho }-\frac{d^{2}W_{3}}{%
d\varphi ^{2}}\left( 10\phi _{\lambda \sigma }\bar{\xi}^{\mu \nu \rho
\lambda \sigma }-K^{\mu \nu \rho }\bar{\xi}\right) \right. \right.  \notag \\
&&\left. -2\varepsilon _{\alpha \beta \gamma \delta \varepsilon }\frac{%
d^{2}W_{4}}{d\varphi ^{2}}K^{\mu \nu \rho }\bar{\xi}^{\alpha \beta \gamma
\delta \varepsilon }-\tfrac{1}{6}\varepsilon ^{\mu \nu \rho \lambda \sigma }%
\frac{d^{2}W_{5}}{d\varphi ^{2}}\phi _{\lambda \sigma }\bar{\xi}\right]
\notag \\
&&+2\frac{d^{2}W_{6}}{d\varphi ^{2}}\left( 3\varepsilon _{\rho \alpha \beta
\gamma \delta }K^{\mu \nu \rho }\bar{\epsilon}^{\alpha \beta \gamma \delta
}+\varepsilon _{\alpha \beta \gamma \delta \varepsilon }B^{\mu \nu }\bar{\xi}%
^{\alpha \beta \gamma \delta \varepsilon }\right)  \notag \\
&&\left. +2\frac{d^{2}W_{2}}{d\varphi ^{2}}\left( 6\phi _{\rho \lambda }\bar{%
\epsilon}^{\mu \nu \rho \lambda }-B^{\mu \nu }\bar{\xi}\right) \right\}
\notag \\
&&+6\lambda \frac{\delta S^{\mathrm{L}}}{\delta B^{\nu \rho }}\left[ \frac{%
dW_{1}}{d\varphi }\bar{\epsilon}^{\mu \nu \rho }-\frac{dW_{3}}{d\varphi }%
\left( 10\phi _{\lambda \sigma }\bar{\xi}^{\mu \nu \rho \lambda \sigma
}-K^{\mu \nu \rho }\bar{\xi}\right) \right.  \notag \\
&&\left. -2\varepsilon _{\alpha \beta \gamma \delta \varepsilon }\frac{dW_{4}%
}{d\varphi }K^{\mu \nu \rho }\bar{\xi}^{\alpha \beta \gamma \delta
\varepsilon }-\tfrac{1}{6}\varepsilon ^{\mu \nu \rho \lambda \sigma }\frac{%
dW_{5}}{d\varphi }\phi _{\lambda \sigma }\bar{\xi}\right]  \notag \\
&&-2\lambda \frac{\delta S^{\mathrm{L}}}{\delta V_{\mu }}\left( \frac{dW_{2}%
}{d\varphi }\bar{\xi}-\varepsilon _{\alpha \beta \gamma \delta \varepsilon }%
\frac{dW_{6}}{d\varphi }\bar{\xi}^{\alpha \beta \gamma \delta \varepsilon
}\right)  \notag \\
&&+\lambda \frac{\delta S^{\mathrm{L}}}{\delta K^{\nu \rho \lambda }}\left[
-V_{\sigma }\left( 60\frac{dW_{3}}{d\varphi }\bar{\xi}^{\mu \nu \rho \lambda
\sigma }+\varepsilon ^{\mu \nu \rho \lambda \sigma }\frac{dW_{5}}{d\varphi }%
\bar{\xi}\right) \right.  \notag \\
&&\left. +12\frac{dW_{2}}{d\varphi }\bar{\epsilon}^{\mu \nu \rho \lambda }%
\right] ++6\lambda \frac{\delta S^{\mathrm{L}}}{\delta \phi _{\mu \nu }}%
\left[ \varepsilon _{\nu \alpha \beta \gamma \delta }\frac{dW_{6}}{d\varphi }%
\bar{\epsilon}^{\alpha \beta \gamma \delta }\right.  \notag \\
&&\left. +V_{\nu }\left( \frac{dW_{3}}{d\varphi }\bar{\xi}-2\varepsilon
_{\alpha \beta \gamma \delta \varepsilon }\frac{dW_{4}}{d\varphi }\bar{\xi}%
^{\alpha \beta \gamma \delta \varepsilon }\right) \right] ,  \label{r1def2}
\end{eqnarray}%
\begin{equation}
\bar{\delta}_{\Omega \left( \bar{\Omega}\right) }V_{\mu }=2\lambda \frac{%
\delta S^{\mathrm{L}}}{\delta H^{\mu }}\left( \frac{dW_{2}}{d\varphi }\bar{%
\xi}-\varepsilon _{\alpha \beta \gamma \delta \varepsilon }\frac{dW_{6}}{%
d\varphi }\bar{\xi}^{\alpha \beta \gamma \delta \varepsilon }\right) ,
\label{r1def3}
\end{equation}%
\begin{eqnarray}
\bar{\delta}_{\Omega \left( \bar{\Omega}\right) }B^{\mu \nu } &=&6\lambda
\frac{\delta S^{\mathrm{L}}}{\delta H^{\rho }}\left[ -\frac{dW_{1}}{d\varphi
}\bar{\epsilon}^{\mu \nu \rho }+10\frac{dW_{3}}{d\varphi }\phi _{\lambda
\sigma }\bar{\xi}^{\mu \nu \rho \lambda \sigma }\right.  \notag \\
&&\left. -K^{\mu \nu \rho }\left( \frac{dW_{3}}{d\varphi }\bar{\xi}-2\frac{%
dW_{4}}{d\varphi }\varepsilon _{\alpha \beta \gamma \delta \varepsilon }\bar{%
\xi}^{\alpha \beta \gamma \delta \varepsilon }\right) +\tfrac{1}{6}%
\varepsilon ^{\mu \nu \rho \lambda \sigma }\frac{dW_{5}}{d\varphi }\phi
_{\lambda \sigma }\bar{\xi}\right]  \notag \\
&&+\lambda \frac{\delta S^{\mathrm{L}}}{\delta K^{\rho \lambda \sigma }}%
\left( 60W_{3}\bar{\xi}^{\mu \nu \rho \lambda \sigma }+\varepsilon ^{\mu \nu
\rho \lambda \sigma }W_{5}\bar{\xi}\right)  \notag \\
&&+6\lambda \frac{\delta S^{\mathrm{L}}}{\delta \phi _{\mu \nu }}\left( W_{3}%
\bar{\xi}-2\varepsilon _{\alpha \beta \gamma \delta \varepsilon }W_{4}\bar{%
\xi}^{\alpha \beta \gamma \delta \varepsilon }\right) ,  \label{r1def4}
\end{eqnarray}%
\begin{eqnarray}
\bar{\delta}_{\Omega \left( \bar{\Omega}\right) }\phi _{\mu \nu }
&=&-3\lambda \frac{\delta S^{\mathrm{L}}}{\delta H^{[\mu }}V_{\nu ]}\left(
\frac{dW_{3}}{d\varphi }\bar{\xi}-2\varepsilon _{\alpha \beta \gamma \delta
\varepsilon }\frac{dW_{4}}{d\varphi }\bar{\xi}^{\alpha \beta \gamma \delta
\varepsilon }\right)  \notag \\
&&-6\lambda \frac{\delta S^{\mathrm{L}}}{\delta B^{\mu \nu }}\left( W_{3}%
\bar{\xi}-2\varepsilon _{\alpha \beta \gamma \delta \varepsilon }W_{4}\bar{%
\xi}^{\alpha \beta \gamma \delta \varepsilon }\right)  \notag \\
&&-3\lambda \frac{dW_{6}}{d\varphi }\frac{\delta S^{\mathrm{L}}}{\delta
H^{[\mu }}\varepsilon _{\nu ]\alpha \beta \gamma \delta }\bar{\epsilon}%
^{\alpha \beta \gamma \delta },  \label{r1def5}
\end{eqnarray}%
\begin{eqnarray}
\bar{\delta}_{\Omega \left( \bar{\Omega}\right) }K^{\mu \nu \rho }
&=&\lambda \frac{\delta S^{\mathrm{L}}}{\delta H^{\lambda }}\left[
-V_{\sigma }\left( 60\frac{dW_{3}}{d\varphi }\bar{\xi}^{\mu \nu \rho \lambda
\sigma }+\varepsilon ^{\mu \nu \rho \lambda \sigma }\frac{dW_{5}}{d\varphi }%
\bar{\xi}\right) \right.  \notag \\
&&\left. +12\frac{dW_{2}}{d\varphi }\bar{\epsilon}^{\mu \nu \rho \lambda }%
\right] -\lambda \frac{\delta S^{\mathrm{L}}}{\delta B^{\lambda \sigma }}%
\left( 60W_{3}\bar{\xi}^{\mu \nu \rho \lambda \sigma }+\varepsilon ^{\mu \nu
\rho \lambda \sigma }W_{5}\bar{\xi}\right) ,  \label{r1def6}
\end{eqnarray}%
\begin{equation}
\bar{\delta}_{\Omega \left( \bar{\Omega}\right) }t_{\mu \nu |\alpha }=0.
\label{r1def7}
\end{equation}

Regarding the second-order reducibility, the transformations (\ref{red2par})
take the concrete form
\begin{gather}
\bar{\epsilon}^{\mu \nu \rho }\left( \check{\Omega}\right) =4D_{\lambda }%
\check{\epsilon}^{\mu \nu \rho \lambda }-\lambda \left( 10\frac{dW_{2}}{%
d\varphi }\phi _{\lambda \sigma }\check{\epsilon}^{\mu \nu \rho \lambda
\sigma }+\varepsilon _{\alpha \beta \gamma \delta \varepsilon }\frac{dW_{6}}{%
d\varphi }K^{\mu \nu \rho }\check{\epsilon}^{\alpha \beta \gamma \delta
\varepsilon }\right) ,  \label{tpar1r1} \\
\bar{\epsilon}^{\mu \nu \rho \lambda }\left( \check{\Omega}\right)
=-5\partial _{\sigma }\check{\epsilon}^{\mu \nu \rho \lambda \sigma
}-2\lambda W_{1}\check{\epsilon}^{\mu \nu \rho \lambda },\qquad \bar{\xi}%
\left( \check{\Omega}\right) =-3\lambda \varepsilon _{\alpha \beta \gamma
\delta \varepsilon }W_{6}\check{\epsilon}^{\alpha \beta \gamma \delta
\varepsilon },  \label{tpar1r2} \\
\bar{\xi}^{\mu \nu \rho \lambda \sigma }\left( \check{\Omega}\right)
=-3\lambda W_{2}\check{\epsilon}^{\mu \nu \rho \lambda \sigma },\qquad \bar{%
\theta}_{\mu }\left( \check{\Omega}\right) =0,  \label{tpar1r3}
\end{gather}%
such that the second-order reducibility relations (\ref{red2def}) become
\begin{eqnarray}
\epsilon ^{\mu \nu }\left( \bar{\Omega}\left( \check{\Omega}\right) \right)
&=&3\lambda \frac{\delta S^{\mathrm{L}}}{\delta H^{\rho }}\left( 4\frac{%
d^{2}W_{1}}{d\varphi ^{2}}V_{\lambda }\check{\epsilon}^{\mu \nu \rho \lambda
}+10\frac{d^{2}W_{2}}{d\varphi ^{2}}\phi _{\lambda \sigma }\check{\epsilon}%
^{\mu \nu \rho \lambda \sigma }\right.  \notag \\
&&\left. +\varepsilon _{\alpha \beta \gamma \delta \varepsilon }\frac{%
d^{2}W_{6}}{d\varphi ^{2}}K^{\mu \nu \rho }\check{\epsilon}^{\alpha \beta
\gamma \delta \varepsilon }\right) +12\lambda \frac{dW_{1}}{d\varphi }\frac{%
\delta S^{\mathrm{L}}}{\delta B^{\rho \lambda }}\check{\epsilon}^{\mu \nu
\rho \lambda }  \notag \\
&&+30\lambda \frac{dW_{2}}{d\varphi }\frac{\delta S^{\mathrm{L}}}{\delta
K^{\rho \lambda \sigma }}\check{\epsilon}^{\mu \nu \rho \lambda \sigma
}-3\lambda \varepsilon _{\alpha \beta \gamma \delta \varepsilon }\frac{dW_{6}%
}{d\varphi }\frac{\delta S^{\mathrm{L}}}{\delta \phi _{\mu \nu }}\check{%
\epsilon}^{\alpha \beta \gamma \delta \varepsilon },  \label{r2def2}
\end{eqnarray}%
\begin{equation}
\epsilon \left( \bar{\Omega}\left( \check{\Omega}\right) \right) =0,
\label{r2def1}
\end{equation}%
\begin{equation}
\epsilon ^{\mu \nu \rho }\left( \bar{\Omega}\left( \check{\Omega}\right)
\right) =-8\lambda \frac{dW_{1}}{d\varphi }\frac{\delta S^{\mathrm{L}}}{%
\delta H^{\lambda }}\check{\epsilon}^{\mu \nu \rho \lambda },  \label{r2def3}
\end{equation}%
\begin{equation}
\xi _{\mu }\left( \bar{\Omega}\left( \check{\Omega}\right) \right)
=-3\lambda \varepsilon _{\alpha \beta \gamma \delta \varepsilon }\frac{dW_{6}%
}{d\varphi }\frac{\delta S^{\mathrm{L}}}{\delta H^{\mu }}\check{\epsilon}%
^{\alpha \beta \gamma \delta \varepsilon },  \label{r2def4}
\end{equation}%
\begin{equation}
\xi ^{\mu \nu \rho \lambda }\left( \bar{\Omega}\left( \check{\Omega}\right)
\right) =15\lambda \frac{dW_{2}}{d\varphi }\frac{\delta S^{\mathrm{L}}}{%
\delta H^{\sigma }}\check{\epsilon}^{\mu \nu \rho \lambda \sigma },
\label{r2def5}
\end{equation}%
\begin{equation}
\theta _{\mu \nu }\left( \bar{\Omega}\left( \check{\Omega}\right) \right)
=0,\qquad \chi _{\mu \nu }\left( \bar{\Omega}\left( \check{\Omega}\right)
\right) =0.  \label{r2def6}
\end{equation}

Finally, we investigate the third-order reducibility, for which the
transformations (\ref{red3par}) can be written as%
\begin{equation}
\check{\epsilon}^{\mu \nu \rho \lambda }\left( \hat{\Omega}\right)
=-5D_{\sigma }\hat{\epsilon}^{\mu \nu \rho \lambda \sigma },\qquad \check{%
\epsilon}^{\mu \nu \rho \lambda \sigma }\left( \hat{\Omega}\right) =2\lambda
W_{1}\hat{\epsilon}^{\mu \nu \rho \lambda \sigma },  \label{tpar2r2}
\end{equation}%
while that the third-order reducibility relations (\ref{red3def}) are listed
below
\begin{gather}
\bar{\epsilon}^{\mu \nu \rho }\left( \check{\Omega}\left( \hat{\Omega}%
\right) \right) =20\lambda \left( \frac{\delta S^{\mathrm{L}}}{\delta
H^{\lambda }}\frac{d^{2}W_{1}}{d\varphi ^{2}}V_{\sigma }+\frac{\delta S^{%
\mathrm{L}}}{\delta B^{\lambda \sigma }}\frac{dW_{1}}{d\varphi }\right) \hat{%
\epsilon}^{\mu \nu \rho \lambda \sigma },  \label{r3def1} \\
\bar{\epsilon}^{\mu \nu \rho \lambda }\left( \check{\Omega}\left( \hat{\Omega%
}\right) \right) =-10\lambda \frac{\delta S^{\mathrm{L}}}{\delta H^{\sigma }}%
\frac{dW_{1}}{d\varphi }\hat{\epsilon}^{\mu \nu \rho \lambda \sigma },
\label{r3def2} \\
\bar{\xi}\left( \check{\Omega}\left( \hat{\Omega}\right) \right) =0,\qquad
\bar{\xi}^{\mu \nu \rho \lambda \sigma }\left( \check{\Omega}\left( \hat{%
\Omega}\right) \right) =0,\qquad \bar{\theta}_{\mu }\left( \check{\Omega}%
\left( \hat{\Omega}\right) \right) =0.  \label{r3def3}
\end{gather}


\begin{thebibliography}{99}
\bibitem{birmingham91} D. Birmingham, M. Blau, M. Rakowski and G. Thompson,
Topological field theory, \textit{Phys. Rept.} \textbf{209} (1991) 129--340.

\bibitem{labastida97} J. M. F. Labastida and C. Lozano, Lectures on
topological quantum field theory, in \textit{Proceedings of La
Plata-CERN-Santiago de Compostela Meeting on Trends in Theoretical Physics,
La Plata, Argentina, April-May 1997,} eds. H. Falomir, R. E. Gamboa Sarav%
\'{\i}, F. A. Schaposnik (AIP, New York 1998), AIP Conference Proceedings
vol. \textbf{419}, 54--93 [arXiv:hep-th/9709192].

\bibitem{stroblspec} P. Schaller and T. Strobl, Poisson structure induced
(topological) field theories, \textit{Mod. Phys. Lett.} \textbf{A9} (1994)
3129--3136 [arXiv:hep-th/9405110].

\bibitem{psmikeda94} N. Ikeda, Two-dimensional gravity and nonlinear gauge
theory, \textit{Annals Phys.} \textbf{235} (1994) 435--464
[arXiv:hep-th/9312059].

\bibitem{psmstrobl95} A. Yu. Alekseev, P. Schaller and T. Strobl,
Topological $G/G$ WZW model in the generalized momentum representation,
\textit{Phys. Rev.} \textbf{D52} (1995) 7146--7160 [arXiv:hep-th/9505012].

\bibitem{psmstroblCQG961} T. Kl\"{o}sch and T. Strobl, Classical and quantum
gravity in 1+1 dimensions: I. A unifying approach, \textit{Class. Quantum
Grav.} \textbf{13} (1996) 965--983 [arXiv:gr-qc/9508020]; Erratum-ibid.
\textbf{14} (1997) 825.

\bibitem{psmstroblCQG962} T. Kl\"{o}sch and T. Strobl, Classical and quantum
gravity in 1+1 dimensions. II: The universal coverings, \textit{Class.
Quantum Grav.} \textbf{13} (1996) 2395--2421 [arXiv:gr-qc/9511081].

\bibitem{psmstrobl97} T. Kl\"{o}sch and T. Strobl, Classical and quantum
gravity in 1+1 dimensions: III. Solutions of arbitrary topology, \textit{%
Class. Quantum Grav.} \textbf{14} (1997) 1689--1723 [arXiv:hep-th/9607226].

\bibitem{psmcattaneo2000} A. S. Cattaneo and G. Felder, A path integral
approach to the Kontsevich quantization formula, \textit{Commun. Math. Phys.}
\textbf{212} (2000) 591--611 [arXiv:math/9902090].

\bibitem{psmcattaneo2001} A. S. Cattaneo and G. Felder, Poisson sigma models
and deformation quantization, \textit{Mod. Phys. Lett.} \textbf{A16} (2001)
179--189 [arXiv:hep-th/0102208].

\bibitem{grav2teit83} C. Teitelboim, Gravitation and hamiltonian structure
in two spacetime dimensions, \textit{Phys. Lett.} \textbf{B126} (1983)
41--45.

\bibitem{grav2jackiw85} R. Jackiw, Lower dimensional gravity, \textit{Nucl.
Phys.} \textbf{B252} (1985) 343--356.

\bibitem{grav2katanaev86} M. O. Katanayev and I. V. Volovich, String model
with dynamical geometry and torsion, \textit{Phys. Lett.} \textbf{B175}
(1986) 413--416 [arXiv:hep-th/0209014].

\bibitem{grav2brown88} J. Brown, \textit{Lower Dimensional Gravity,} World
Scientific, Singapore 1988.

\bibitem{grav2katanaev90} M. O. Katanaev and I. V. Volovich, Two-dimensional
gravity with dynamical torsion and strings, \textit{Annals Phys.} \textbf{197%
} (1990) 1--32.

\bibitem{grav2schmidt} H.-J. Schmidt, Scale-invariant gravity in two
dimensions, \textit{J. Math. Phys.} \textbf{32} (1991) 1562--1566.

\bibitem{grav2solod} S. N. Solodukhin, Topological 2D Riemann-Cartan-Weyl
gravity, \textit{Class. Quantum Grav.} \textbf{10} (1993) 1011--1021.

\bibitem{grav2ikedaizawa90} N. Ikeda and K. I. Izawa, General form of
dilaton gravity and nonlinear gauge theory, \textit{Prog. Theor. Phys.}
\textbf{90} (1993) 237--245 [arXiv:hep-th/9304012].

\bibitem{grav2strobl94} T. Strobl, Dirac quantization of gravity-Yang-Mills
systems in 1+1 dimensions, \textit{Phys. Rev.} \textbf{D50} (1994)
7346--7350 [arXiv:hep-th/9403121].

\bibitem{grav2grumvassil02} D. Grumiller, W. Kummer and D. V. Vassilevich,
Dilaton gravity in two dimensions, \textit{Phys. Rept.} \textbf{369} (2002)
327--430 [arXiv:hep-th/0204253].

\bibitem{grav2strobl00} T. Strobl, \textit{Gravity in two space-time
dimensions,} Habilitation thesis RWTH Aachen, May 1999, arXiv:hep-th/0011240.

\bibitem{ezawa} K. Ezawa, Ashtekar's formulation for N = 1, 2 supergravities
as ``constrained'' BF theories, \textit{Prog. Theor. Phys. }\textbf{95}
(1996) 863--882 [arXiv:hep-th/9511047].

\bibitem{freidel} L. Freidel, K. Krasnov and R. Puzio, BF description of
higher-dimensional gravity theories, \textit{Adv. Theor. Math. Phys.}
\textbf{3} (1999) 1289--1324 [arXiv:hep-th/9901069].

\bibitem{smolin} L. Smolin, Holographic formulation of quantum general
relativity, \textit{Phys. Rev.} \textbf{D61} (2000) 084007
[arXiv:hep-th/9808191].

\bibitem{ling} Y. Ling and L. Smolin, Holographic formulation of quantum
supergravity, \textit{Phys. Rev.} \textbf{D63} (2001) 064010
[arXiv:hep-th/0009018].

\bibitem{defBFizawa2000} K.-I. Izawa, On nonlinear gauge theory from a
deformation theory perspective, \textit{Prog. Theor. Phys.} \textbf{103}
(2000) 225--228 [arXiv:hep-th/9910133].

\bibitem{defBFmpla} C. Bizdadea, Note on two-dimensional nonlinear gauge
theories, \textit{Mod. Phys. Lett.} \textbf{A15} (2000) 2047--2055
[arXiv:hep-th/0201059].

\bibitem{defBFikeda00} N. Ikeda, A deformation of three dimensional BF
theory, \textit{J.High Energy Phys.} JHEP\textbf{11}(2000)009
[arXiv:hep-th/0010096].

\bibitem{defBFikeda01} N. Ikeda, Deformation of BF theories, topological
open membrane and a generalization of the star deformation, \textit{J. High
Energy Phys.} JHEP\textbf{07}(2001)037 [arXiv:hep-th/0105286].

\bibitem{defBFijmpa} C. Bizdadea, E. M. Cioroianu and S. O. Saliu,
Hamiltonian cohomological derivation of four-dimensional nonlinear gauge
theories, \textit{Int. J. Mod. Phys.} \textbf{A17} (2002) 2191--2210
[arXiv:hep-th/0206186].

\bibitem{defBFjhep} C. Bizdadea, C. C. Ciob\^{\i}rc\u{a}, E. M. Cioroianu,
S. O. Saliu and S. C. S\u{a}raru, Hamiltonian BRST deformation of a class of
n-dimensional BF-type theories, \textit{J. High Energy Phys.} JHEP\textbf{01}%
(2003)049 [arXiv:hep-th/0302037].

\bibitem{juviBFD5} E. M. Cioroianu and S. C. S\u{a}raru, Self-interactions
in a topological BF-type model in D = 5, \textit{J. High Energy Phys.} JHEP%
\textbf{07}(2005)056 [arXiv:hep-th/0508035].

\bibitem{defBFijmpajuvi06} E. M. Cioroianu and S. C. S\u{a}raru, PT-symmetry
breaking Hamiltonian interactions in BF models, \textit{Int. J. Mod. Phys.}
\textbf{A21} (2006) 2573--2599 [arXiv:hep-th/0606164].

\bibitem{defBFikeda03} N. Ikeda, Chern--Simons gauge theory coupled with BF
theory, \textit{Int. J. Mod. Phys.} \textbf{A18} (2003) 2689--2702
[arXiv:hep-th/0203043].

\bibitem{defBFijmpajuvi04} E. M. Cioroianu and S. C. S\u{a}raru,
Two-dimensional interactions between a BF-type theory and a collection of
vector fields, \textit{Int. J. Mod. Phys.} \textbf{A19} (2004) 4101--4125
[arXiv:hep-th/0501056].

\bibitem{defBFepjc} C. Bizdadea, E. M. Cioroianu, S. O. Saliu and S. C. S%
\u{a}raru, Couplings of a collection of BF models to matter theories,
\textit{Eur. Phys. J.} \textbf{C41} (2005) 401--420 [arXiv:hep-th/0508037].

\bibitem{defBFjhep06} C. Bizdadea, E. M. Cioroianu, I. Negru, S. O. Saliu
and S. C. S\u{a}raru, On the generalized Freedman-Townsend model, \textit{J.
High Energy Phys.} JHEP\textbf{10}(2006)004 [arXiv:0704.3407(hep-th)].

\bibitem{defBFRSepjc} C. Bizdadea, E. M. Cioroianu, S. O. Saliu, S. C. S\u{a}%
raru and M. Iordache, Four-dimensional couplings among BF and massless
Rarita--Schwinger theories: a BRST cohomological approach, \textit{Eur.
Phys. J.} \textbf{C58} (2008) 123--149 [arXiv:0812.3810(hep-th)].

\bibitem{deflag} G. Barnich and M. Henneaux, Consistent couplings between
fields with a gauge freedom and deformations of the master equation, \textit{%
Phys. Lett.} \textbf{B311} (1993) 123--129 [arXiv:hep-th/9304057].

\bibitem{17and5} M. Henneaux, Consistent interactions between gauge fields:
the cohomological approach, \textit{Contemp. Math.} \textbf{219} (1998) 93
[arXiv:hep-th/9712226].

\bibitem{defham} C. Bizdadea, Consistent interactions in the Hamiltonian
BRST formalism, \textit{Acta Phys. Polon.} \textbf{B32} (2001) 2843--2862
[arXiv:hep-th/0003199].

\bibitem{gen1} G. Barnich, F. Brandt and M. Henneaux, Local BRST cohomology
in the antifield formalism: I. General theorems, \textit{Commun. Math. Phys.}
\textbf{174} (1995) 57--91 [arXiv:hep-th/9405109].

\bibitem{gen2} G. Barnich, F. Brandt and M. Henneaux, Local BRST cohomology
in the antifield formalism: II. Application to Yang-Mills theory, \textit{%
Commun. Math. Phys.} \textbf{174} (1995) 93--116 [arXiv:hep-th/9405194].

\bibitem{gen} G. Barnich, F. Brandt and M. Henneaux, Local BRST cohomology
in gauge theories, \textit{Phys. Rept.} \textbf{338} (2000) 439--569
[arXiv:hep-th/0002245].

\bibitem{otherBFikeda02} N. Ikeda, Topological field theories and geometry
of Batalin-Vilkovisky algebras, \textit{J. High Energy Phys.} JHEP\textbf{10}%
(2002)076 [arXiv:hep-th/0209042].

\bibitem{otherBFikedaizawa04} N. Ikeda and K.-I. Izawa, Dimensional
reduction of nonlinear gauge theories, \textit{J. High Energy Phys.} JHEP%
\textbf{09}(2004)030 [arXiv:hep-th/0407243].

\bibitem{curt2} T. Curtright and P. G. O. Freund, Massive dual fields,
\textit{Nucl. Phys.} \textbf{B172} (1980) 413--424.

\bibitem{curt1} T. Curtright, Generalized gauge fields, \textit{Phys. Lett.}
\textbf{B165} (1985) 304--308.

\bibitem{aul} C. S. Aulakh, I. G. Koh and S. Ouvry, Higher spin fields with
mixed symmetry, \textit{Phys. Lett.} \textbf{B173} (1986) 284--288.

\bibitem{labast1} J. M. Labastida and T. R. Morris, Massless mixed-symmetry
bosonic free fields, \textit{Phys. Lett.} \textbf{B180} (1986) 101--106.

\bibitem{labast2} J. M. Labastida, Massless particles in arbitrary
representations of the Lorentz group, \textit{Nucl. Phys.} \textbf{B322}
(1989) 185--209.

\bibitem{burd} C. Burdik, A. Pashnev and M. Tsulaia, On the mixed symmetry
irreducible representations of the Poincar\'{e} group in the BRST approach,
\textit{Mod. Phys. Lett.} \textbf{A16} (2001) 731--746
[arXiv:hep-th/0101201].

\bibitem{zinov1} Yu. M. Zinoviev, On massive mixed symmetry tensor fields in
Minkowski space and $(A)dS$ [arXiv:hep-th/0211233].

\bibitem{dualsp1} C. M. Hull, Duality in gravity and higher spin gauge
fields, \textit{J. High Energy Phys.} JHEP\textbf{09}(2001)027
[arXiv:hep-th/0107149].

\bibitem{dualsp2a} X. Bekaert and N. Boulanger, Massless spin-two field
S-duality, \textit{Class. Quantum Grav.} \textbf{20} (2003) S417--S423
[arXiv:hep-th/0212131].

\bibitem{dualsp2b} X. Bekaert and N. Boulanger, On geometric equations and
duality for free higher spins, \textit{Phys. Lett.} \textbf{B561} (2003)
183--190 [arXiv:hep-th/0301243].

\bibitem{dualsp3} H. Casini, R. Montemayor and L. F. Urrutia, Duality for
symmetric second rank tensors. II. The linearized gravitational field,
\textit{Phys. Rev.} \textbf{D68} (2003) 065011 [arXiv:hep-th/0304228].

\bibitem{dualsp4} N. Boulanger, S. Cnockaert and M. Henneaux, A note on
spin-s duality, \textit{J. High Energy Phys. }JHEP\textbf{06}(2003)060
[arXiv:hep-th/0306023].

\bibitem{dualsp5} P. de Medeiros and C. Hull, Exotic tensor gauge theory and
duality, \textit{Commun. Math. Phys.} \textbf{235} (2003) 255--273
[arXiv:hep-th/0208155].

\bibitem{dualsp2} X. Bekaert and N. Boulanger, Tensor gauge fields in
arbitrary representations of $GL(D,\mathbb{R})$. Duality and Poincar\'{e}
Lemma, \textit{Commun. Math. Phys.} \textbf{245} (2004) 27--67
[arXiv:hep-th/0208058].

\bibitem{dualsp2II} X. Bekaert and N. Boulanger, Tensor gauge fields in
arbitrary representations of $GL(D,\mathbb{R})$: II. Quadratic actions,
\textit{Commun. Math. Phys.} \textbf{271} (2007) 723--773
[arXiv:hep-th/0606198].

\bibitem{lingr} X. Bekaert, N. Boulanger and M. Henneaux, Consistent
deformations of dual formulations of linearized gravity: A no-go result,
\textit{Phys. Rev.} \textbf{D67} (2003) 044010 [arXiv:hep-th/0210278].

\bibitem{zinov2} Yu. M. Zinoviev, First order formalism for mixed symmetry
tensor fields [arXiv:hep-th/0304067].

\bibitem{zinov3} Yu. M. Zinoviev, First order formalism for massive mixed
symmetry tensor fields in Minkowski and $(A)dS$ spaces
[arXiv:hep-th/0306292].

\bibitem{boulangerCQG} N. Boulanger and L. Gualtieri, An exotic theory of
massless spin-2 fields in three dimensions, \textit{Class. Quantum Grav.}
\textbf{18} (2001) 1485--1502 [arXiv:hep-th/0012003].

\bibitem{ancoPRD} S. C. Anco, Parity violating spin-two gauge theories,
\textit{Phys. Rev.} \textbf{D67} (2003) 124007 [arXiv:gr-qc/0305026].

\bibitem{high1} A. K. Bengtsson, I. Bengtsson and L. Brink, Cubic
interaction terms for arbitrary spin, \textit{Nucl. Phys.} \textbf{B227}
(1983) 31--40.

\bibitem{high2} M. A. Vasiliev, Cubic interactions of bosonic higher spin
gauge fields in $AdS_5$, \textit{Nucl. Phys.} \textbf{B616} (2001) 106--162
[arXiv:hep-th/0106200]; Erratum-ibid. \textbf{B652} (2003) 407.

\bibitem{high3} E. Sezgin and P. Sundell, $7D$ bosonic higher spin gauge
theory: symmetry algebra and linearized constraints, \textit{Nucl. Phys.}
\textbf{B634} (2002) 120--140 [arXiv:hep-th/0112100].

\bibitem{high4} D. Francia and A. Sagnotti, Free geometric equations for
higher spins, \textit{Phys. Lett.} \textbf{B543} (2002) 303--310
[arXiv:hep-th/0207002].

\bibitem{ccjhep} C. Bizdadea, C. C. Ciob\^{\i}rc\u{a}, E. M. Cioroianu, I.
Negru, S. O. Saliu and S. C. S\u{a}raru, Interactions of a single massless
tensor field with the mixed symmetry (3,1). No-go results, \textit{J. High
Energy Phys.} JHEP\textbf{10}(2003)019.

\bibitem{9} N. Boulanger and S. Cnockaert, Consistent deformations of
[p,p]--type gauge field theories, \textit{J. High Energy Phys.} JHEP\textbf{%
03}(2004)031 [arXiv:hep-th/0402180].

\bibitem{kk} C. C. Ciob\^{\i}rc\u{a}, E. M. Cioroianu and S. O. Saliu,
Cohomological BRST aspects of the massless tensor field with the mixed
symmetry (k,k), \textit{Int. J. Mod. Phys.} \textbf{A19} (2004) 4579--4619
[arXiv:hep-th/0403017].

\bibitem{ccepjc} C. Bizdadea, C. C. Ciob\^{\i}rc\u{a}, E. M. Cioroianu, S.
O. Saliu and S. C. S\u{a}raru, Interactions of a massless tensor field with
the mixed symmetry of the Riemann tensor. No-go results, \textit{Eur. Phys.
J.} \textbf{C36} (2004) 253--270 [arXiv:hep-th/0306154].

\bibitem{7} X. Bekaert, N. Boulanger and S. Cnockaert, No self-interaction
for two-column massless fields, \textit{J. Math. Phys.} \textbf{46} (2005)
012303 [arXiv:hep-th/0407102].

\bibitem{3} N. Boulanger, S. Leclercq and S. Cnockaert, Parity-violating
vertices for spin-3 gauge fields, \textit{Phys.Rev.} \textbf{D73} (2006)
065019 [arXiv:hep-th/0509118].

\bibitem{4} X. Bekaert, N. Boulanger and S. Cnockaert, Spin three gauge
theory revisited, \textit{J. High Energy Phys. }JHEP\textbf{01}(2006)052
[arXiv:hep-th/0508048].

\bibitem{ccprd} C. Bizdadea, C. C. Ciob\^{\i}rc\u{a}, I. Negru and S. O.
Saliu, Couplings between a single massless tensor field with the mixed
symmetry (3,1) and one vector field, \textit{Phys.Rev.} \textbf{D74} (2006)
045031 [arXiv:0705.1048(hep-th)].

\bibitem{ccjpa} C. Bizdadea, C. C. Ciob\^{\i}rc\u{a}, E. M. Cioroianu and S.
O. Saliu, Interactions between a massless tensor field with the mixed
symmetry of the Riemann tensor and a massless vector field, \textit{J. Phys.
A: Math. Gen.} \textbf{39} (2006) 10549--10564 [arXiv:0705.1054(hep-th)].

\bibitem{dcjpa} C. Bizdadea, D. Cornea and S. O. Saliu, No
cross-interactions among different tensor fields with the mixed symmetry (3,
1) intermediated by a vector field, \textit{J. Phys. A: Math. Theor.}
\textbf{41} (2008) 285202 [arXiv:0901.4059(hep-th)].
\end{thebibliography}
\end{document}